\def\blankline{\vskip \baselineskip}
\def\DF{{\tenrm DF}}
\def\Msun{{M$_\odot$}}

\def\per{{$^{-1}$}}
\def\degrees{{$^\circ$}}
\def\half{{\leavevmode\kern.1em\raise.5ex\hbox{\the\scriptfont0 1}\kern-.1em
/\kern-.1em\lower.25ex\hbox{\the\scriptfont0 2}}} 
\def\quarter{{\leavevmode\kern.1em\raise.5ex\hbox{\the\scriptfont0 1}\kern-.1em
/\kern-.1em\lower.25ex\hbox{\the\scriptfont0 4}}}

\def\spose#1{\hbox to 0pt{#1\hss}} 
\def\ltsim{\mathrel{\spose{\lower.5ex\hbox{$\mathchar"218$}}
     \raise.4ex\hbox{$\mathchar"13C$}}}
\def\gtsim{\mathrel{\spose{\lower.5ex \hbox{$\mathchar"218$}}
     \raise.4ex\hbox{$\mathchar"13E$}}}
\def\gtlt{\mathrel{\spose{\lower.5ex\hbox{$\mathchar"13E$}}
     \raise.5ex\hbox{$\mathchar"13C$}}}

\def\pmb#1{\setbox0=\hbox{$#1$}%
  \kern-0.25em\copy0\kern-\wd0
  \kern.05em\copy0\kern-\wd0
  \kern-0.025em\raise.0433em\box0}
\def\spmb#1{\setbox1=\hbox{${\scriptstyle #1}$}%
  \kern-0.25em\copy1\kern-\wd1
  \kern.05em\copy1\kern-\wd1
  \kern-0.025em\raise.0433em\box1}

\def\today{\count99=\day
           \ifnum\count99>20 \count98=\day
                             \divide\count98 by 10
                             \multiply\count98 by 10
                             \advance\count99 by -\count98 \fi
           \number\day\ifcase\count99 th\or st\or nd\or rd\else th\fi
           ~\ifcase\month none\or January\or February\or March\or April\or
                  May\or June\or July\or August\or September\or October\or
                  November\or December\fi
           ~\number\year}

\long\def\Ignore#1{\relax}%

\newdimen\digitwidth
\setbox0=\hbox{0}
\digitwidth=\wd0
%
%

\def\cf{{\it cf.}}
\def\eg{{\it e.g.}}
\def\etal{{\it et al.}}
\def\etc{{\it etc.}}
\def\ie{{\it i.e.}}
\def\nb{{\it n.b.}}

\newcount\linespacingstep
\newcount\linespacing
\linespacingstep=1
\def\multiplelines{\linespacing=\linespacingstep
   \advance\linespacing by 2
   \multiply\normalbaselineskip by \linespacing
   \advance\normalbaselineskip by 2pt 
   \divide\normalbaselineskip by 3}

\font\fiverm=cmr5	\font\sixrm=cmr6	\font\sevenrm=cmr7
\font\eightrm=cmr8 	\font\ninerm=cmr9	\font\tenrm=cmr10
\font\twelverm=cmr12 	

\font\fivei=cmmi5 	\font\sixi=cmmi6 	\font\seveni=cmmi7
\font\eighti=cmmi8 	\font\ninei=cmmi9 	\font\teni=cmmi10
\font\twelvei=cmmi12

\font\fivesy=cmsy5  	\font\sixsy=cmsy6 	\font\sevensy=cmsy7
\font\eightsy=cmsy8 	\font\ninesy=cmsy9 	\font\tensy=cmsy10
\font\magnifiedtensy=cmsy10 at 12pt

\font\fivebf=cmbx5  	\font\sixbf=cmbx6 	\font\sevenbf=cmbx7
\font\eightbf=cmbx8 	\font\ninebf=cmbx9 	\font\tenbf=cmbx10
\font\twelvebf=cmbx12

	\font\eightit=cmti8 	\font\nineit=cmti9
\font\tenit=cmti10	\font\twelveit=cmti12

\font\eightsl=cmsl8 	\font\ninesl=cmsl9	\font\tensl=cmsl10
\font\twelvesl=cmsl12

\font\eighttt=cmtt8 	\font\ninett=cmtt9 	\font\tentt=cmtt10
\font\twelvett=cmtt12

\font\tenex=cmex10

\def\eightpoint{\def\rm{\fam0\eightrm}
\textfont0=\eightrm\scriptfont0=\sixrm\scriptscriptfont0=\fiverm
\textfont1=\eighti\scriptfont1=\sixi\scriptscriptfont1=\fivei
\textfont2=\eightsy\scriptfont2=\sixsy\scriptscriptfont2=\fivesy
\textfont3=\tenex\scriptfont3=\tenex\scriptscriptfont3=\tenex
\textfont\itfam=\eightit\def\it{\fam\itfam\eightit}
\textfont\slfam=\eightsl\def\sl{\fam\slfam\eightsl}
\textfont\ttfam=\eighttt\def\tt{\fam\ttfam\eighttt}
\textfont\bffam=\eightbf\scriptfont\bffam=\sixbf
\scriptscriptfont\bffam=\fivebf\def\bf{\fam\bffam\eightbf}
\normalbaselineskip=9pt
\ifnum\linespacingstep>1\multiplelines\fi
\setbox\strutbox=\hbox{\vrule height7pt depth2pt width0pt}
\let\sc=\sixrm\let\big=\eightbig\normalbaselines\rm}

\def\ninepoint{\def\rm{\fam0\ninerm}
\textfont0=\ninerm\scriptfont0=\sixrm\scriptscriptfont0=\fiverm
\textfont1=\ninei\scriptfont1=\sixi\scriptscriptfont1=\fivei
\textfont2=\ninesy\scriptfont2=\sixsy\scriptscriptfont2=\fivesy
\textfont3=\tenex\scriptfont3=\tenex\scriptscriptfont3=\tenex
\textfont\itfam=\nineit\def\it{\fam\itfam\nineit}
\textfont\slfam=\ninesl\def\sl{\fam\slfam\ninesl}
\textfont\ttfam=\ninett\def\tt{\fam\ttfam\ninett}
\textfont\bffam=\ninebf\scriptfont\bffam=\sixbf
\scriptscriptfont\bffam=\fivebf\def\bf{\fam\bffam\ninebf}
\normalbaselineskip=11pt
\ifnum\linespacingstep>1\multiplelines\fi
\setbox\strutbox=\hbox{\vrule height8pt depth3pt width0pt}
\let\sc=\sevenrm\let\big=\ninebig\normalbaselines\rm}

\def\tenpoint{\def\rm{\fam0\tenrm}
\textfont0=\tenrm\scriptfont0=\sevenrm\scriptscriptfont0=\fiverm%
\textfont1=\teni\scriptfont1=\seveni\scriptscriptfont1=\fivei%
\textfont2=\tensy\scriptfont2=\sevensy\scriptscriptfont2=\fivesy%
\textfont3=\tenex\scriptfont3=\tenex\scriptscriptfont3=\tenex%
\textfont\itfam=\tenit\def\it{\fam\itfam\tenit}%
\textfont\slfam=\tensl\def\sl{\fam\slfam\tensl}%
\textfont\ttfam=\tentt\def\tt{\fam\ttfam\tentt}%
\textfont\bffam=\tenbf\scriptfont\bffam=\sevenbf%
\scriptscriptfont\bffam=\fivebf\def\bf{\fam\bffam\tenbf}%
\normalbaselineskip=12pt%
\ifnum\linespacingstep>1\multiplelines\fi
\setbox\strutbox=\hbox{\vrule height8.5pt depth3.5pt width0pt}%
\let\sc=\eightrm\let\big=\tenbig\normalbaselines\rm}

\def\twelvepoint{\def\rm{\fam0\twelverm}
\textfont0=\twelverm\scriptfont0=\eightrm\scriptscriptfont0=\sixrm
\textfont1=\twelvei\scriptfont1=\eighti\scriptscriptfont1=\sixi
\textfont2=\magnifiedtensy\scriptfont2=\eightsy\scriptscriptfont2=\sixsy
\textfont3=\tenex\scriptfont3=\tenex\scriptscriptfont3=\tenex
\textfont\itfam=\twelveit\def\it{\fam\itfam\twelveit}
\textfont\slfam=\twelvesl\def\sl{\fam\slfam\twelvesl}
\textfont\ttfam=\twelvett\def\tt{\fam\ttfam\twelvett}
\textfont\bffam=\twelvebf\scriptfont\bffam=\eightbf
\scriptscriptfont\bffam=\sixbf\def\bf{\fam\bffam\twelvebf}
\tt 
\normalbaselineskip=14pt
\ifnum\linespacingstep>1\multiplelines\fi
\setbox\strutbox=\hbox{\vrule height10pt depth5pt width0pt}
\let\sc=\eightrm\let\big=\twelvebig\normalbaselines\rm}

\twelvepoint

\vsize=22.6 true cm \hsize=17 true cm 
\clubpenalty=5000
\widowpenalty=5000

\font\headfont=cmti12
\font\titlefont=cmbx10 at 14.4pt
\font\authorfont=cmr12

\font\lonefont=cmbx12
\font\ltwofont=cmti12
\font\lthreefont=cmti12

\def\textpar{\nobreak\medskip\noindent\ignorespaces}

\def\sect#1{\par \goodbreak\bigskip
\noindent{\lonefont\nextsect.~#1}\par \medskip
\textpar}

\def\subsect#1{\par\goodbreak\bigskip
\noindent{\ltwofont\nextsub.~#1}\par \smallskip
\textpar}

\def\subsubsect#1{\par\goodbreak\bigskip
\noindent{\lthreefont\nextssub.~#1.} \hskip3ex \ignorespaces}

\countdef\sectcount=11
\countdef\subcount=12
\countdef\ssubcount=13
\def\nextsect{\global\advance\sectcount by 1
        \number\sectcount \global\subcount=0}
\def\nextsub{\global\advance\subcount by 1 \number\sectcount.\number\subcount
        \global\ssubcount=0}
\def\nextssub{\global\advance\ssubcount by 1
\number\sectcount.\number\subcount.\number\ssubcount}
\sectcount=0

\countdef\eqcount=14
\eqcount=0
\def\equno{\global\advance\eqcount by 1 \number\eqcount}
\def\equat#1{\count99=\eqcount \advance\count99 by #1 \number\count99}

\countdef\figcount=15
\figcount=0
\def\nextfig{\global\advance\figcount by 1 Figure~\number\figcount}
\def\figno#1{\count99=\figcount \advance\count99 by #1 Figure~\number\count99}

\countdef\footcount=16
\footcount=0
\def\nextfoot#1{\global\advance\footcount by 1
   \footnote{$^{\number\footcount}$}{\tenpoint #1\par}}

\def\refs{\parskip=0pt\par\goodbreak\bigskip
\parindent=0pt\everypar{\hangindent 1cm}
\lonefont References\par\medskip
\nobreak\vskip 6pt\tenpoint

\def\AAp{{\it Astron. Astrophys.} }
\def\AApS{{\it Astron. Astrophys. Suppl.} }
\def\AJ{{\it Astron. J.} }
\def\AnnRev{{\it Ann. Rev. Astron. and Astrophys.} }
\def\ApJ{{\it Astrophys. J.} }
\def\ApJL{{\it Astrophys. J. Lett.} }
\def\ApJS{{\it Astrophys. J. Suppl.} }
\def\ApSS{{\it Astrophys. Sp. Sci.} }
\def\BAN{{\it Bull. Astron. Inst. Netherlands} }
\def\CelMech{{\it Celestial Mechanics} }
\def\JCP{{\it J. Comp. Phys.} }
\def\Kluwer{(Dordrecht: Kluwer)}
\def\Messenger{{\it ESO Messenger} }
\def\MNRAS{{\it Mon. Not. R. Astron. Soc.} }
\def\Nature{{\it Nature\/} }
\def\Obs{{\it Observatory\/} }
\def\PASJ{{\it Publs. Astron. Soc. Japan\/} }
\def\PASP{{\it Publs. Astron. Soc. Pacific\/} }
\def\PhD{{\it PhD thesis\/} }
\def\PhilTrans{{\it Phil. Trans. R. Soc. London A\/} }
\def\PhysFl{{\it Phys. Fluids\/} }
\def\PhysRep{{\it Phys. Reports} }
\def\Reidel{(Dordrecht: Reidel)}
\def\RMP{{\it Rev. Mod. Phys.} }
\def\RPP{{\it Rep. Prog. Phys.} }
\def\SovAst{{\it Soviet Astr.} }
\def\SovAstL{{\it Soviet Astr. Letters} }
\def\Vistas{{\it Vistas Astron.} }

}

\vsize=22.6 true cm \hsize=17 true cm 

\def\EJ{E_{\rm J}}
\def\Ja{J_{\rm a}}
\def\Jr{J_{\rm r}}
\def\kappaz{\kappa_{\rm z}}
\def\Omegaa{\Omega_{\rm a}}
\def\Omegac{\Omega_{\rm c}}
\def\Omegap{\Omega_{\rm p}}
\def\Omegaf{\Omega_{\rm frame}}
\def\Omegar{\Omega_{\rm r}}
\def\Phieff{\Phi_{\rm eff}}
\def\rc{r_{\rm c}}
\def\wa{w_{\rm a}}
\def\wr{w_{\rm r}}

\def\bJ{\;\pmb{\mit J}}
\def\bm{\;\pmb{\mit m}}
\def\br{\;\pmb{\mit r}}
\def\bv{\;\pmb{\mit v}}
\def\bw{\;\pmb{\mit w}}
\def\bx{\;\pmb{\mit x}}
\def\bOmega{{\bf \Omega}}
\def\sbm{\;\spmb{\mit m}}
\def\sbw{\;\spmb{\mit w}}

\def\DF{{\tenrm DF}}
\def\ILR{{\tenrm ILR}}
\def\OLR{{\tenrm OLR}}
\def\ZVC{{\tenrm ZVC}}

\footline{\hfil}
\headline{\ifnum\pageno=-1
  \noindent {\tenpoint Reports on Progress in Physics {\bf 56} (1993) 173-255} \hfil
\else
  \ifodd\pageno
    \hskip 2cm {\headfont Barred galaxies} \hfil \tenrm\folio
  \else
    \hbox to 2cm{\tenrm\folio \hfil}{\headfont Sellwood and Wilkinson} \hfil
  \fi
\fi}

\input psfig
\def\caption#1{\par\narrower\tenpoint\noindent{\bf Figure #1.}\enspace}

%
%

\pageno=-1

\topinsert \vskip 1 cm \endinsert

{\parindent=0pt
{\titlefont Dynamics of Barred Galaxies}

\blankline
\blankline
{\authorfont J A Sellwood$^{1,2}$ and A Wilkinson$^3$}

\smallskip
{\tenpoint $^1$Space Telescope Science Institute, 3700 San Martin Drive, 
Baltimore, MD~21218, USA

$^2$Department of Physics and Astronomy, Rutgers University, PO Box 849, 
Piscataway, NJ~08855-0849, USA

$^3$Department of Astronomy, The University, Manchester M13~9PL, England \par}

\blankline
\blankline
{\bf Abstract} \par}

\medskip
\noindent Some 30\% of disc galaxies have a pronounced central bar feature in 
the disc plane and many more have weaker features of a similar kind.
Kinematic data indicate that the bar constitutes a major
non-axisymmetric component of the mass distribution and that the bar
pattern tumbles rapidly about the axis normal to the disc plane.  The
observed motions are consistent with material within the bar streaming
along highly elongated orbits aligned with the rotating major axis.  A
barred galaxy may also contain a spheroidal bulge at its centre,
spirals in the outer disc and, less commonly, other features such as a
ring or lens.  Mild asymmetries in both the light and kinematics are
quite common.

We review the main problems presented by these complicated dynamical
systems and summarize the effort so far made towards their solution,
emphasizing results which appear secure.  Bars are probably formed
through a global dynamical instability of a rotationally supported
galactic disc.  Studies of the orbital structure seem to indicate that
most stars in the bar follow regular orbits but that a small fraction
may be stochastic.  Theoretical work on the three-dimensional
structure of bars is in its infancy, but first results suggest that
bars should be thicker in the third dimension than the disc from which
they formed.  Gas flow patterns within bars seem to be reasonably well
understood, as are the conditions under which straight offset dust
lanes are formed.  However, no observation so far supports the widely
held idea that the spiral arms are the driven response to the bar,
while evidence accumulates that the spiral patterns are distinct
dynamical features having a different pattern speed.  Both the gaseous
and stellar distributions are expected to evolve on a time-scale of
many bar rotation periods.

\vskip 1cm
\noindent Submitted: July 1992, accepted November 1992, appeared February 1993 
\vfill\eject

{\parindent = 0pt
\nopagenumbers
\global\sectcount=0

\def\sect#1#2{\line{\hbox to 0.5cm{\hss \nextsect. ~}#1 \dotfill \hbox to 
1cm{\hss #2}}}
\def\subsect#1#2{\line{\hbox to 1.5cm{\hss \nextsub. ~}#1 \dotfill \hbox to 
1cm{\hss #2}}}
\def\subsubsect#1#2{\line{\hbox to 2.5cm{\hss \nextssub. ~}#1 \dotfill \hbox 
to 1cm{\hss #2}}}

{\bf Table of Contents}

\blankline
\sect{Introduction}{1}

\sect{Observed properties of bars}{7}
\subsect{Components of the light distribution}{7}
\subsect{Fraction of total luminosity in the bar}{7}
\subsect{The light distribution within the bar}{8}
\subsect{Tri-axial bulges and/or nuclear bars}{9}
\subsect{Kinematic properties of bars}{9}
\subsect{Bar angular velocity or pattern speed}{10}
\subsect{Velocity dispersions}{11}

\sect{Stellar dynamics of galaxies}{11}
\subsect{Relaxation time}{12}
\subsect{Dynamical equations}{12}
\subsect{Analytic distribution functions}{13}
\subsect{Near-integrable systems}{14}
\subsect{Linear programming}{15}
\subsect{$N$-body techniques}{15}

\sect{Two-dimensional bar models}{16}
\subsect{Thin disc approximation}{16}
\subsect{Linear theory}{17}
\subsect{Strong bars}{25}
\subsect{Periodic orbits}{29}
\subsect{Periodic orbit families}{30}
\subsect{Non-periodic orbits}{25}
\subsect{Onset of chaos}{37}
\subsect{Actions}{38}
\subsect{Self-consistency}{39}

\sect{Three-dimensional bar models}{41}
\subsect{Vertical resonances}{41}
\subsect{Periodic orbits in three-dimensions}{42}
\subsect{Structure of a three-dimensional $N$-body model}{45}
\subsect{Stochasticity in three-dimensions}{46}

\sect{Gas and dust}{47}
\subsect{Observations of gas in barred galaxies}{48}
\subsect{Modelling the ISM}{49}
\subsect{Streamlines and periodic orbits}{51}
\subsect{Strong bars}{51}
\subsect{Driven spiral arms?}{53}
\subsect{Angular momentum changes}{53}
\subsect{Comparison with observations}{54}

\sect{Rings and lenses}{55}
\subsect{Observed properties of rings}{55}
\subsect{Lenses and oval distortions}{56}
\subsect{Formation of rings}{57}
\subsect{Formation of lenses}{58}

\sect{Asymmetries}{58}
\subsect{Observed properties}{58}
\subsect{Models}{59}

\sect{Origin of bars}{59}
\subsect{Global analysis}{59}
\subsect{Bar-forming modes}{63}
\subsect{Properties of the resulting bars}{64}
\subsect{Mechanism for the mode}{64}
\subsect{Controlling the bar instability}{65}
\subsect{Meta-stability and tidal triggering}{65}
\subsect{An alternative theory of bar formation}{66}

\sect{Evolution of the bar}{66}
\subsect{Buckling instability}{67}
\subsect{Peanut growth}{68}
\subsect{Continuing interactions with the disc}{69}
\subsect{Interations with spheroidal components}{69}
\subsect{Destruction of bars}{70}

\sect{Conclusions}{71}

\line{\hbox to 0.5cm{\hss  ~}Acknowldegments \dotfill \hbox to 1cm{\hss 72}}
\line{\hbox to 0.5cm{\hss  ~}References \dotfill \hbox to 1cm{\hss 72}}

\global\sectcount=0

}

\vfill\eject

\def\equno{\number\eqcount \global\advance\eqcount by 1}
\global\eqcount=1

\pageno=1

\sect{Introduction}
Galaxies are beautiful objects, and the graceful symmetry of many
barred galaxies is particularly striking.  One of the best examples
close enough for us to examine in detail is NGC~1365 in the Fornax
cluster (\nextfig).  Such objects are doubly pleasing to the
astro-physicist because they also present a number of very challenging
dynamical problems.

\pageinsert{
\centerline{\psfig{file=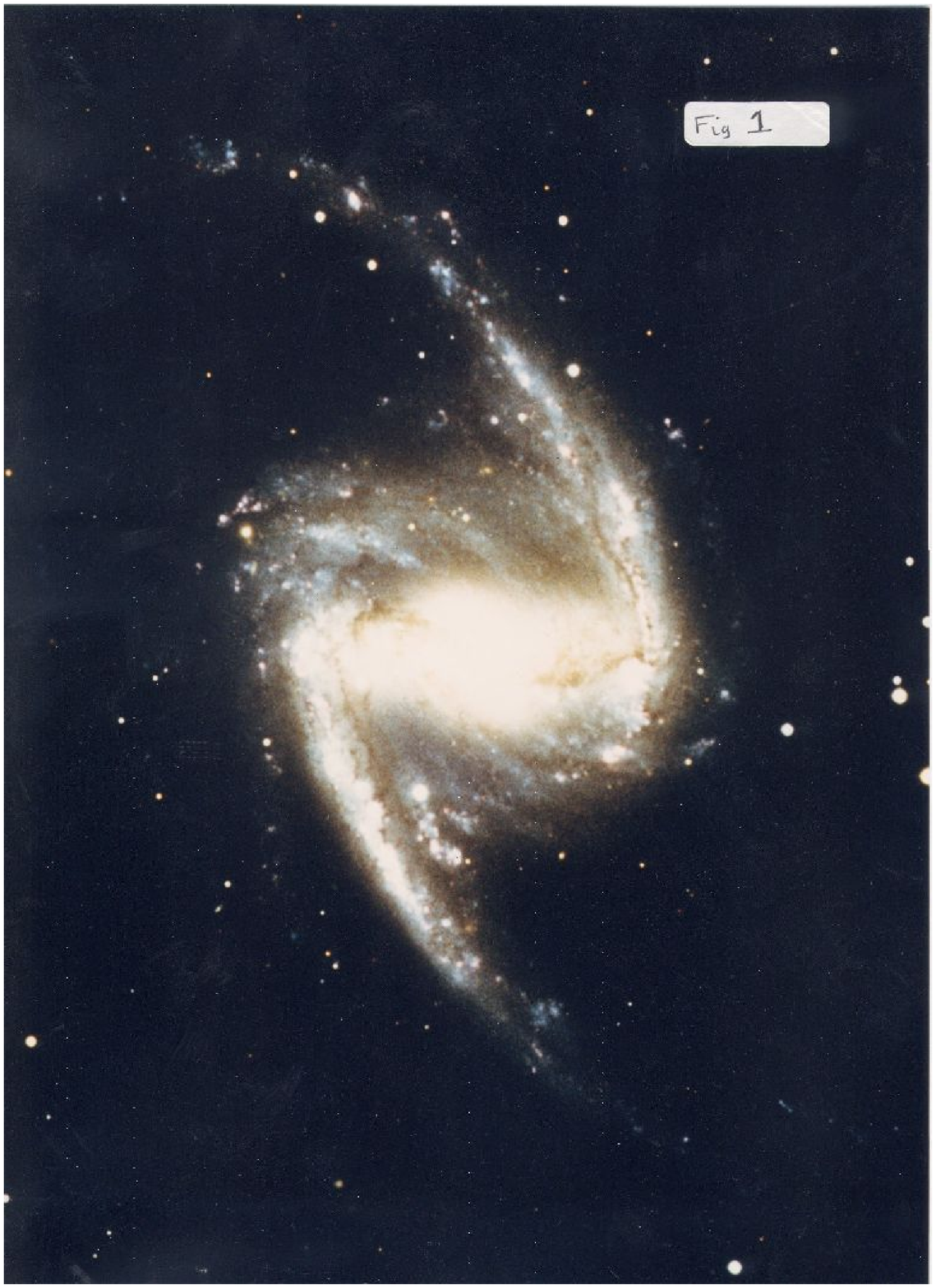,width=.83\hsize,clip=}}
\caption{1}
NGC~1365, a magnificent barred spiral in Fornax.  The galaxy has a
very bright amorphous central bulge, a strong bar joined to a nearly
symmetrical pair of spiral arms.  Pronounced brown dust lanes can be
seen in the spiral arms and on the leading edge of the bar (assuming
the arms to be trailing).  Sites of star formation are tinged in pink
while the youngest stars are blue and old stars are yellow.  As in
most galaxies, the spiral arms are superposed on a much lower surface
brightness disc, which is too faint to be distinguishable from the sky
in this photograph.  (Photograph courtesy of David Malin,
Anglo-Australian Observatory.) \par
\vfill
}\endinsert

\pageinsert{
\centerline{\psfig{file=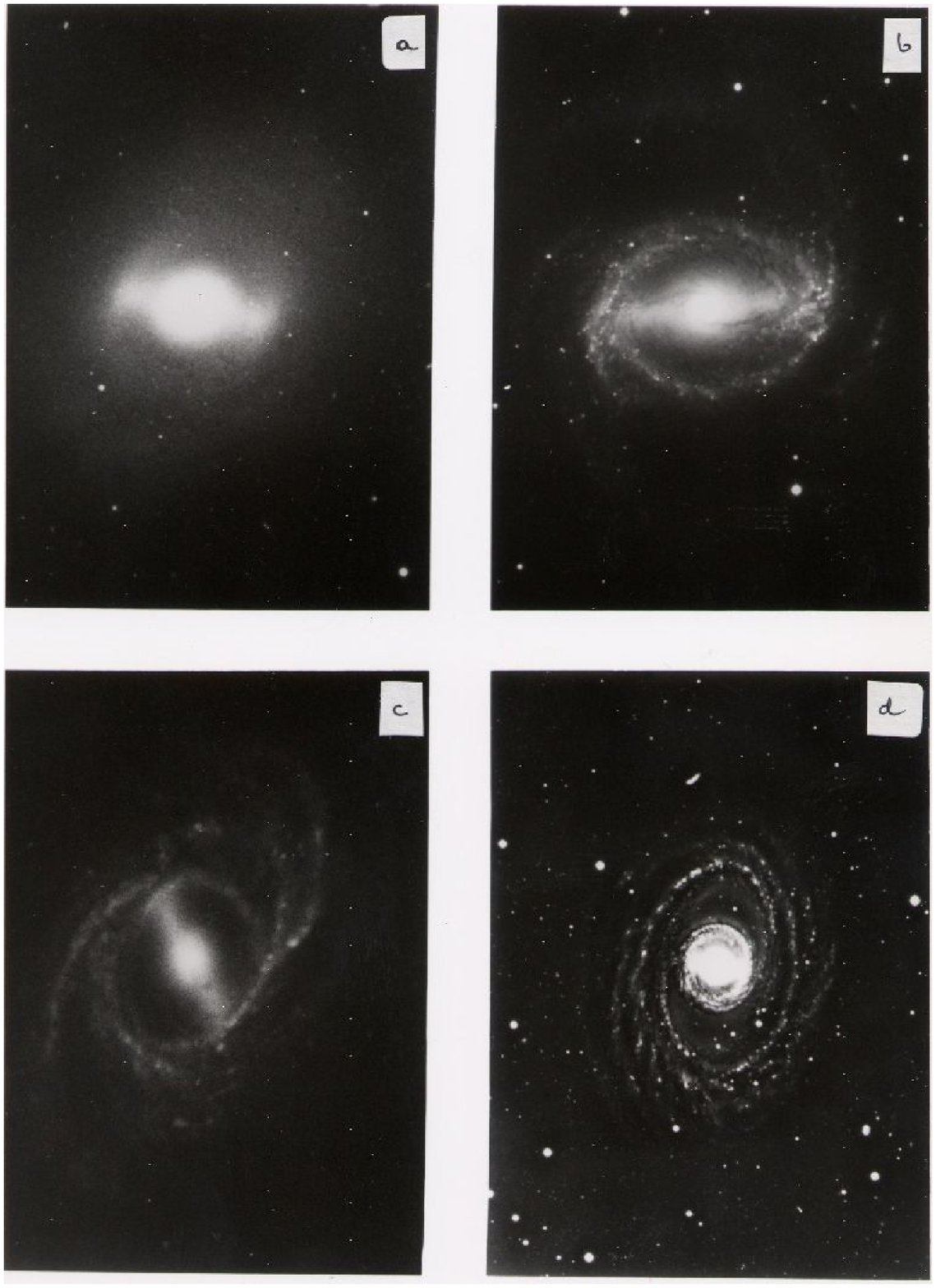,width=.87\hsize,clip=}}
\caption{2}
Eight further examples chosen to illustrate the variety of galaxies
classified as barred.  They are (a) NGC~936, (b) NGC~1433, (c)
NGC~2523, (d) NGC~1398, (e) NGC~2217, (f) NGC~5236 (M83), (g)
NGC~5383, (h) NGC~1313.  Seven photographs are courtesy of Allan
Sandage and the Carnegie Institution.  The eighth, NGC~1398 (d) which
was supplied by David Malin and has been processed to bring out
small-scale structure. \par
}\endinsert

\pageinsert{
\centerline{\psfig{file=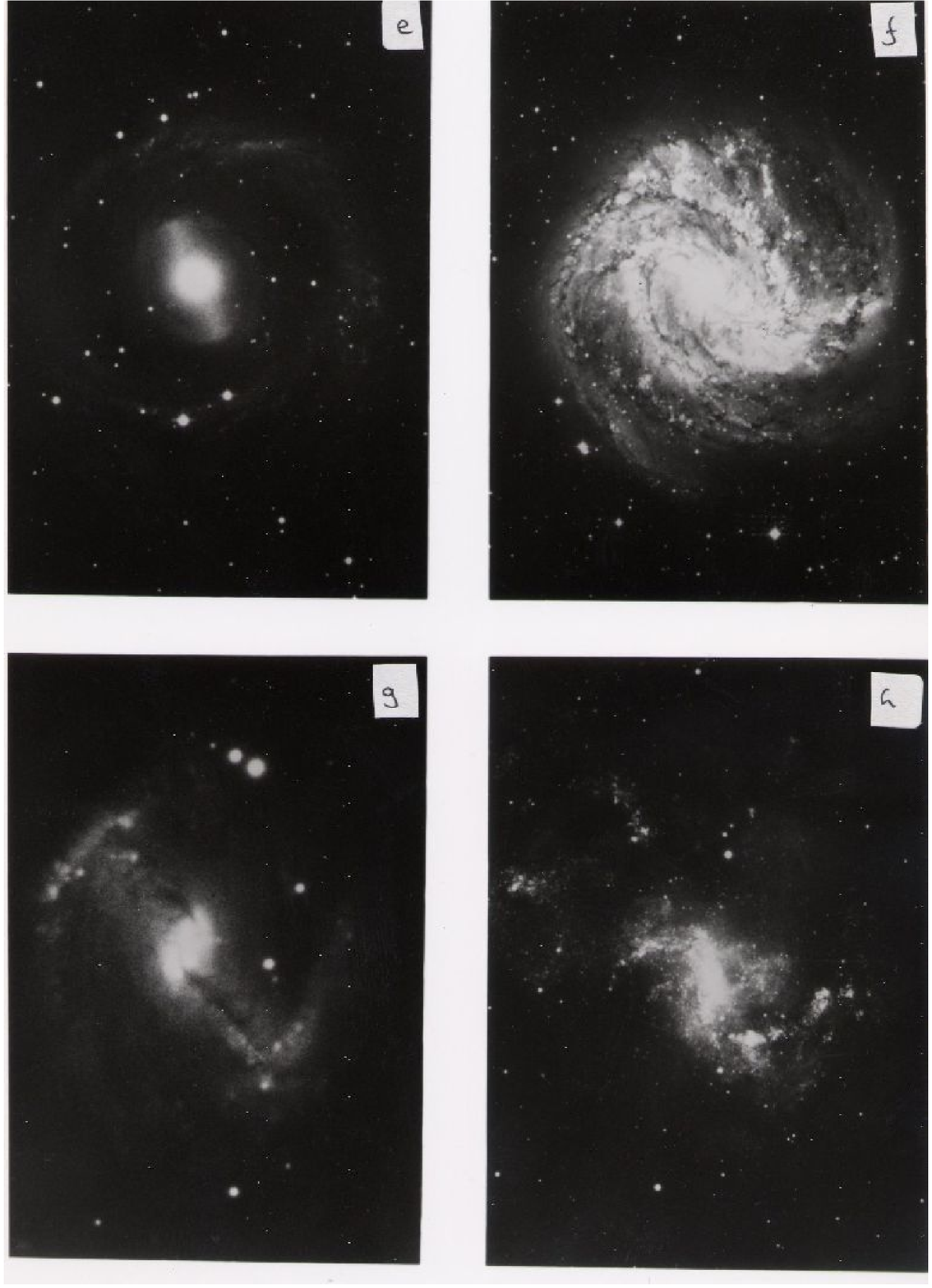,width=.85\hsize,clip=}}
\caption{2}
(Continued)
\vfill
}\endinsert

\topinsert{
\centerline{\psfig{file=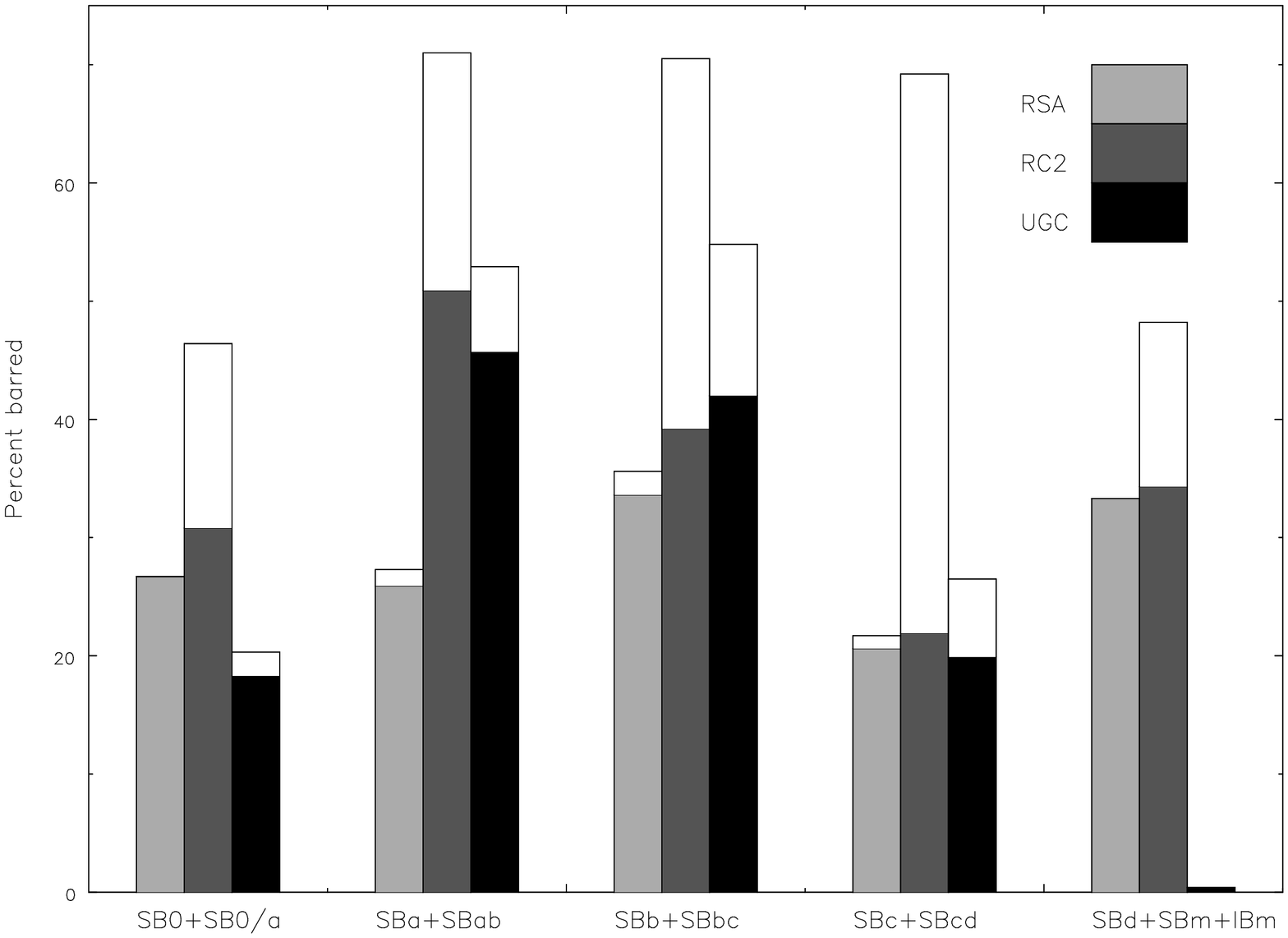,width=.75\hsize,angle=270,clip=}}
\caption{3}
The fractions of barred (SB, shaded) and intermediate (SAB, unshaded)
types at different stages along the Hubble sequence of spiral galaxies
identified in three independent morphological classifications.  The
statistics from the RSA are based upon 987 objects (only the first
classification was used here).  The RC2 sample contains 1339 objects,
after excluding peculiar, uncertain and ``spindle'' types, edge-on
objects (those with axis ratios $>3$) and those with diameters less
than 1 arcmin -- unless large-scale plate material was available.  The
UGC sample contains 4169 galaxies when selection criteria similar to
those for the RC2 were applied.  The statistics are insensitive to the
selection criteria, provided they remain within sensible ranges.\par
}\endinsert

Barred galaxies are a heterogeneous class of objects, as can be seen
from the selection assembled in \nextfig.  The bar component ranges
from a minor non-axisymmetric perturbation to a major feature in the
light distribution.  Other properties, such as the size of the bar
relative to the host galaxy, the degree of overall symmetry, the
existence of rings and the numbers (and position relative to the bar)
of spiral arms in the outer disc, the gas and dust content, \etc, vary
considerably from galaxy to galaxy.

Bars can be found in all types of disc galaxies, from the earliest to
the latest stages of the Hubble sequence.\nextfoot{Galaxies are
classified by their morphological appearance in a system based upon
that devised by Hubble (1926): the relative luminosity of the central
bulge and openness of the spirals are used to define a sequence (S0,
Sa, Sb, Sc, Sd) for both barred and unbarred galaxies.  Early type
galaxies (S0, Sa) have bright bulges and weak, tightly wrapped
spirals; late types (Sc, Sd) have little or no bulge and open, ragged
spirals.}  Because there is a continuum of apparent bar strengths from
very weak oval distortions to major features, it becomes more a matter
of taste to decide what strength of bar in a galaxy is sufficient to
merit a barred classification.  \nextfig\ shows the fraction of each
spiral type judged to contain bars, compiled from {\it A Revised
Shapley Ames Catalogue\/} (Sandage and Tammann 1981, hereafter RSA),
{\it The Second Reference Catalogue\/} (de Vaucouleurs \etal\ 1976,
hereafter RC2) and the {\it Uppsala General Catalogue\/} (Nilson 1973,
hereafter UGC).  Apart from the very latest types, there is rough
agreement over the fractions containing strong bars (shaded); when
combined over all stages, the SB family constitutes between 25\% and
35\% of all disc galaxies.  However, there is considerably less
agreement over intermediate cases, the SAB family: the morphological
classifications assigned in the RC2 indicate a combined fraction for
the SAB family as high as 26.4\% -- substantially higher than in
either of the other two catalogues.  Notwithstanding these variations,
it is clear that barred galaxies constitute a major fraction of all
disc galaxies.

A small number of galaxies which appear unbarred at visual wavelengths
have been found to be barred when observed in the near infra-red.  The
three clearest cases are NGC~1566 (Hackwell and Schweizer 1983),
NGC~1068 (Scoville \etal\ 1988, Thronson \etal\ 1989) and NGC~309
(Block and Wainscoat 1991); much shorter or weaker bar-like features
have shown up in many others (\S2.4).  Many more such cases could yet
be discovered, as IR cameras are in their infancy; if so, the fraction
of barred galaxies could turn out to be much higher than indicated by
\figno0.

Elliptical galaxies are sometimes described as stellar bars, which is
apt for certain purposes.  However, we exclude such galaxies from this
review because, unlike barred galaxies, they appear to be single
component systems with no surrounding disc or distinct central bulge.
Moreover, the stellar distribution is not thought to be as flattened,
or to rotate as fast as bars in disc systems.  De Zeeuw and Franx
(1991) have surveyed the literature on the dynamics of these objects.

We are still far from a complete understanding of the dynamical
structure of barred galaxies.  We would really like to know the
three-dimensional density distribution of each galaxy in order to be
able to calculate the gravitational potential.  This information,
combined with the rotation rate of the bar, would enable us to
calculate the motion of stars and gas.  We could claim we understood
the dynamical equilibrium if we were able to propose a self-consistent
model in which the various orbits were populated so as to reconstruct
the observed mass distribution.  We should then ask whether the
equilibrium model would be dynamically stable, or how it would evolve,
how the object was formed in the first place, \etc~ Here we will be
able to do no more than scratch the surface of the majority of these
problems.

There are many excuses for our ignorance:

\smallskip
\item{i} The distance of these objects, combined with a surface brightness 
which declines to below that of the night sky, makes it very hard to
make measurements of the quality required.

\item{ii} We see these objects only in projection.  As we believe they are 
disc-like, we might na\"\i vely hope that there is a plane of
symmetry, but because barred galaxies are intrinsically non-circular,
determination of the inclination angle is more difficult than in their
unbarred counterparts.

\item{iii} Worse still, it is unlikely that a single plane of symmetry exists 
in many galaxies.  Those seen edge-on frequently exhibit warps mainly
in the outer parts, yet this is precisely where the problem of point
(2) is best avoided.

\item{iv} We have very little knowledge of the thickness along the line of 
sight; it is reasonable to believe that the thickness of discs seen
edge-on is typical, but some edge-on systems have box-shaped bulges --
we have no way of telling whether these are strongly barred galaxies,
or whether the box shape is uncorrelated with the morphology in the
plane.

\item{v} The line-of-sight velocity component, which is all we can measure, 
also gives us no more than the average integrated through the object,
or to the point at which it becomes opaque (itself a controversial
question; \eg\ Disney \etal\ 1989, Valentijn 1990, Huizinga and van
Albada 1992, \etc).

\item{vi} Our ``snapshot'' view of each galaxy prevents our making direct 
measurements of the rotation rate (or pattern speed) of the bar; the
angular speed of the non-axisymmetric features inferred from the
observed motions of gas and stars are highly uncertain.

\item{vii} Many features, \eg\ dust lanes and rings, result from dissipative 
processes in the gaseous component.  Unfortunately, we lack a good
theoretical description for the large-scale behaviour of gas, which is
stirred and damped on scales of a few parsecs, while we wish to model
gaseous features traceable over many kiloparsecs.

\item{viii} We are uncertain of the extent to which the internal dynamics are 
affected by dark matter.  At large radii, the orbital motion in barred
galaxies resembles the nearly circular rotation pattern typical of a
normal spiral galaxy and, as emphasized by Bosma (1992), the mass
again appears to have a different distribution from the light.  There
is some evidence that the luminous matter dominates the dynamics of
the inner parts of unbarred galaxies (\eg\ Casertano and van Albada,
1990) and we shall proceed by assuming that this is the case in the
bar region.  This, perhaps rash, assumption is buttressed by the
kinematic evidence (\S\S2.5 \& 6) that the more prominent bars are
associated with large departures from simple circular motion in the
disc plane, indicating that the bar contains a significant fraction of
the mass in the inner galaxy.

\item{ix} The weakest excuse is that our mathematical ability is so limited 
that we are unable to solve the equations which should describe the
structure of such objects, except in a few highly idealized cases.
Our understanding of the dynamics has therefore to be laboriously
pieced together using numerical techniques, which themselves have
limitations.

\smallskip
It is customary to break the light of a barred galaxy into a number of
different components, or building blocks: a disc, a bulge, a bar and,
sometimes, a lens and/or rings (\eg\ Kormendy 1979).  Most of these
components are evident from Figures 1 \& 2, only the lens requires
description: if present, it is a comparatively bright, oval part of
the inner disc surrounding the bar.  It is distinguished by a
moderately sharp edge, which causes a locally steep gradient in the
photometric profile.  Although the working hypothesis of separate
components seems justified by the fact that the individual features
are readily distinguished by the eye, it is much more difficult to
separate the components quantitatively (\S2.1).  The idea of
morphologically distinct components has been taken much further,
however, and each component is frequently assumed to be a separate
dynamical entity.  This fundamental assumption is rarely discussed or
even stated.  Because it is obviously much easier to understand the
dynamics of each apparent component separately, we also treat them as
dynamically independent for the first part of the review, and examine
the validity of this assumption only towards the end.

Most of the mass in the inner bright regions of a galaxy is in stars;
the gas mass is rarely sufficient to affect the gravitational
potential and we have already indicated that we assume dark matter
begins to dominate only in the faint outer parts.  We therefore
consider the dynamics of the inner galaxy to be that of a purely
stellar system in which any gas present acts as tracer material (\S6).

Galaxies are thought to be of intermediate dynamical age; a typical
star might have completed some 50 orbits around the centre.  This is
neither so old that the galaxy must be in settled equilibrium (the
existence of spiral structure and on-going star formation show this is
not the case) nor so young that its morphological features just
reflect initial transients.  It is most appropriate, therefore, to
consider the slow evolution of a nearly equilibrium model.
Accordingly we discuss equilibrium models before we go on to consider
how they might have arisen and will evolve.  We conclude by
considering interactions between the different components.

\sect{Observed properties of bars}
We begin by summarizing the observational information relating to the
dynamical structure of the stellar bar, and leave data on gas to \S6
and on rings to \S7, where we discuss these other phenomena.

Unfortunately we cannot make direct measurements of even the most
basic properties, and are forced to make indirect inferences from the
observables.  Our requirements fall into three general areas:

\smallskip
\item{i} The distribution of mass, in order to determine the gravitational 
potential.  This has to be deduced from measurements of (a) the light
distributions and (b) the velocity field.
\item{ii} The bar rotation velocity, or pattern speed, which can be inferred 
only with considerable uncertainty from the velocity field or, still
less directly, through modelling the gas flow pattern.
\item{iii} The full three-dimensional velocity dispersion which, jointly
with orbital streaming, determines the stellar dynamical equilibrium.
This can be constructed from the projected velocity dispersion data
only with the aid of a mass model.

\subsect{Components of the light distribution}
Deprojection of the light distribution of barred galaxies is more
difficult than for nearly axisymmetric galaxies since it is far from
clear that any isophote should be intrinsically round.  The
inclination is generally inferred by assuming that the outer faint
isophotes are projected circles, but it should be borne in mind that
even far out the shape could be intrinsically elliptical, especially
if an outer ring is present (\S7), or the plane is warped.

The major axis of the bar is always less than the diameter of the host
galaxy.  In general, bars in late-type systems are shorter relative to
the total galaxy size ($D_{\rm bar}/D_{25} = 0.2$ to 0.3) than those
in early-type galaxies ($D_{\rm bar}/D_{25} = 0.3$ to 0.6), where
$D_{25}$ is the diameter at which the surface brightness of the galaxy
falls below 25 mag arcsec$^{-2}$ (\eg\ Athanassoula and Martinet 1980,
Elmegreen and Elmegreen 1985, Duval and Monnet 1985).

There also appears to be a correlation between the length of the bar
and the size of the bulge.  Athanassoula and Martinet (1980) suggest
that the deprojected length of the bar scales as $\sim 2.34$ times the
bulge diameter, while Baumgart and Peterson (1986) estimate $2.6\pm
0.7$.

\subsect{Fraction of total luminosity in the bar}
Estimates of the luminosity fraction in the bar depend not only on the
radius to which the total light is measured but also on how the
decomposition is performed.  Luminosity fractions are usually quoted
as a fraction of either the light out to the end of the bar, or
integrated out to some faint isophotal level, \eg\ $D_{25}$.

Decompositions were originally performed by attempting to fit the
components with idealized models: prolate bars, exponential discs,
\etc\ (\eg\ Crane 1975, Okamura 1978, Duval and Athanassoula 1983,
Duval and Monnet 1985).  Blackman (1983) used breaks, or changes of
slope, in the photometric profile to define the boundaries between the
different components.  Others have tried Fourier analysis of the light
distribution (\eg\ Elmegreen and Elmegreen 1985, Buta 1986b, 1987,
Ohta \etal\ 1990, Athanassoula and Wozniak 1992), which imposes no
prejudice as to the form of the components and gives direct
measurements of the strengths of the different non-axisymmetric
features, once the inclination is determined.  Probably the most
successful decomposition technique for a barred galaxy was proposed by
Kent and Glaudell (1989) who devised an iterative method to separate
an oblate bulge model from the bar.

It is not surprising therefore that the estimates for the same object
vary widely from author to author.  For example, for the SBc galaxy
NGC~7479 they range from the unrealistically low value of 8\% by the
modelling technique (Duval and Monnet 1985) to a more likely 40\% by
direct estimation of the light within an approximate isophote at the
edge of the bar (Blackman 1983) and 38\% by Fourier decomposition
(Elmegreen and Elmegreen 1985).  Ohta \etal\ (1990) give values in the
range $\sim 25$ to $\sim50$\% in the ``bar region'' to $\sim10$ to
$\sim30$\% out to D$_{25}$ for their sample of early-type galaxies.

\subsect{The light distribution within the bar}
Generally, the shapes, strengths and lengths of bars seem to vary 
systematically from early to late-type systems.

\subsubsect{Early-type galaxies}
Many bars in galaxies of types SB0 and SBa have a pronounced
rectangular shape seen in projection (Ohta \etal\ 1990, Athanassoula
\etal\ 1990) with axial ratios between 0.3 to 0.1.  The surface
brightness decreases slowly along the bar major axis, in some cases as
a shallow exponential but in others it is almost constant until close
to the end of the bar (\eg\ Elmegreen and Elmegreen 1985, Kent and
Glaudell 1989).  There is no difference between the leading and
trailing sides (Ohta \etal).

The surface brightness contrast between bars and the axisymmetric
component can range from 2.5 to 5.5 (Ohta \etal).  Bars are
considerably brighter on the major axis, and usually fainter on the
minor axis, than the inwardly extrapolated disc profile.  Azimuthally
averaged, however, the radial profiles are no more varied than those
of ordinary spirals; Wozniak and Pierce (1991) fit an exponential to
the disc profile outside the bar but ring features, which are much
more common in barred galaxies, can make this a poor approximation
(\eg\ Buta 1986b).

\subsubsect{Late-type galaxies}
The light distribution in late-type galaxies is generally less smooth
owing to greater dust obscuration and more intense knots of young
stars.  The difficulties created by both these problems are lessened
by making the observations at wavelengths as far into the infra-red as
possible (\eg\ Adamson \etal\ 1987).

Bars in galaxies of types SBbc to SBm are generally more elliptical
(\eg\ Duval and Monnet 1985), shorter and weaker than their earlier
counterparts.  They also begin to show quite pronounced asymmetries,
\eg\ one end may appear squarer than the other (see \S8).

The surface brightness distribution of weak bars in late-type systems
is much more centrally peaked and falls off exponentially along the
bar, sometimes even more steeply than the disk (Elmegreen and
Elmegreen 1985, Baumgart and Peterson 1986).  Across the bar, the
light profile is close to Gaussian (\eg\ Blackman 1983).

\subsubsect{Light distribution normal to the galactic plane}
Although we cannot see them, we expect bars to be as common in edge-on
systems as in all disc galaxies.  It has therefore been argued that
since we do not see a large fraction of unusually thick discs, bars
must be as thin as the rest of the disc.  A range of values has been
suggested for the bar thickness: Kormendy (1982) suggests typical axis
ratios of 1:2:10 and extremes of perhaps 1:3:15 or 1:$3\over2$:4 (\eg\
Burstein 1979, Tsikoudi 1980), 1:$x$:10 (Wakamatsu and Hamabe 1984).

However, this argument could be fallacious.  The bulges at the centres
of a significant fraction of nearly edge-on galaxies do not have a
simple spheroidal or ellipsoidal shape, but are squared-off, boxy or
even indented peanut-shaped (\eg\ Sandage 1961 plate 7, Jarvis 1986,
Shaw 1987, de Souza and dos Anjos 1987).  It has been suggested
(Combes and Sanders 1981, Raha \etal\ 1991) that such bulges may be
the signature of a bar seen edge-on, and if so bars are much thicker
than discs.  We discuss this point further in \S10.2.

Bottema (1990) has measured the velocity dispersion profile of a bar
which we see nearly face on.  He finds that the dispersion, which is
dominated by the vertical component, remains approximately constant at
$\sim 55$ km s\per\ from just outside the bulge to the end of the bar.
As the light profile declines strongly along this late-type bar, we
might expect the surface mass density to do so also.  Data of this
kind should be examined more closely to see whether they imply a
flaring bar.

\subsect{Tri-axial bulges and/or nuclear bars}
The position angle of the major axes of isophotes in the inner regions
of many barred galaxies twists away from the bar major axis near the
centre.  The twist angles are substantial and may, when deprojected,
be consistent with a central elongation perpendicular to the bar.
Such features have been seen mainly in the optical (\eg\ de
Vaucouleurs 1974, Sandage and Brucato 1979, Buta 1986b, Jarvis \etal\
1988) but also in the infrared (Baumgart and Peterson 1986, Pierce
1986, Shaw \etal\ 1992), in molecular gas (Ball \etal\ 1985, Canzian
\etal\ 1988, Ishizuki \etal\ 1990, Kenney 1991) and even radio
continuum (Hummel \etal\ 1987a \& b).

It has frequently been suggested that such features indicate a
tri-axial, rather than an axially symmetric, bulge (\eg\ de
Vaucouleurs 1974, Kormendy 1979, 1982, Gerhard and Vietri 1986, and
many others).  Others have claimed it indicates a small nuclear bar,
in both barred and unbarred galaxies.  As there are theoretical
reasons to expect twisted isophotes near the bar centre (see \S\S4.5.1
\& 6.4) it is of some importance to determine whether the inner
feature is perpendicular to the bar major axis when deprojected to
face-on.  While this difficult question is not yet settled, there is
evidence that the features are frequently almost perpendicular, but
Gerhard and Louis (1988) argue that this cannot be so in all cases.

\subsect{Kinematic properties of bars}
Material in galaxies lacking a bar, or other strong non-axisymmetric
feature, seems to follow nearly circular orbits, and the velocity
field is simply that of rotational motion seen in projection.  Barred
galaxies, on the other hand, are more complicated for two reasons:
firstly, the velocities, especially in the barred region, manifest
strong non-circular streaming motions, and secondly, the viewing
geometry is much more difficult to determine.  Both the inclination
angle to the plane of the sky and the position angle of the projection
axis -- the line of nodes -- are more difficult to determine when the
light distribution is intrinsically non-axisymmetric and the motions
depart systematically from circular.

Many early observations consisted of a slit spectrum at a single
position angle along the bar or projected major axis.  Such data
yields very incomplete information; several slit positions, or ideally
a full two-dimensional map, are required to determine the velocity
field.

\subsubsect{Velocity fields}
Kormendy (1983) presented the first clear evidence for non-circular
{\it stellar\/} streaming motions in a barred galaxy.  The
non-circular motions in the SB0 galaxy NGC~936 (\figno{-1}) are about
20\% of the circular streaming velocity and are consistent with orbits
being elongated along the bar, with the circulation in the forward
sense along the bar major axis (assuming the weak spiral arms trail).
More recent data for the same galaxy (Kent 1987, Kent and Glaudell
1989) have confirmed this interpretation.

Other examples are: NGC~6684 which shows deviations of 100 km/sec from
circular motion (Bettoni \etal\ 1988), NGC~1543, 1574, 4477 for which
velocities on the minor axis of $\sim 100$ km s\per\ have been
detected (Jarvis \etal\ 1988) and NGC~4596 (Kent 1990).  Only galaxies
for which the bar is inclined to both the projected major and minor
axes show non-circular motions clearly (Pence and Blackman 1984b, Long
1992).

Bettoni and Galletta (1988) argued that the picture is more
complicated than simple streaming in the same sense as the direction
of bar rotation because the apparent streaming velocity does not
always rise monotonically when measured along the bar major axis and
may even dip back to near zero at a point about one third of the
distance from the centre to the end of the bar.  They suggested that
this behaviour might possibly be due to retrograde stellar streaming
within the bar.  However, Sparke showed that it can quite naturally be
explained as being due to the finite width of the slit, and a possible
very small misalignment with the bar major axis, allowing light from
stars streaming along the bar to influence the measurement.  Her
argument is reported by Bettoni (1989).

These measurements establish that the bar distorts the axisymmetric
potential of the disc sufficiently strongly to force the stars to
stream on highly elliptical orbits inside the bar.  Additional
evidence which helps to determine the strength of the non-axisymmetric
potential comes from observations of the gas kinematics, though these
usually need to be interpreted using a model (see \S6).

\subsubsect{Azimuthally averaged velocity profiles}
Since the potential of a barred galaxy is strongly non-axisymmetric,
it is even more difficult than for a nearly axisymmetric galaxy to
determine the distribution of mass.  Although an approximation to a
``rotation curve'' found by crudely averaging tangential streaming
velocities can be significantly in error (Long 1992), this practice is
widespread.

The error is unlikely to affect the comment (Elmegreen and Elmegreen
1985) that many early-type galaxies have bars which extend beyond the
rising part of the rotation curve, while late-type bars end before the
rotation curve flattens off.  However, the positions of resonances
within the bar (see \S4.2) deduced from an axially symmetrized
rotation curve and an estimate of the bar pattern speed, are highly
uncertain.  Not only could the axisymmetric rotation curve be wrong,
but a resonance predicted by linear theory for a weak perturbation
could be absent in a strong bar.  Orbit integrations are the only
reliable way to identify resonances in a strong bar (\S4.5).

\subsect{Bar angular velocity or pattern speed}
The rate of rotation of the bar is one of the most important
parameters determining its dynamical structure, but unfortunately it
is very difficult to measure directly.  Tremaine and Weinberg (1984b)
have shown how the continuity equation could in principle be used to
determine this quantity from observables alone.  The method requires
high signal-to-noise long-slit spectra and a photometric light profile
both taken along cuts offset from the nucleus and parallel to the
major axis.  In practice, the pattern speed inferred is sensitive to
centering and alignment errors, warps, and the presence of any
non-bisymmetric perturbation and is therefore uncertain.

Using this technique, Kent (1987) estimated $\Omega_p \sin i = 5.4 \pm
1.9$ km s\per arcsec\per ($\sim 104\pm37$ km s\per kpc\per) for NGC
936.  Combining his value with Kormendy's rotation curve, Kent's
measurement suggests that co-rotation lies within the bar, although a
radius just beyond the end of the bar is within the
error.\nextfoot{Notwithstanding his measurement, Kent and Glaudell
(1989) preferred a pattern speed 64 $\pm$ 15 km s\per kpc\per\ in
their dynamical study of this galaxy (\S4.7.3), in order to place
co-rotation just outside the end of the bar.}

Application of this method to galaxies other than SB0s is even more
problematic, because of the presence of dust patches and bright knots.
Tremaine and Weinberg tried to apply it to the HI observations of
NGC~5383 (Sancisi \etal\ 1979) but were unable to obtain a conclusive
result.  The method failed because of noise, an extended HI envelope
and because any single easily observed species probably does not obey
the equation of continuity (\eg\ atomic to molecular transitions, star
formation, stellar mass loss, \etc).

The very limited success of this method is disappointing;
nevertheless, Kent's measurement does at least furnish independent
evidence that the bar in NGC~936 rotates rapidly.  We are therefore
forced to rely upon less direct alternative techniques.  The most
reliable method, described in \S6, attempts to match the observed gas
flow pattern to a set of hydrodynamical models in which the bar
pattern speed is varied.  Some authors also attempt to associate
individual morphological and kinematic features with resonances for
the pattern; we note examples in \S\S6.7 \& 7.

\subsect{Velocity dispersions}
Velocity dispersion profiles have also been determined for some
galaxies, though frequently only along the bright bar major axis.  The
central values of the velocity dispersion are found to be similar to
that of the lens (\S7) -- typically about 150 to 200~km~s\per\ (\eg\
Kormendy 1982 for the SB0 galaxies NGC 936 and NGC 3945).  The ratio
of the maximum rotation velocity to the central velocity dispersion,
$V_{\rm max}/\sigma_0$, gives a measure of the relative importance of
streaming and random motion.  Values are typically between $0.4$ and
$0.5$, implying that there are significant contributions from both
streaming motions along the bar and random motions.  More generally,
the bar velocity dispersion profiles can vary from flat to sharply
falling with increasing radius (\eg\ Jarvis \etal\ 1988).

\sect{Stellar dynamics of galaxies}
In this section we summarize the basic equations of stellar dynamics
and some of the techniques used to solve them.  The following two
sections discuss the structure of steady bars in two and three
dimensions respectively.  An excellent introductory text has been
provided by Binney and Tremaine (1987).

\subsect{Relaxation time}
Since the density of stars throughout the main body of a galaxy is
very low, the orbit of an individual star is governed by the
large-scale gravitational field of the galaxy and is not appreciably
affected by the attraction of the relatively few nearby stars.  The
gravitational impulses received by a star as it moves through a random
distribution of scattering stars nevertheless accumulate over time.
The {\it relaxation time\/} is the time taken for these cumulative
random deflections to change the velocity components along the orbit
of a typical star by an amount equal to the stellar velocity
dispersion.  Chandrasekhar's (1941) formula yields a value of $\sim
10^{13}$ years, or $\sim 1000$ times the age of the universe, for
star-star encounters in the neighbourhood of the Sun.  Since the
relaxation time varies inversely as the stellar density, considerably
shorter timescales apply in the centres of galaxies, where the star
density is higher by several orders of magnitude.

Chandrasekhar's formula has not required revision, but the
impressively long relaxation time it predicts is believed to be an
overestimate.  We now know that the distribution of mass in disc
galaxies is not always as smooth as he assumed -- gas is accumulated
into giant molecular cloud complexes, ranging in mass up to $10^5$
\Msun\ or even $10^6$ \Msun, and some bound star clusters are known to
contain almost as much mass in stars.  Although these objects are much
more diffuse than point masses, clumps in this mass range shorten the
relaxation time considerably.  It is even further reduced by the
tendency for the gravitational attraction of massive objects to raise
the stellar density near themselves.  The accumulated material, known
as a polarization cloud, can easily exceed the mass of the perturber
when the stellar motions are highly ordered -- as in a rotationally
supported disc (Julian and Toomre 1966).  Nevertheless, the relaxation
time in the objects we consider here is never shorter than the orbital
period, and is usually considerably longer.

\subsect{Dynamical equations}
The long relaxation time implies that the distribution of stars in a
galaxy approximates a collisionless fluid and therefore obeys
dynamical equations similar to those governing a Vlasov plasma.  The
closest parallel is with a single species plasma (\eg\ Davidson \etal\
1991).

To describe such a fluid mathematically, we introduce a distribution
function, $F$, which is the density of particles in an element of
phase space -- a multi-dimensional space with (generally) three
spatial dimensions and three velocity dimensions.  The definition of
$F(\bx, \bv, t)$ is the mass within a volume element of phase
space, divided by the volume of that element, and we would like to
take this to the limit of infinitesimal volume.  As the number of
stars in a galaxy is finite, however, $F$ can be defined only for
volume elements large enough to contain many stars.  Since this can be
very inconvenient mathematically, we generally imagine that the number
of stars is greatly (infinitely) increased, while their individual
masses are correspondingly reduced, so that $F$ can be meaningfully
defined for infinitesimal volume elements.  (Because encounters are
negligible, the masses of individual stars do not enter into the
equations.)

The motion of a fluid element in this phase space is governed by the
collisionless Boltzmann (or Vlasov) equation $$ {\partial F \over
\partial t} + \bv \cdot {\bf \nabla}F - {\bf
\nabla}\Phi \cdot {\partial F \over \partial \bv} = 0. \eqno(\equno)
$$ Since the left hand side is the convective derivative in this
space, equation (\equat0) simply states that the phase space density
is conserved at a point which moves with the flow.

The quantity $\Phi$ in (\equat0) is the smooth gravitational potential
of the galaxy, which is related to the volume density, $\rho$, through
Poisson's equation $$
\nabla^2\Phi(\bx,t) = 4\pi G\rho(\bx,t), \eqno(\equno)
$$ where $G$ is Newton's gravitational constant.  The volume density
of stars at any point is the integral of the phase space density over
all velocities: $$
\rho(\bx,t) = \int_{-\infty}^\infty F({\bf x},{\bf v},t)  d^3\bv. 
\eqno(\equno) $$

The right hand side of equation (\equat{-2}) is zero only for a
collisionless fluid.  Rapid spatial or temporal variations about the
mean potential, of whatever origin, can scatter stars in phase space
and lead to a non-zero Fokker-Planck term here (\eg\ Binney and Lacey
1988).  We assume, for the purposes of this review, that no such term
exists.

The field of galactic dynamics is, to a large extent, concerned with
solutions to this coupled set of integro-differential equations.  Only
a few analytic equilibrium solutions are known in cases where a
particularly simple symmetry (\eg\ spherical, axial, planar, \etc) has
been assumed.  These, and in some cases the stability of the resulting
models, are discussed in the monograph by Fridman and Polyachenko
(1984).

\subsect{Analytic distribution functions}
The classical approach to solving these equations for a particular
mass distribution is first to identify the integrals; \ie\ quantities
that are conserved by a test particle pursuing any orbit in the
potential.  In a steady, non-rotating potential, the particle
conserves its energy, and if the potential has an axis of symmetry,
the component of angular momentum about that axis is conserved.
Jeans's theorem tells us that the distribution function for an
equilibrium model is a function of the integrals of the motion (\eg\
Lynden-Bell 1962a).

The identification of integrals with simple physical quantities such
as energy or angular momentum is possible only for very simple
potentials, and the approach rapidly becomes too hard for realistic,
rotating, non-axisymmetric potentials.  Vandervoort (1979) and
Hietarinta (1987) discuss the possible forms an extra integral could
take in a bar-like potential.  One possibility had already been
exploited by Freeman (1966), who proposed a family of two-dimensional
bar models in a rotating potential consisting of two uncoupled
harmonic oscillators.  His highly idealized models have some
interesting properties, but all have retrograde streaming of the stars
inside the bar, which conflicts with the observed situation (\S2.5).

Vandervoort (1980) developed a set of stellar dynamical analogues to
the Maclaurin and Jacobi sequences of gaseous polytropes.  These are
uniformly rotating, non-axisymmetric, isotropic bodies for which the
distribution function is a function of one integral only.  These
models are not centrally concentrated enough for elliptical galaxies;
but this may not be such a drawback for strong bars, in which the
density is less centrally concentrated.

The most promising recent bar-like models for which some headway has
been, and still is being, made are the so called St\"ackel models,
which emerged from Hamilton-Jacobi theory (\eg\ Lynden-Bell 1962b).
St\"ackel (1890) identified a class of potentials in which the
Hamilton-Jacobi equation separates in confocal ellipsoidal coordinates
and in which the density is stratified on concentric tri-axial
ellipsoids.  Their main advantage for theoretical work is that they
have the most general form known to be always integrable (see \S3.4).
This implies that {\it once the distribution function is known}, all
the observable properties of the model, such as the streaming
velocities and velocity dispersion fields are predictable.  The
principal disadvantage for our subject is that almost all such models
so far discussed are non-rotating.

{\tolerance = 2000 
These models have been extensively explored by Eddington (1915),
Kuz'min (1956), Lynden-Bell (1962b) and de Zeeuw (1985) and his
co-workers as models for galaxies.  The central dense part can have a
surprisingly wide variety of shapes (de Zeeuw \etal\ 1986), ranging
from simple ellipsoids to more rectangular bodies with squared-off
ends.
\par}

Unfortunately, the distribution functions required to construct
self-consistent models are not so easy to obtain.  Statler (1987)
obtained a variety of approximate solutions using Schwarz\-schild's
method (\S3.5) for several three-dimensional ellipsoidal
models.\nextfoot{Teuben (1987) provides solutions for the
two-dimensional analogue, the elliptic disc.}  Hunter \etal\ (1990)
derived exact expressions for the distribution function for
arbitrarily flattened prolate models populated by thin tube orbits.
Their solutions can be tailored to allow maximal or minimal streaming
around the symmetry axis.  Hunter and de Zeeuw (1992) have given
solutions for the tri-axial case; again they found a wide range of
solutions, but were able to show that self-consistent models require
some members from all four of the main orbit families.

One integrable bar model noted by Vandervoort (1979) as a curiosity
was later shown to be the first known rotating St\"ackel model
(Contopoulos and Vandervoort 1992).  The centrifugal term is exactly
balanced by a term in this highly contrived potential, which has the
unphysical features of two singularities at the co-ordinate foci and
negative mass density at large distances.  The orbital structure of
the model, studied in depth by Contopoulos and Vandervoort, is quite
different from that found in more realistic bar models.

Very little is known about the stability of St\"ackel models, though
Merritt and his collaborators have initiated a programme to explore
this issue.  Merritt and Stiavelli (1990) found that lop-sided
instabilities developed in {\it all\/} oblate models with
predominantly shell orbits.  Prolate models, on the other hand, were
unstable to buckling modes only when highly elongated (Merritt and
Hernquist 1991).  This numerical approach complements stability
analyses which are possible only for extremely simple models (\eg\
Vandervoort 1991).

\subsect{Near-integrable systems}
The basic assumption underlying the approach just outlined is that
bars are integrable systems, \ie\ that all stars possess the number of
integrals equal to the number of spatial dimensions.  This now seems
most unlikely, as we show in \S\S4 \& 5; galactic bars furnish an
excellent practical example of the kind of near-integrable Hamiltonian
system first studied by H\'enon and Heiles (1964), also in an
astronomical context.  The study of near-integrable systems opens up
the modern topic of non-linear dynamics.  Unfortunately, even a highly
simplified introduction would represent too large a digression for
this review and we refer the reader to a standard text (\eg\
Lichtenberg \& Lieberman 1983).

The identification of bars as non-integrable systems not only raises
the theoretical problem of why they are not integrable, but also a
number of practical questions.  The most important are: what is the
fraction of chaotic orbits within the bar? how much does their density
distribution contribute to the bar potential? and what are the
long-term consequences of their presence?  These questions remain
largely unanswered, and it should be clear that they represent a
formidable obstacle to progress in our understanding of bar dynamics.

The field has therefore resorted to the less elegant numerical methods 
described in the next two subsections in order to make further progress.

\subsect{Linear programming}
Schwarzschild (1979) achieved a major breakthrough when he
successfully constructed the first numerical models of a
self-consistent, non-rotating, tri-axial stellar ellipsoid, without
requiring any knowledge of the nature, or even the number, of
integrals supported by the potential.

He began by computing a large number of orbits in a potential arising
from an {\it assumed\/} tri-axial ellipsoidal density distribution and
noted the time-averaged density along each orbit in a lattice of cells
spanning the volume of the model.  He then used linear programming
techniques to find the non-negative weights to assign to each orbit in
order that the total density in each cell added up to that in his
initially assumed model.  As he was able to find more than one
solution, Schwarzschild added a ``cost function'' to seek solutions
with maximal streaming, for example.  However, the method yields only
an approximate equilibrium model, even when a fine lattice of cells is
used, and gives no indication of its stability.

Newton and Binney (1984) proposed a variation to this strategy based
upon Lucy's iterative deconvolution technique.  Its main advantages
are that it is easy to program, fast and the resulting distribution
function is smoother than that yielded by linear programming.  Other
iterative algorithms are based upon non-negative least-squares
(Pfenniger 1984b) and maximum entropy (Richstone 1987).

\subsect{$N$-body techniques}
Because of the intractability of the full-blown self-consistency
problem, many workers in the field have turned to $N$-body
simulations.  Not only does a quasi-steady $N$-body model represent a
self-consistent equilibrium, but it also demonstrates that the model
is not catastrophically unstable.  Such models have yielded many
important results, and have contributed inestimably to our
understanding of barred galaxies.  We present these results at
appropriate points in this review, and confine ourselves here to
alerting the reader to some of the essential limitations of the
$N$-body techniques.

Simple restrictions of computer time (and to a lesser extent memory)
limit the numbers of particles that can be used in the simulations.
Inevitably, therefore, the potential in simulations is less smooth
than in the system being modelled.  Statistical fluctuations in the
particle density, on all scales from the inter-particle separation up
to the size of the system, are larger than they ought to be by the
square-root of the ratio of the numbers of particles; in some of the
most efficient high quality simulations, which employ perhaps a few
times $10^5$ particles, the noise level remains about one thousand
times higher than in a galaxy of stars.\nextfoot{Density fluctuations
of this magnitude may be present in the discs of gas-rich galaxies,
however.}

The various techniques available to combat this problem have been
reviewed by Sellwood (1987) but only one, the ``quiet start'' method,
removes fluctuations on the largest scales.  The technique involves
choosing the initial coordinates of particles in some symmetric
configuration and filtering out the components of the potential which
arise from the imposed symmetry (Sellwood 1983).  Unfortunately, the
symmetric configurations themselves are unstable and disintegrate
quite rapidly; the respite from noise is therefore only temporary.

The main effect of these irrepressible fluctuations is to scatter
particles away from the paths they would have pursued in a noise free
potential.  The integrals which should be invariant along the orbit of
each particle are no longer preserved.  There have been comparatively
few studies of orbit quality in simulations, but enough to raise
substantial worries (\eg\ van Albada 1986, Hernquist and Barnes 1990).
The simulations are thought to be useful for longer than the
scattering time of a typical orbit because the individual scattering
events, when averaged over the whole population of particles,
accumulate only statistically.  However, the long-term stability and
any possible secular effects cannot be studied by these techniques.

For obvious reasons of computational economy, the particles in a large
majority of bar-unstable disc simulations have been confined to a
plane.  A few early attempts to include full three-dimensional motion
(Hockney and Brownrigg 1974, Hohl 1978, Sellwood 1980) revealed
behaviour which differed only slightly from models having equivalent
resolution in two-dimensions, but the grids used in all these
calculations were so coarse that the out-of-plane motions were
virtually unresolved.  [The results obtained by Combes and Sanders
(1981) did reveal an important difference, which was not understood at
the time.]  As extra resolution in two-dimensions seemed essential
(Sellwood 1981), full three-dimensional simulations for this problem
were considered unnecessary for several years.  However, recent
three-dimensional models have rediscovered Combes and Sanders' old
result and show that the two-dimensional models miss some of the
essential physics (\S\S5.1 \& 10.1).

\sect{Two-dimensional bar models}
We begin our detailed discussion of the dynamics of bars by making the
great simplifying approximation that the motion of stars normal to the
galactic symmetry plane can be ignored.  The usual justification for
this is summarized in \S4.1, though it is now clear that this
approximation is inadequate for bars; the dynamical structure of a
three-dimensional bar is more than just a simple addition of
independent vertical oscillations to orbits in the plane.
Nevertheless, as many properties of orbits in three-dimensional bars
can be recognized as generalizations from two-dimensions, a
preliminary discussion of the simpler, but by no means
straightforward, two-dimensional case remains an appropriate starting
point.

As we have no analytic models for rapidly rotating bars, even in
two-dimensions, we might hope to construct an approximate one using
Schwarzschild's method (\S3.5). Despite the vast literature on
two-dimensional bars, this programme has been carried through in only
one case (Pfenniger 1984b); most papers confine themselves to a
discussion of orbits in arbitrarily selected bar-like potentials
having simple functional forms.  Clearly, it is necessary to learn
which of the many possible orbit families could contribute to a
self-consistent model and to understand how they are affected by
changes in the potential: a good grounding in both these aspects is an
essential pre-requisite for a search for a fully self-consistent
model.  It is also sensible to explore many types of model in the hope
of finding some that might be more nearly integrable (\S3.4).
However, in two-dimensions at least, the field is now sufficiently
mature that such arguments are wearing thin.

\subsect{Thin disc approximation}
There are three principal requirements which must be fulfilled before
three-dimensional motion can be neglected as an unnecessary
complication.  First, the scale of the non-axisymmetric structures,
spirals, bars, \etc, should be large compared with the thickness of
the disc.  Second, the oscillations of stars in the direction normal
to the plane must not couple to motion in the plane.  Third, there can
be no instabilities which would cause the plane to warp or corrugate.

These three conditions are widely believed to be fulfilled for nearly
{\it axisymmetric\/} discs.  The first is doubtful only in the case of
flocculent spiral patterns.  The second seems valid because the
frequency of $z$-motion is higher than that of any perturbing forces
which might arise from disturbances in the plane, and therefore the
motion normal to the plane should be adiabatically invariant (see
\S5.1).  Finally, the principal instability which could give rise to
corrugations of the plane, the fire-hose instability, is known to be
suppressed by a modest degree of pressure normal to the plane, and
``galaxy discs seem to be well clear of this stability boundary''
(Toomre 1966).  However, the second and third requirements are
violated for strong bars (Combes and Sanders 1981, Pfenniger 1984a,
Raha \etal\ 1991) and introduce the further aspects to the dynamics we
discuss in \S5.

\subsect{Linear theory}
With the exception of the St\"ackel models, orbits in all strong
bar-like potentials must be calculated numerically.  The bewildering
variety of orbit types found in most numerical studies of a strong
bar-like potential is difficult for the newcomer to assimilate.
Moreover, the {\it characteristic diagrams\/} and {\it surfaces of
section\/} drawn in many papers, require considerable explanation
before the wealth of information they contain can be comprehended.
Before plunging into numerically calculated orbits, therefore, we
first explore linear theory for a weakly perturbed case.  Although a
very crude approximation for bars, the results give useful insight to
those from more realistic strongly barred models.  We also introduce
characteristic curves for orbits of arbitrary eccentricity in the
simplest possible case of an {\it axisymmetric\/} potential.

The equations of motion of a test particle (\eg\ a star or planet) in
the symmetry plane $(r,\theta,z=0)$ of a potential, $\Phi$, are $$
\eqalign{
\ddot r - r\dot\theta^2 & = -{\partial\Phi \over \partial r}, \cr
\dot J & = -{\partial\Phi \over \partial\theta}, \cr} \eqno(\equno)
$$ where a dot denotes a time derivative and $J \,(= r^2\dot\theta)$,
is the specific angular momentum of a star.  It is useful to divide
the potential into an unperturbed part, which is axisymmetric, and a
non-axisymmetric perturbation: $\Phi = \Phi_0 + \phi$.

In an axisymmetric potential ($\phi\equiv0$), $J$ is conserved and we
define a home radius, $\rc$, at which a star would pursue a circular
orbit: $$
{\partial\Phi_0 \over \partial r} = \rc\Omegac^2 = {J^2 \over \rc^3},
\eqno(\equno)
$$ where $\Omegac$ is the angular frequency of circular motion at
$\rc$.  A star possessing more energy than the minimum required for
circular motion for a given $J$, oscillates about its home radius.  As
the period of the radial oscillation is generally incommensurable with
the orbital period, the orbit does not form a closed figure in an
inertial frame, except for a few special cases such as harmonic or
Keplerian potentials.  Successive apo-galactica of the rosette-like
orbit are between $\pi$ and $2\pi$ radians apart in all realistic
galactic potentials.

\subsubsect{Lindblad epicycles}
When the star's orbit is far from circular, the radial motion is
anharmonic and asymmetric about its home radius.  For nearly circular
orbits, on the other hand, the star's motion approximates an harmonic
epicycle about a guiding centre which orbits at $\Omegac$, as first
exploited by Lindblad (1927) in a galactic context.

To see this, we describe the motion in terms of displacements,
$(\xi,\eta)$, from uniform circular motion; $\xi$ in the outward
radial and $\eta$ in the forward tangential, directions.  Assuming the
displacements $(\xi,\eta)$ to remain small, the equations of motion
(\equat{-1}) can be approximated as (Hill 1878) $$ \eqalign{
\ddot \xi + 2\rc\Omegac{d\Omegac \over dr}\xi - 2\Omegac\dot \eta &= 
-{\partial \phi \over \partial r} \cr
\ddot \eta + 2\Omegac\dot \xi &= -{1\over \rc}{\partial\phi \over 
\partial\theta}, \cr} \eqno(\equno)
$$ where the potential derivatives on the right hand sides are to be 
evaluated along the circular orbit.

Setting the perturbing potential $\phi=0$ for the moment, the
solutions to the homogeneous equations are: $$
\xi = a e^{i(\kappa t+\psi_0)} \qquad {\rm and} \qquad \eta = ia{2\Omegac 
\over \kappa } e^{i(\kappa t+\psi_0)}, \eqno(\equno)
$$ where $a$ is the maximum radial excursion of the particle and the
Lindblad epicyclic frequency is $$
\kappa = \left[ 4\Omegac^2 + r{d\Omegac^2 \over dr} \right]^{1/2}. 
\eqno(\equno)
$$ These equations describe an elliptic epicycle about the guiding
centre.  Since $2\Omegac > \kappa$ in all realistic potentials, the
major axis of the ellipse lies along the direction of rotation.

In linear theory, we can treat the individual Fourier components of
the potential perturbation separately.  When a sinusoidal perturbation
rotates at a uniform rate $\Omegap$, we can write it as
$\phi(r,\theta,t) = P(r) e^{im(\theta-\Omegap t)}$, where $m$ is the
azimuthal wavenumber of the perturbation: $m=2$ is the fundamental
wave for a bar or bi-symmetric spiral.  The function $P$ describes the
radial variation of amplitude and phase of the perturbation; for a
spiral it is necessarily complex, but it can be purely real at all
radii for a bar.  (As usual with complex notation, the physical
quantity corresponds to the real part.)

With this form for $\phi$, equations (\equat{-2}) become $$ \eqalign{
\ddot \xi + \rc{d\Omegac^2 \over dr}\xi - 2\Omegac\dot \eta &= -{\partial P 
\over \partial r} e^{im[\theta+(\Omegac-\Omegap)t]} \cr
\ddot \eta + 2\Omegac\dot \xi &= -{im \over \rc} P 
e^{im[\theta+(\Omegac-\Omegap)t]}. \cr} \eqno(\equno)
$$ It is convenient to write the angular frequency at which the
guiding centre overtakes the perturbation as $\omega = m(\Omegac -
\Omegap)$.  The particular solution is $$
\xi = {-1 \over \kappa^2 - \omega^2} \left[ {2m\Omegac \over \rc\omega} P + 
{\partial P \over \partial r} \right] e^{i(m\theta+\omega t)}, \eqno(\equno 
{\rm a})
$$ and $$
\eta = {-i \over \kappa^2 - \omega^2} \left[ {2\Omegac \over \omega} 
{\partial P \over \partial r} + {4\Omegac^2 - \kappa^2 + \omega^2 \over 
\omega^2} {m \over \rc} P \right] e^{i(m\theta+\omega t)}, \eqno(\equat0 {\rm 
b})
$$ which describes only the distorted path of the guiding centre.  The
full solution, which includes epicycles about the new distorted path,
is obtained by adding the complementary function [the solution
(\equat{-3}) to the homogeneous equations].

A resonant condition occurs wherever the denominators in equations
(\equat0) pass through zero.  This occurs at co-rotation, where
$\omega = 0$, and at the Lindblad resonances, where $\omega = \pm
\kappa$.  The resonant denominators also imply that, for a fixed
amplitude $P$, the displacements increase without limit as the
resonance is approached; our linearized analysis must therefore break
down near these radii.

\nextfig\ shows the angular frequencies $\Omegac$ and $\Omegac \pm \kappa/m$, 
with $m=2$ as appropriate for a bar, plotted as functions of radius in
a reasonably realistic galaxy model.  (The axisymmetric mass
distribution used is described in \S4.3.1.)  Resonances for nearly
circular orbits occur wherever the horizontal line at the bar pattern
speed $\Omegap$ intersects one of these curves.  The pattern speed
chosen for this illustration has one outer Lindblad resonance (\OLR)
and two inner Lindblad resonances (\ILR s) in this mass model.  Clearly
there would be no \ILR s if $\Omegap$ exceeded the maximum value of
$\Omegac - \kappa/2$; moreover, this peak would be much lower if the
galactic potential had a more extensive harmonic core.  \ILR s need not
be present in every barred galaxy, therefore.

\topinsert{
\centerline{\psfig{file=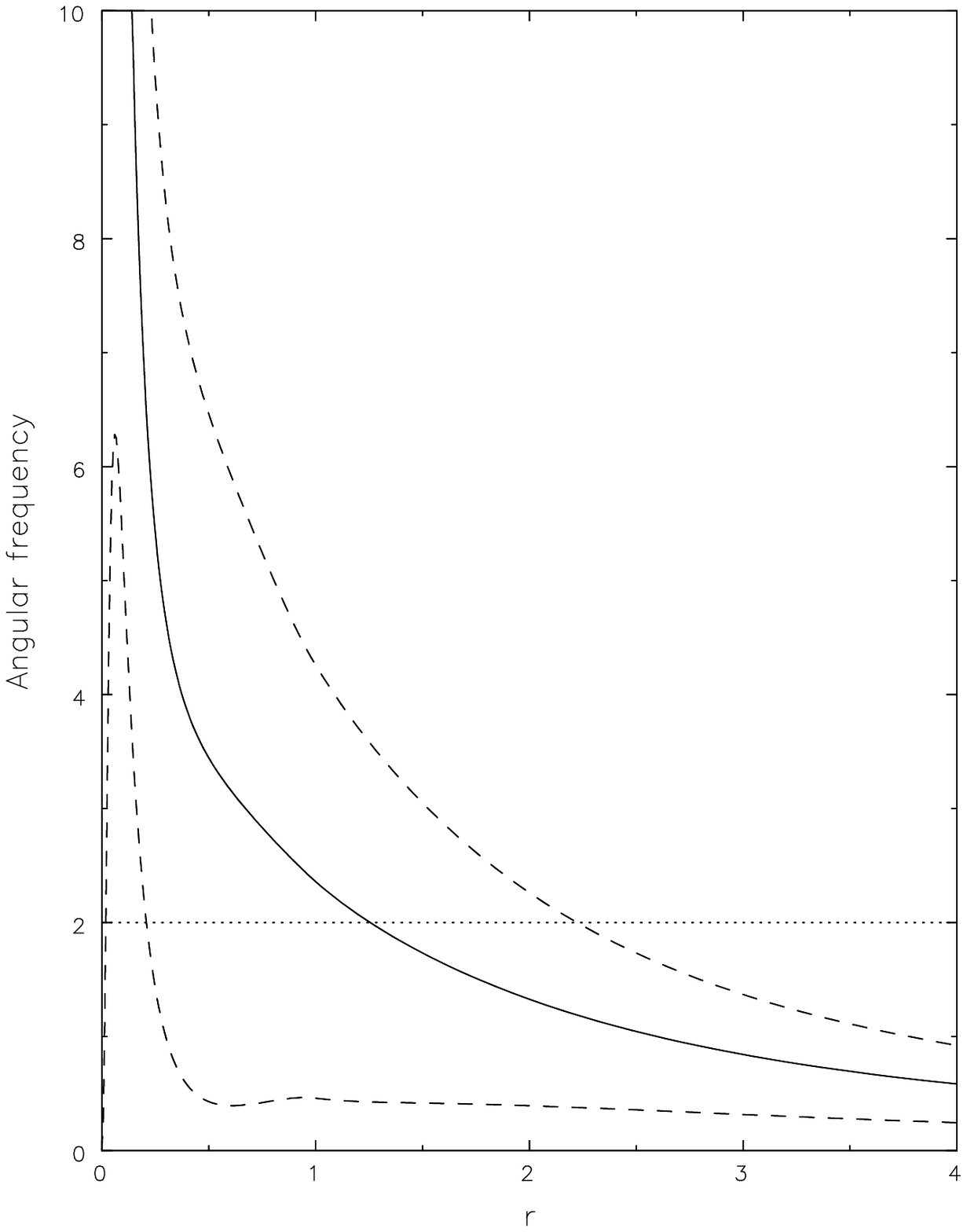,width=.5\hsize,clip=}}
\caption{4}
The angular frequency of circular motion, $\Omegac$ (full drawn curve)
and $\Omegac\pm\kappa/2$ (dashed curves) for our {\it axisymmetric\/}
mass model.  The horizontal dotted line shows the angular frequency of
the frame adopted in \S4.2.4; the intersections of this line with the
dashed curves mark points where very nearly circular orbits would form
closed bi-symmetric figures in this frame.\par
}\endinsert

\subsubsect{Orbit orientations}
Equations (\equat0) give the displacements of the guiding centre when
subjected to a mild, uniformly rotating, sinusoidal potential
perturbation.  The guiding centre moves (at a non-uniform rate) around
a centred ellipse which is elongated either parallel or perpendicular
to the potential minimum (Sanders and Huntley 1976).

In order to examine the sense of alignment more easily, we neglect the
radial derivative, $\partial P / \partial r$, on the grounds that the
perturbed potential is likely to vary only slowly with radius.
Dropping the exponential factor, the radial displacement of the
guiding centre is therefore approximately $$
\xi \simeq -{2m\Omegac \over \rc\omega (\kappa^2 - \omega^2)}P. \eqno(\equno)
$$ If we choose the line $\theta = 0$ for the bar major axis, $P$ is
real and negative and $\xi$ has the same sign as $\omega$ between the
Lindblad resonances, but takes the opposite sign when $|\omega| >
\kappa$, \ie\ further from co-rotation than the Lindblad resonances.
Thus the orientation of the orbit changes across every one of the
principal resonances.  Between the \ILR\ (or the centre, if none is
present) and co-rotation, the orbit is aligned with the bar, but it is
anti-aligned beyond co-rotation.  The orientation reverts to parallel
alignment beyond the \OLR.

The orbit therefore responds as a harmonic oscillator of natural
frequency $\kappa$ driven at frequency $\omega$, with a negative
forcing frequency being interpreted as its absolute value but $\pi$
out of phase.  The switch of alignment across the Lindblad resonances
occurs for the familiar reason that any driven harmonic oscillator is
in phase with the forcing term when the driving frequency, $\omega$,
is below its natural frequency, $\kappa$, but is exactly $\pi$ out of
phase when the forcing frequency exceeds the natural frequency.

Even though equations (\equat{-1}) were derived assuming infinitesimal
perturbations, most studies in a variety of reasonable potentials have
found that the more nearly circular periodic orbits are aligned with
the bar inside co-rotation, and that stable perpendicularly oriented
orbits are often found inside the \ILR.\nextfoot{Contopoulos and
Mertzanides (1977) found an example where a steep radial gradient in
the perturbation, which we neglected in equation (\equat0), caused a
local reversal of these normal orbit orientations.}

\subsubsect{Action-angle variables}
The epicyclic viewpoint remains valid for arbitrarily eccentric orbits
in an axisymmetric potential, but the motion can no longer be
described by equations (\equat{-5}).  Kalnajs (1965, 1971) was the
first to use action-angle coordinates, $(\Jr,\Ja,\wr,\wa)$, to
describe galactic orbits of arbitrary eccentricity.  In an
axisymmetric potential, the radial velocity of a star of energy (per
unit mass) $E$ and specific angular momentum $J$ is $$
\dot r = \left\{ 2 \Big[ E-\Phi(r) \Big] - {J^2 \over r^2} \right\}^\half. 
\eqno(\equno)
$$ The period of its radial oscillation is $$
\oint \dot r^{-1} dr = \tau_{\rm r} = {2\pi \over \Omegar}, \eqno(\equno)
$$ and the frequency $\Omegar$ thus defined is constant for that star.
We define the angle $\wr$ to be the phase of this oscillation; thus
$\dot\wr = \Omegar$.  The action conjugate to this angle, known as the
{\it radial action,} is the area of the oscillation in phase space
(divided by $2\pi$ in the normalization used in this field), \ie $$
\Jr = {1 \over 2\pi } \oint \dot r dr. \eqno(\equno)
$$ The radial action has the dimensions of angular momentum and is a
measure of the amplitude of a star's radial oscillation.

Because a star conserves its angular momentum during this radial
oscillation, it does not progress around the galactic centre at a
uniform angular rate (except for circular orbits), but we introduce an
angle $\wa$ that does.  If $\Delta\theta$ is the change in a star's
azimuth during one radial period, we define this angle to change at
the uniform rate $$
\dot\wa \equiv \Omegaa = {\Delta\theta \over \tau_{\rm r}}. \eqno(\equno)
$$ Thus $\Omegaa$ is the orbital frequency of the guiding centre and
$\wa$ its phase angle.  The action $\Ja$ conjugate to $\wa$ is simply
$J$, the specific angular momentum.

For less eccentric orbits, the motion tends towards the epicyclic
description of \S4.2.1, and these new coordinates behave as: $$
\eqalign{ \Jr &\rightarrow {\textstyle{1\over2}}\kappa a^2 \cr \Omegar 
&\rightarrow \kappa \cr} \qquad
\eqalign{ \Ja &\equiv J \cr \Omegaa &\rightarrow \Omegac \cr}. \eqno(\equno)
$$ For eccentric orbits in most reasonable potentials, $\Omegaa$ and
$\Omegar$ are respectively less than $\Omegac$ and $\kappa$ evaluated
at the guiding centre.

\subsubsect{Orbits in rotating frames}
Because bars (and perhaps spirals) are believed to be steadily
rotating features, it is sensible to study orbits in a frame which
rotates with the non-axisymmetric pattern.  A number of their
properties in perturbed potentials can be understood more readily by
first considering orbits in an axisymmetric potential viewed from
rotating axes.

\topinsert{
\centerline{\psfig{file=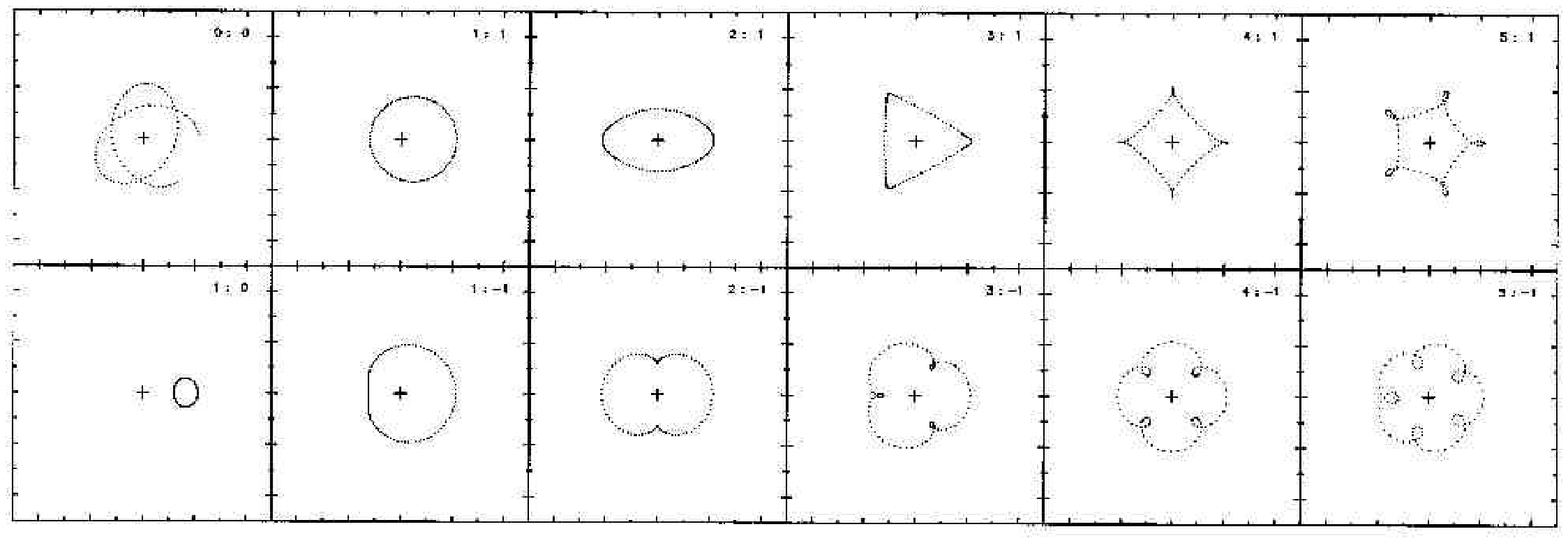,width=\hsize,angle=270,clip=}}
\caption{5}
A single orbit drawn in frames rotating at many different rates.  The
view marked 0:0 is from an inertial frame ($\Omegaf=0$) and all the
other views are seen when $\Omegaf$ is given by equation (17) with the
values of $m$:$l$ marked in each.  The 1:0 frame shows the Lindblad
epicycle.  In all frames with $l$ negative, the star orbits less
rapidly than the frame, whereas the reverse is true when $l$ is
positive.  The only frame which rotates counter to the star's orbit is
that marked 1:1.  The 100 dots in each panel are separated by equal
time intervals for that frame. \par
}\endinsert

Although the orbit of a general star in an axisymmetric potential does
not close when viewed from an inertial frame, it will appear to close
to an observer in a coordinate system which rotates at certain
frequencies.  The angular rotation rate of the frame in which the
orbit will appear to close must be $$
\Omegaf = \Omegaa - {l \over m}\Omegar, \eqno(\equno)
$$ for integer $l$ and $m$.  \nextfig\ shows the {\it same\/} orbit
drawn in several rotating frames, with the values of $l$ and $m$
marked for each.  The figure closes after the orbit has made $m$
radial oscillations and $|l|$ turns about the centre of the potential.

Since the frequencies vary from orbit to orbit, most orbits will not
close in one arbitrary rotating frame.  However, as both $\Omegaa$ and
$\Omegar$ vary with both the specific energy, $E$, and angular
momentum, $J$, (\equat0) can be satisfied for more than a few isolated
orbits, for a single $\Omegaf$.  In fact, there will be an infinite
number of 1-D sequences in the two-dimensional $(E,J)$ space along
which the orbits appear to close.

To illustrate this important concept, \nextfig\ shows many different
orbits which close in the {\it same\/} rotating frame.  (These orbits
were found numerically in the axisymmetric mass model described in
\S4.3.1.)  We have grouped these selected orbits by shape, and it
should be clear that each group represents a continuous sequence of
orbits which close in this frame.  Obviously in this axisymmetric
potential, the orientation of the shapes is arbitrary; we have simply
chosen to draw them so that the axis of reflection symmetry is in the
$y$ direction.

\topinsert{
\centerline{\psfig{file=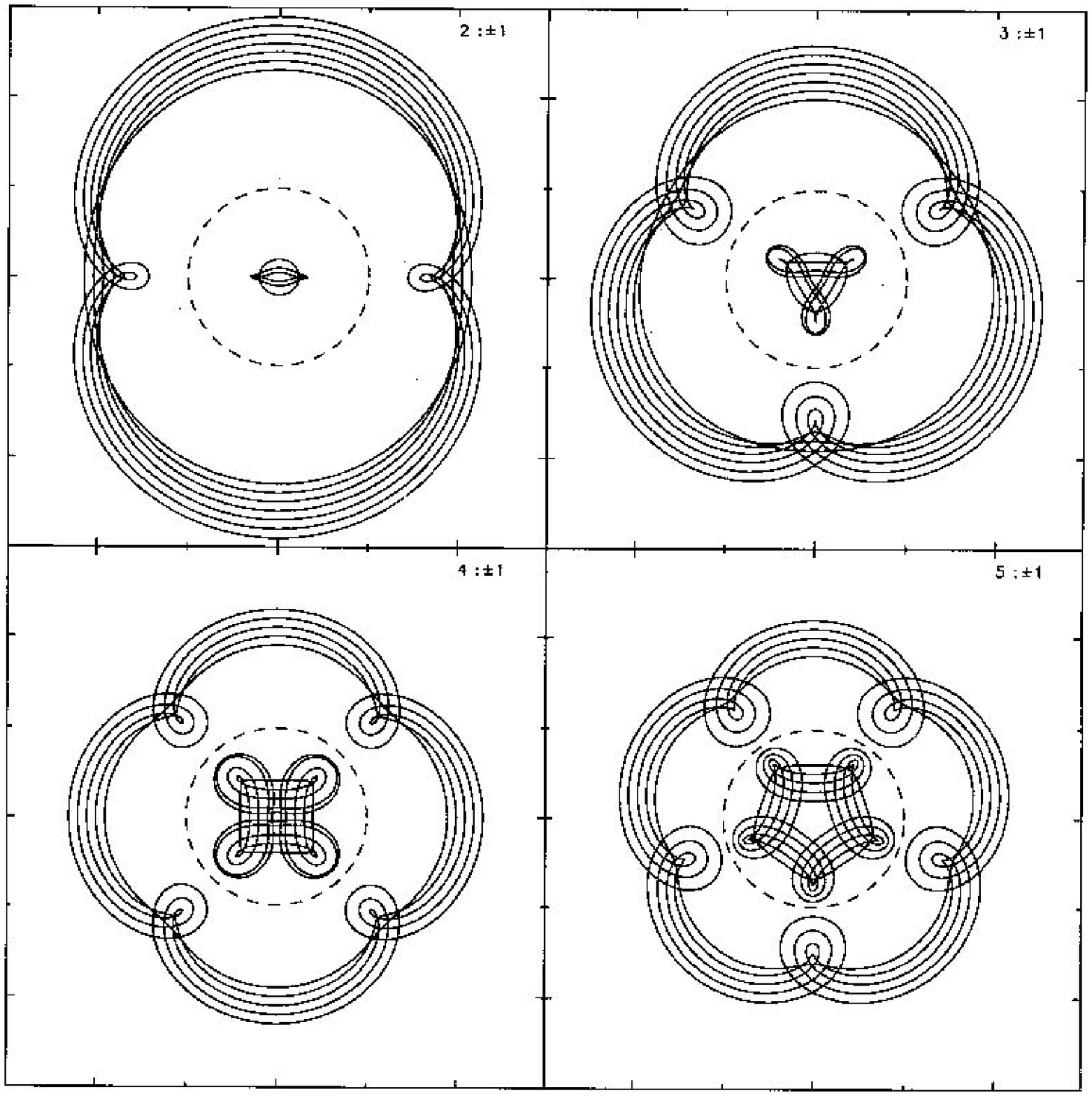,width=.6\hsize,clip=}}
\caption{6}
Many orbits which all close in the same rotating frame.  Most orbits
are highly eccentric, but as the order of the symmetry rises, the more
nearly circular orbits lie progressively closer to the co-rotation
circle (dashed). \par
}\endinsert

\topinsert{
\centerline{\psfig{file=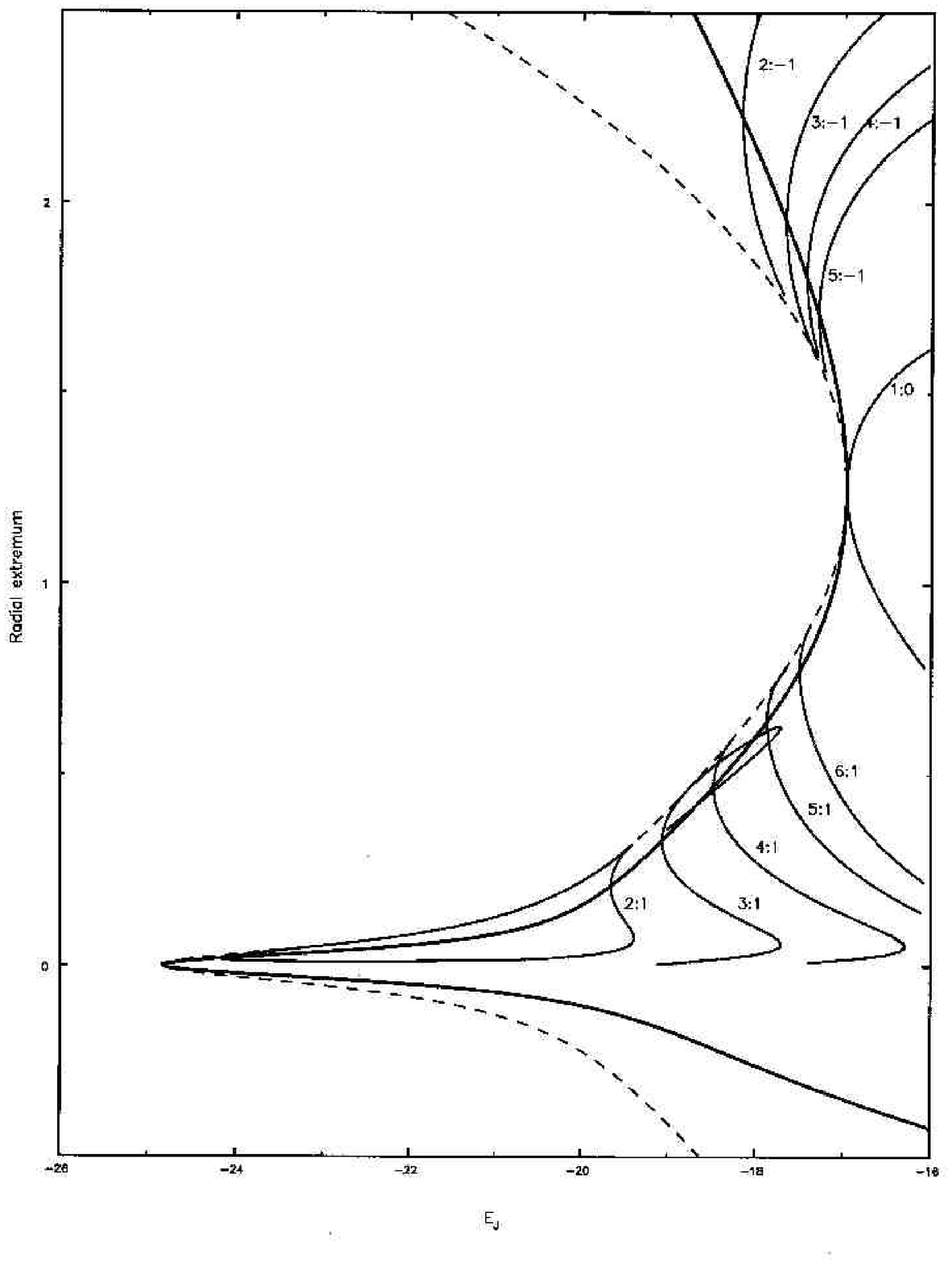,width=.6\hsize,clip=}}
\caption{7}
Characteristic curves showing the loci of the radial extrema of
families of orbits which all close in one rotating frame; negative
values are used for orbits which rotate in a sense counter to the
frame.  The thick line marks the circular orbits as a function of the
Jacobi integral, and the dashed curve marks the \ZVC\ which no orbit can
cross.  Each characteristic curve is marked with the $m$:$l$ value for
which the orbits close. \par
}\endinsert

\nextfig\ shows {\it characteristic curves\/} for sequences of orbits in this 
potential which all close in the same frame.  The radial extrema of
closed orbits are plotted as a function of the quantity $\EJ = E -
J\Omegaf$ (with $\Omegaf$ held constant), which is known as Jacobi's
integral (see \S4.3.2), and we plot these as negative values for stars
with $J<0$.  The heavily drawn line shows circular orbits, which
trivially close in any frame, and the dashed {\it zero velocity
curve\/} (or \ZVC) marks the radii at which a particle would appear
momentarily at rest in this frame.  These two curves touch at their
maximum values of $\EJ$ near the ordinate $r=1.25$, at which a star on
a circular orbit appears at rest -- \ie\ it co-rotates with the frame.

All other lines on this Figure show parts of several characteristic
curves.  Each is marked with the ratio of radial to orbital periods
before the orbit closes in the rotating frame.  We have selected only
orbits for which $l=\pm1$ or 0 and low values of $m$ and the majority
of sequences are incomplete -- only in the case of the 3:1 family have
we drawn the sequence as far as almost radial orbits.  The 1:0 family
are simply the extension of Lindblad epicycles (\S4.2.1) to more
eccentric orbits.

Each $m$:$\pm1$ sequence crosses the circular orbits at a point known
as a {\it bifurcation\/}.  In an axisymmetric potential, each
bifurcation occurs where $\Omegaf = \Omegac(r) - {l \over m}\kappa(r)$
for the infinitesimally eccentric orbits discussed above.  (The 3:1
sequence crosses the circular orbit line at three points inside
co-rotation; the bifurcation occurs only at the point where the orbits
are nearly circular.)  There is one bifurcation outside co-rotation
($l=-1$) and one inside ($l=1$) for all sequences except for the 2:1
orbits.  In this exceptional case, the two sequences starting from the
two 2:1 bifurcations are connected and form a closed loop around the
circular orbits, called a bubble by Contopoulos (1983a).  The 1:1
bifurcations occur outside the diagram; the $l=1$ lies at large $\EJ$
for retrograde orbits and the $l=-1$ at large $r$.

The two branches of each sequence, on either side of the bifurcation,
are not independent, since each traces one of the two extremal radii
for the same orbit.  Each sequence also touches the \ZVC, where the
orbit develops cusps.  More eccentric orbits loop back on themselves
for part of the time (\figno{-1}), but we have drawn this continuation
only for the 3:1 family in order not to clutter the diagram with too
many crossing lines.

Similar sequences of higher $m$ (still with $|l|=1$) can be found
closer to co-rotation and are ever more densely packed as they
approach this point.  More sequences occur for higher values of $|l|$,
and correspond to orbits which close after $l$ rotations.  For small
$m$, they are found either far outside co-rotation or are retrograde
(if they exist at all), but the higher $m$ sequences could also be
drawn within the boundaries of \figno0\ -- in fact sequences for all
$m$:$l$ would densely fill the plane, except for the region excluded
by the \ZVC.

\topinsert{
\centerline{\psfig{file=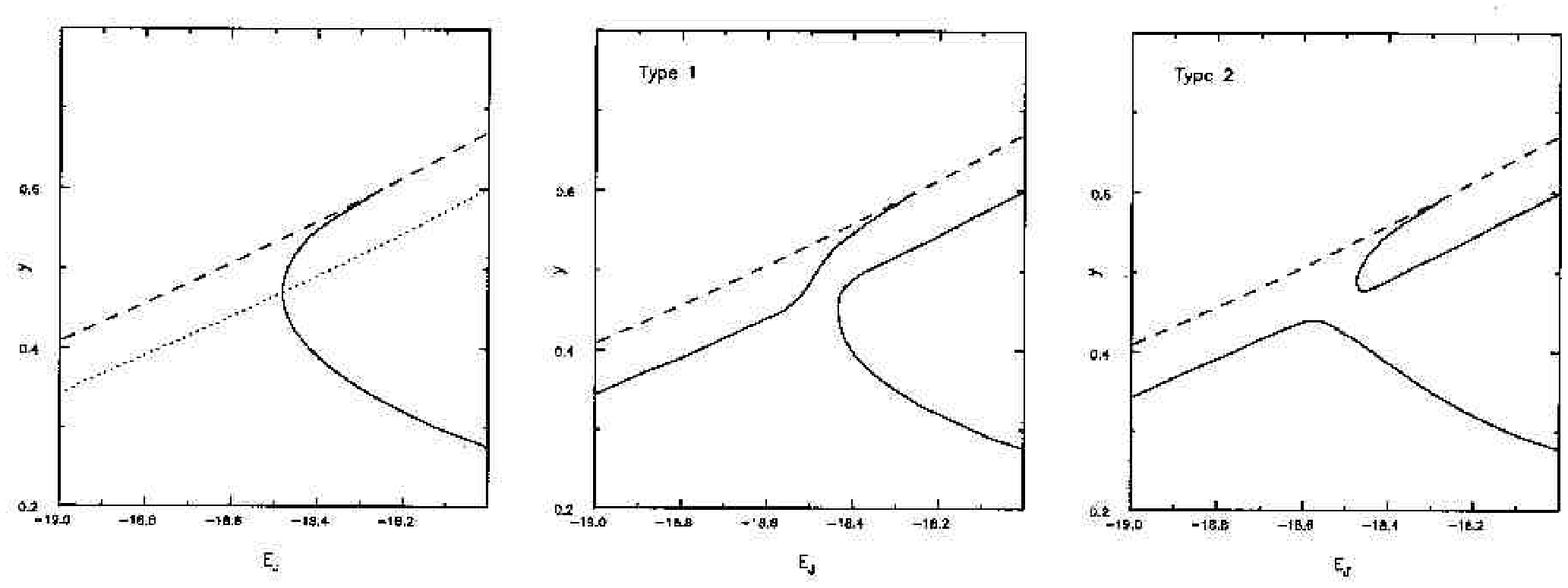,width=\hsize,angle=90,clip=}}
\caption{8}
A schematic representation of the two principal types of resonance
gaps.  The dashed curve in each panel is the \ZVC; the left hand panels
show the bifurcation with no potential perturbation, and the other
illustrate the two different types of gap that a perturbation can
create.  These diagrams are drawn for the situation inside
co-rotation. \par
}\endinsert

\subsubsect{Effect of a weak non-axisymmetric perturbation}
If we again add a weak perturbing potential co-rotating with this
frame (\ie\ so $\Omegaf = \Omegap$), all these closed non-circular
orbits resonate with the perturbation.  We should therefore expect
them to be substantially altered in a perturbed system.

We have already calculated the effect on the circular orbits in
\S4.2.2.  An infinitesimal, pure $m$-fold, symmetric potential
perturbation introduces Lindblad resonances at the $m$:$\pm1$
bifurcations.  No closed orbits can exist at the resonant point,
because a particle there would experience a monotonic acceleration
until it is driven off resonance.  Linear theory cannot tell us any
more, but exact calculations show that the four branches at a
bifurcation separate into two, non-crossing pairs; the resonance
imposes a gap, both in the circular orbits and in the $m$:1 sequence.
The stronger the potential perturbation, the wider the gap becomes.

We can see which pairs must join from the direction in which the
perturbation moves the circular orbits in the characteristic diagram
(\figno0), a small part of which is shown enlarged in the first panel
of \nextfig.  It is customary in these diagrams to plot the intercept
of the distorted orbit on the {\it minor\/} axis of the potential
perturbation; thus the circular orbits are displaced in the direction
given by the sign of $-\xi$ in (\equat{-7}a).

When $\partial P/\partial r$ is negligible, $\xi$ has the same sign as
$2m\Omegac / [\rc\omega (\kappa^2 - \omega^2)]$, since $P$ is
negative.  In this case, the circular orbit branch on the co-rotation
side of the Lindblad resonance bends away from the \ZVC, while that on
the other side of the resonance bends towards the \ZVC, which is called
a gap of type 1.  On the other hand, should the perturbation amplitude
decrease rapidly with increasing radius, the term containing $\partial
P/\partial r$ could outweigh the other in (\equat{-7}a); under these
circumstances, the bends occur in the opposite sense creating a gap of
type 2.  Both types of gap are shown schematically in
\figno0.\nextfoot{This is not precisely the convention used by
Contopoulos and Grosb\o l (1989), who define 4 types of gap.}

Linear theory predicts gaps only at the Lindblad ($m$:$\pm1$)
resonances.  Finite perturbations, however, give rise to many more
gaps for two distinct reasons.  Firstly, in most reasonable bar-like
potentials, the perturbation is not a pure $m=2$ sinusoid, but also
contains other Fourier components.  We should therefore expect gaps at
every $m$:$\pm1$ resonance for which the Fourier component $m$ is
non-zero; these will be at even values of $m$ whenever the
perturbation has exact two-fold rotational symmetry.  However, gaps
continue to occur at all even $m$ bifurcations when the perturbation
has a pure $\cos(2\theta)$ form (Contopoulos 1983a).  This is because
the finite radial extent of the distorted path of the guiding centre
breaks the precise $m$-fold symmetry of the potential explored by the
star.  The first such additional resonances for non-linear orbits, the
$2m$:$\pm1$ resonances, are known as ultra- or hyper-harmonic
resonances, and occur at 4:$\pm1$ for a bar.

No gaps develop at the odd $m$:$\pm1$ bifurcations for a purely
symmetric potential perturbation.  Some authors rather confusingly
continue to refer to a resonance gap as a bifurcation; the phrase {\it
pitchfork bifurcation\/} is useful to emphasize a bifurcation without
a gap.

\subsect{Strong bars}
The daunting task of summarizing the major results on the dynamical
structure of two-dimen\-sional stellar bars is made still harder by the
number of rival conventions adopted by the various groups.  The most
obvious is the number of different naming conventions for the orbit
families in current use, but even such simple things as the bar
orientation in diagrams has not been standardized: \eg\ its major axis
lies up the page in some papers and across the page in others and a
recent author has decided it looks nicer at 45\degrees!  While a
careful reader can always infer the orientation, it is frequently left
unstated.  Other inconsistencies abound in the important
characteristic diagrams (\S4.4).

A large number of different mass models has been investigated.  This
has the dual advantages that we can ``separate what is generic from
what is accidental'' (Contopoulos and Grosb\o l 1989) and can begin to
understand how the properties of the orbits change as parameters are
varied.  However, this strategy makes it impossible for a reviewer to
select sample diagrams from the literature to illustrate all aspects
of a single model.  The relation between the three principal diagrams,
showing shapes of periodic orbits, characteristic curves and surfaces
of section, and the reason for constructing all three, is much easier
to comprehend if an illustrative example of each can be shown for the
same model.

\subsubsect{Mass model}
We have therefore anchored this part of our review around a single
mass model which we introduce to illustrate the main results from the
literature.  The model we have selected is constructed from building
blocks in the usual artificial manner.  However, it is among the
simplest which display many of the properties frequently discussed and
its orbital structure is qualitatively similar to the more realistic
$N$-body and photometric models discussed in \S\S4.9.2 \& 4.9.3, for
which the gravitational potential is known only numerically.  We also
make no attempt to ascertain whether the orbital structure of our
model is favourable for complete self-consistency.

Our model, which most closely resembles those used by Sanders and his 
collaborators, has the following three mass components:

\smallskip
\item{B} A uniformly tumbling Ferrers (1877) ellipsoid to model the bar.  
This is an inhomogeneous prolate spheroid of mass $M_{\rm B}$ having a 
density profile $$
\rho_{\rm B}(m) = \cases{ {105 \over 32\pi ac^2} M_{\rm B} (1-\mu^2)^2 & 
$\mu<1$ \cr
0, & $\mu\ge 1$ \cr} \eqno(\equno)
$$ where $$
\mu^2 = {x^2 \over a^2} + {y^2 + z^2 \over c^2}, \eqno(\equno)
$$ and $a>c$.  The major axis of the ellipsoid, $a$, is therefore
aligned with the $x$-axis [following the convention in Binney and
Tremaine (1987)] and the bar and coordinate system rotate about the
$z$-axis.  De Vaucouleurs and Freeman (1972) give the potential of
this mass distribution in the following convenient form: let
$\epsilon^2 = a^2 - c^2$ and define $\psi$ by $$ \eqalign{
y^2 \tan^2\psi + x^2\sin^2\psi = & \; \epsilon^2 \qquad\qquad \mu > 1 \cr
\cos\psi = & \; c/a \;\qquad\quad \mu \le 1 \cr} \eqno(\equno)
$$ and $$
w_{lk}(\psi) = 2 \int_0^\psi \tan^{2l-1}\theta \sin^{2k-1} \theta d\theta.
$$ These integrals are all straightforward.  The potential can now be
written as $$ \eqalign{
\Phi_{\rm B}(x,y) = {105GM_{\rm B} \over 32\epsilon} & \left[ {1\over 
3}w_{10} - {1 \over \epsilon^2}(x^2w_{11} + y^2w_{20}) \right. \cr
& + {1 \over \epsilon^4}(x^4w_{12} + 2x^2y^2w_{21} + y^4w_{30}) \cr
& \left.  - {1 \over 3\epsilon^6}(x^6w_{13} + 3x^4y^2w_{22} + 3x^2y^4w_{31} + 
y^6w_{40}) \right]. \cr} \eqno(\equno) $$

\item{S} A small dense spherically symmetric component to model the 
bulge/spheroid which has the density profile of a Plummer sphere $$
\rho_{\rm S}(r) = {3M_{\rm S} \over 4\pi s^3} \left( 1 + {r^2 \over
s^2} \right)^{-{5 \over 2}}, \eqno(\equno)
$$ where $M_{\rm S}$ is the mass of the spheroid and $s$ is a length
scale.  (Of course, $r$ is here a spherical radius.)

\item{H} An extensive spherically symmetric component to model the halo which 
has the same form as equation (\equat0), but a much lower central
density, larger total mass, $M_{\rm H}$, and longer length scale, $h$.

\smallskip \noindent Although all three components are 3-dimensional, we 
require the potential in the plane $z=0$ only -- its variation out of
the plane is of no importance in this section.  We therefore do not
need to distinguish a separate disc component: component H can be
thought of as representing both the disc and halo.

The advantages of choosing a Ferrers ellipsoid to represent the bar
are that the potential and its derivatives can be written in closed
form and that the potential is continuous up to its third derivative
(because only the second derivative of the density is discontinuous at
the boundary).  The disadvantage is that being elliptical with a
constant eccentricity at all radii, the mass distribution does not
resemble that of real bars (see \S2.3).  We have adopted this model
here because it is easy to program and widely used, and because it
exhibits almost all the generic properties we wish to illustrate.

We set the gravitational constant $G=1$ and adopt the bar major axis
as our unit of length ($a=1$); in this unit $s = 0.05$ and $h = 1.5$.
The masses of the three components are $M_{\rm B} = 1.1852$, $M_{\rm
S} = 0.3$ and $M_{\rm H} = 25$.  The axis ratio of the ellipsoid we
choose to be $a/c = 3$ and we set its angular rotation rate
$\Omegap=2$ in these units.  [The axisymmetric mass model used in
\S4.2 was very similar, differing only in that the Ferrers ellipsoid
was spherical ($a/c=1$) and had a mass of precisely 1 unit.]

\nextfig\ shows the azimuthal variations in both the radial and
tangential forces in our model.  The gravitational forces therefore
vary quite strongly with azimuth within the bar, but quickly approach
axial symmetry beyond it.

\topinsert{
\centerline{\psfig{file=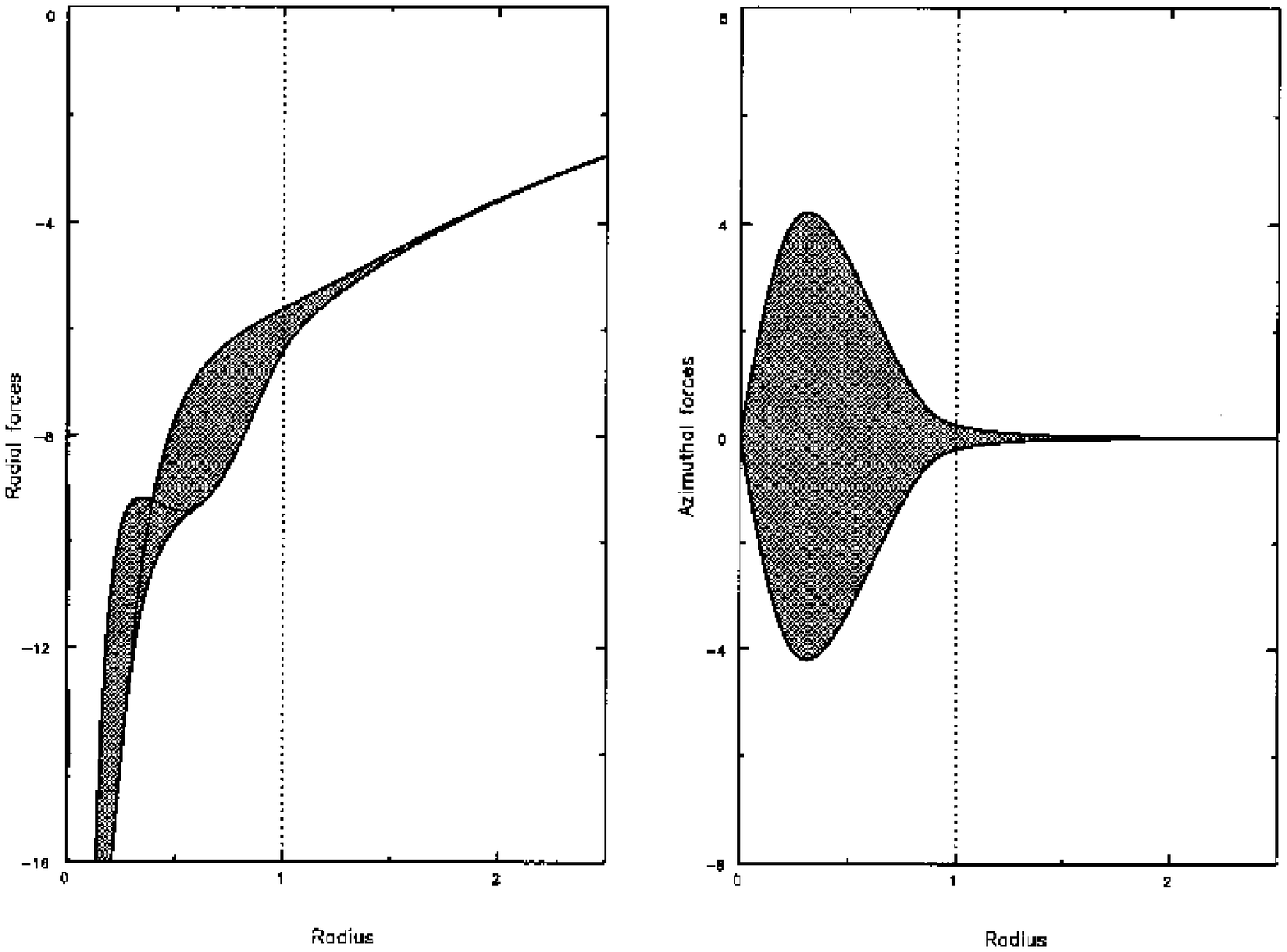,width=.7\hsize,angle=270,clip=}}
\caption{9}
The two force components at various azimuthal angles in our barred
model as functions of radius.  The boundaries to the shaded areas
indicate the largest and smallest values attained around at circle at
that radius.  Within radius 1 (the semi-major axis of the bar) the
forces vary strongly with azimuth, but their range decays rapidly
outside the bar. \par
}\endinsert

\subsubsect{Rotating non-axisymmetric potentials}
The gravitational potential of a uniformly rotating, non-axisymmetric
density distribution, is steady when viewed from axes which co-rotate
with the perturbation.  We define an {\it effective potential} in a
frame rotating at the angular rate $\Omegap$: $$
\textstyle \Phieff = \Phi - {1 \over 2}\Omegap^2 r^2, \eqno(\equno)
$$ where $r$ is the distance from the rotation centre.  The equation
of motion in the rotating frame of reference may then be written $$
\ddot{\br} = -\nabla\Phieff - 2(\Omegap \times \dot{\br}).  \eqno(\equno)
$$ Though the second (Coriolis) term complicates the dynamics, the
gradient of the effective potential at least determines the
acceleration of a particle momentarily at rest in this frame.

\topinsert{
\centerline{\psfig{file=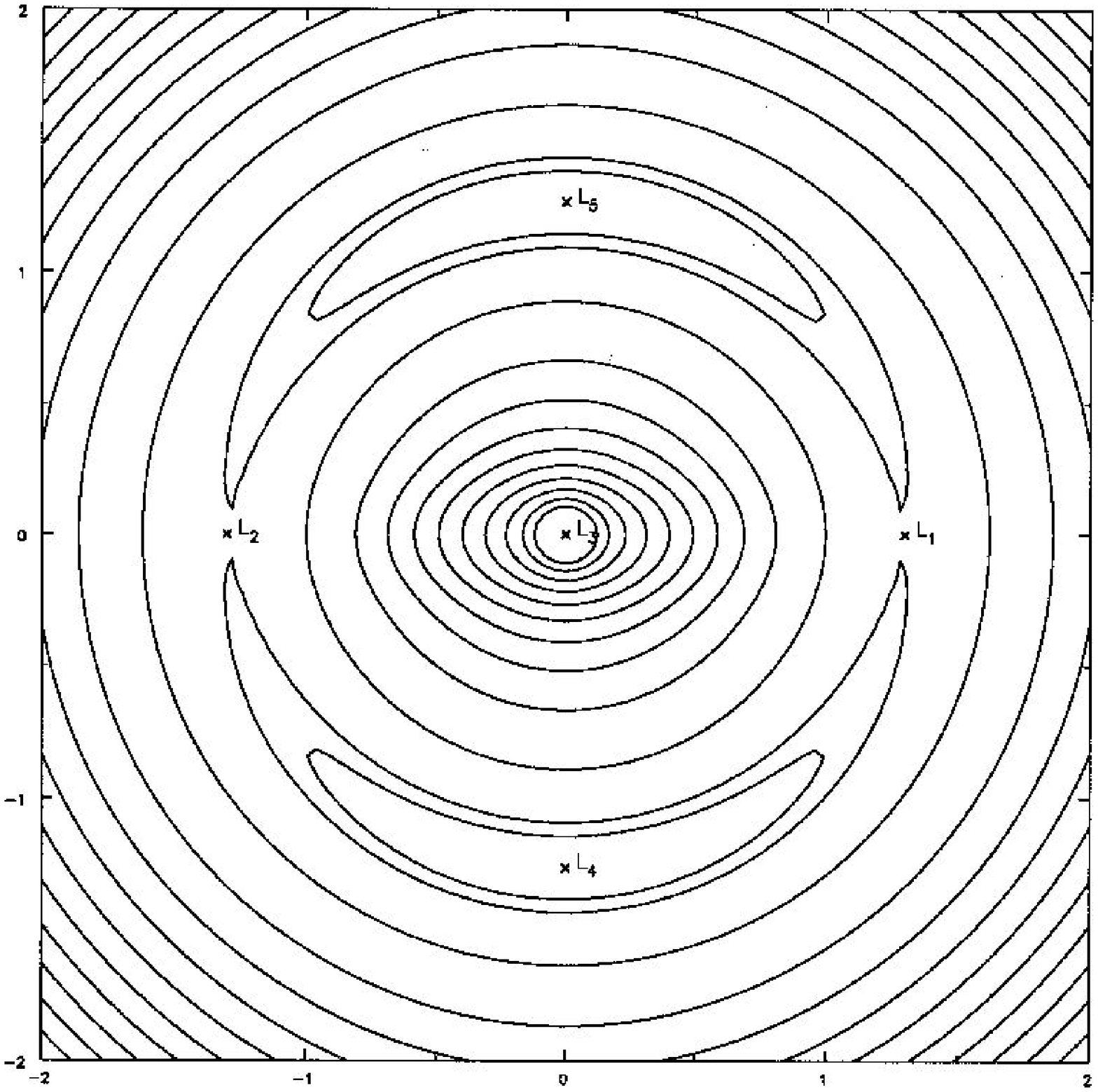,width=.7\hsize,clip=}}
\caption{10}
Contours of the effective potential in our barred model and the
locations of the five Lagrange points.  The L$_3$ point at the centre
is a minimum, the points marked L$_4$ and L$_5$ are equal absolute
maxima and those marked L$_1$ and L$_2$ are saddle points.  Well
beyond this ring of Lagrange points, the effective potential slopes
away parabolically to infinity.  The bar major axis is across the
page. \par
}\endinsert

\nextfig\ shows contours of the effective potential in our model.  The 
topology has been aptly likened by Prendergast (1983) to that of a
volcano; there is a deep minimum at the centre which forms the crater,
a rim whose height varies slightly and steep walls sloping away to
infinity.  There are five {\it Lagrange points\/} altogether at which
the gradient of $\Phieff$ is zero: a minimum at the centre (L$_3$) and
the other four lie at symmetry points on the crater rim: there are two
global maxima on the bar minor axis (L$_4$ and L$_5$), and two saddle
points on the major axis (L$_1$ and L$_2$).  A stationary particle
located at any one of these points would remain at rest in this
rotating frame, or would co-rotate with the bar.  The equilibrium at
the saddle points, L$_1$ and L$_2$, is always unstable, that at the
bar centre, L$_3$, is always stable, but the stability of the minor
axis Lagrange points, L$_4$ and L$_5$, depends upon the details of the
mass distribution (\eg\ Binney and Tremaine
1987);\nextfoot{Unfortunately, their application to the logarithmic
potential has a minor flaw, pointed out by Pfenniger (1990).} they are
stable in many reasonable models, including that we have chosen.

Neither the specific energy, $E$, nor the angular momentum, $J$, of a
particle is separately conserved in a rotating non-axisymmetric
potential, but Jacobi's integral ($\EJ = E - \Omegap J$, introduced in
\S4.2.4) is conserved.  It may also be expressed in terms of the
effective potential: $$
\textstyle \EJ = {1\over2}|\dot{\br}|^2 + \Phieff, \eqno(\equno)
$$ and is therefore the energy with respect to rotating axes.  As a 
consequence, it is often loosely referred to as the energy.

This form for $\EJ$ renders the concept of the effective potential
doubly useful, since it determines whether a particle of a given
energy ($\EJ$) is confined to particular regions of the space.  A
particle for which $\EJ < \Phieff$ at the Lagrange points L$_1$ or
L$_2$ is confined to remain either inside co-rotation or outside it.
Only those particles having $\EJ > \Phieff$ at the Lagrange points
L$_4$ or L$_5$ (the absolute maxima of $\Phieff$) are free
energetically to explore all space.  Note however, that a particle
having a specific $\EJ$ will not necessarily explore the whole region
accessible to it; in particular, particles beyond the Lagrange points
may not be unbound in reality, even though they appear to be so
energetically, because the Coriolis force generally prevents them from
escaping to infinity.

\subsect{Periodic orbits}
A simple periodic orbit is a special orbit which a star would retrace
identically on each passage around the galaxy in the rotating frame of
the perturbation.  More complicated periodic orbits also exist which
close after more than one passage around the galaxy.  All orbits which
close in a steady (rotating) potential are therefore periodic orbits
(\S4.2.4).

Stars in any barred galaxy are most unlikely to follow periodic
orbits, yet the study of periodic orbits is of interest because many
non-periodic orbits are {\it trapped\/} to oscillate about a parent
periodic orbit, in a manner exactly analogous to non-circular orbits
following the path of the guiding centre (which is a periodic orbit)
in an unperturbed or weakly non-axisymmetric potential (\S4.2.1).  A
periodic orbit therefore gives an approximate indication of the shape
of the density distribution of stars trapped about it.

Unlike the infinitesimally perturbed case, however, it is possible to
find orbits which are not trapped about any periodic orbit, but
explore a much larger region of phase space.  This is the
distinguishing property of a near-integrable system (\S3.4).  We may
determine whether an orbit is trapped or not through an examination of
the {\it surface of section} (\S4.6), introduced by Poincar\'e (1892).

\subsubsect{Techniques}
The shooting method is the most widely used to find periodic orbits.
One chooses an initial set of coordinates in phase space, $\bf x_0$,
usually on one of the principal axes of the potential, and integrates
the orbit until it recrosses the same axis at some point $\bf x_1$.
The orbit integration can be considered as a mapping $$
T(\bx_0) = \bx_1, \eqno(\equno)
$$ and a periodic orbit is then defined by $T(\bx_0) = \bx_0$.
The map is, in fact, a Hamiltonian map (\eg\ Lichtenberg and Lieberman
1983) and periodic orbits are also known as {\it fixed points\/} in
the map.  To find such a point, we calculate the mapped points for
other trial orbits in the vicinity of an initial guess $\bf x_0$, to
determine the changes required to $\bf x_0$ in order to move the end
point $\bf x_1$ towards $\bf x_0$.  The calculation proceeds
iteratively and generally converges quickly to a periodic orbit to
high precision.  An alternative relaxation, or Henyey, method is
described by Baker \etal\ (1971), but is less widely used.

Although $\bf x_0$ has four components (for planar motion) there are
only two degrees of freedom for the search, since the Jacobi integral
is conserved by the mapping and a change of starting phase along the
orbit is trivial.  It is customary to search for orbits which cross
the minor ($y$-) axis of the bar with $\dot y = 0$ by adjusting $y_0$
and $\dot x_0$.  All orbits found by this search strategy must be
reflection symmetric about the $y$-axis; searches in which $\dot x_0$
and $\dot y_0$ are varied are required to find asymmetric periodic
orbits.

\subsubsect{Orbit stability}
A star on any periodic orbit will forever retrace exactly the same
path.  Floquet's theorem (\eg\ Mathews and Walker 1970, Binney and
Tremaine 1987) tells us that a star whose starting point is very close
to that of a periodic orbit, will follow a path which either winds
tightly around that of the periodic orbit, or diverges away from it in
an exponential fashion, at least while the departure from the periodic
orbit remains small.  These two types of behaviour imply that the
periodic orbit is, respectively, stable or unstable.

In practice, having found a periodic orbit, one then integrates two
neighbouring orbits, each having the same value of $\EJ$ but displaced
slightly from $\bf x_0$ in different directions, to find the end
points $\bf x_1$ where it recrosses the $y$-axis.  In the
neighbourhood of a periodic orbit, these must be approximately $$
\bx_1 = T(\bx_0+\delta\bx) \simeq \bx_0 +
\left.{\partial T \over \partial \bx}\right|_{\bx_0}\delta\bx.
\eqno(\equno)
$$ The $2\times2$ Jacobian matrix of partial derivatives of $T$ can be
estimated numerically from the displacements of the final point, given
the initial displacements $\delta\bf x$.  Because Hamiltonian maps are
area preserving (\eg\ Lichtenberg and Lieberman 1983), the determinant
of the matrix, and therefore the product of the eigenvalues, is unity.
Since the matrix is real there are just two cases: the orbit is stable
if the eigenvalues are complex and lie on the unit circle, and it is
unstable when the eigenvalues are real and one lies outside the unit
circle.  The two cases are respectively elliptic or hyperbolic fixed
points in mapping parlance.

\subsect{Periodic orbit families}
Numerous orbit families have been described in the literature.  We do
not attempt to enumerate them all here, but confine our discussion to
those which are relevant to the structure of real bars.  Contopoulos
and Grosb\o l (1989) give a more comprehensive review.

A selection of periodic orbits supported by our model (\S4.3.1) is
shown in \nextfig.  In every panel of this figure, the dashed curve
marks the outer boundary of the Ferrers ellipsoid.  Stable periodic
orbits are full drawn, unstable ones are dotted.

There are a number of different naming conventions for the orbit
families.  In \figno0\ we use that developed by Contopoulos, which is
neither descriptive nor easily memorable, simply because it is the
most widely used.  Athanassoula \etal\ (1983) attempted an alternative
system but, like others, theirs is not easily generalized to
three-dimensions, and even those authors have reverted to the
Contopoulos system in their more recent papers.  The $x_1$ family
(\figno0a), also referred to as the {\it main\/} family, is elongated
parallel to the bar within co-rotation.  The $x_2$ family is
represented by the two orbits within the bar in \figno0(b); note that
these are elongated perpendicularly to the bar axis.  We do not show
any $x_3$ orbits which are about as extensive as the $x_2$ family, but
are much more elongated and always unstable.  The retrograde $x_4$
family (\figno0c) are very nearly round, but are slightly extended
perpendicular to the bar.  All four $x_i$ families are 2:1 orbits.
\figno0(d-e) shows both the inner and outer 3:1 and 4:1 resonant
families while \figno0(f) shows the short and long period orbit
families which circulate about the Lagrange points.

We show characteristic curves for most of the main periodic orbit
families in \nextfig.  This diagram, which plots the $y$ (minor axis)
intercepts of the orbits in each family plotted as a function of
Jacobi's integral, should be compared with that for the axisymmetric
potential (\figno{-5}).  Again we differentiate between stable (full
drawn) and unstable (dotted) families; \nb\ all bound orbits in an
axisymmetric potential are stable (\figno{-5}).

There are broad similarities and important differences between Figures
12 and 7.  The most obvious difference is that the circular orbit
sequence in the axisymmetric potential has been broken by numerous
gaps in this strong bar case; these gaps are so large and frequent in
some regions that the sequence becomes hard to trace, especially
inside co-rotation.

\topinsert{
\centerline{\psfig{file=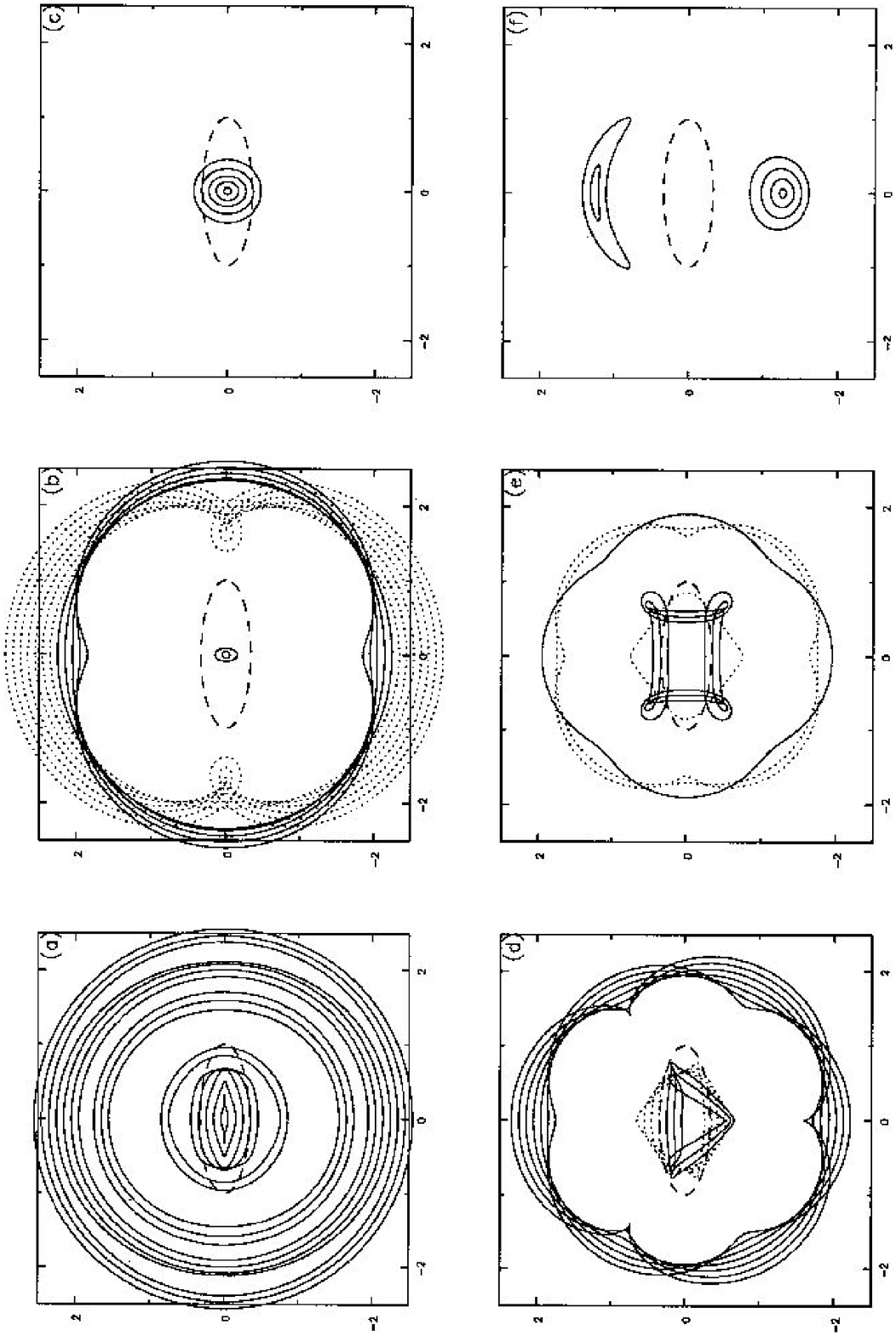,width=\hsize,angle=270,clip=}}
\caption{11}
Some periodic orbits in our barred model; those drawn with full lines
are stable orbits, unstable orbits are drawn as dotted curves.  The
boundary of the elliptical bar is marked with a dashed curve.  Orbits
from the main $x_1$ family are shown in (a).  The small orbits near
the bar centre in (b) are $x_2$ while the others are members of the
outer 2:1 families.  (c) shows the retrograde family $x_4$, (d) \& (e)
show inner and outer 3:1 and 4:1 orbits respectively and (f) shows two
kinds of orbits about the Lagrange points; the elongated banana orbits
are known as long period orbits (LPO), while the rounder are members
of the short period orbit (SPO) family.  \nb\ both SPO and LPO orbits
occur around both Lagrange points. \par
}\endinsert

\subsubsect{Bi-symmetric families}
Starting with retrograde orbits, we see that the $x_4$ family in
\figno0\ closely corresponds to the retrograde circular orbits in
\figno{-5}.  Both the characteristic curve and the orbit shape
(\figno{-1}c) indicate that these counter rotating orbits are not much
affected by the bar potential, being only slightly elongated
perpendicularly to the major axis.  These seem to be generic
properties of the retrograde orbits in every potential investigated.

The direct 2:1 families are substantially affected by the bar;
however, those beyond co-rotation are more easily related to their
counterparts in \figno{-5}.  The orbital eccentricity of the
bi-symmetric families outside co-rotation rises in the vicinity of the
\OLR.  The orbits of the outer nearly circular family are elongated
parallel to the bar (\figno{-1}a) while the inner family are
perpendicular, creating a typical gap of type 1.  All this is
consistent with the linear theory predictions of \S4.2, even though
the eccentricity rises well beyond the linear regime.

\topinsert{
\centerline{\psfig{file=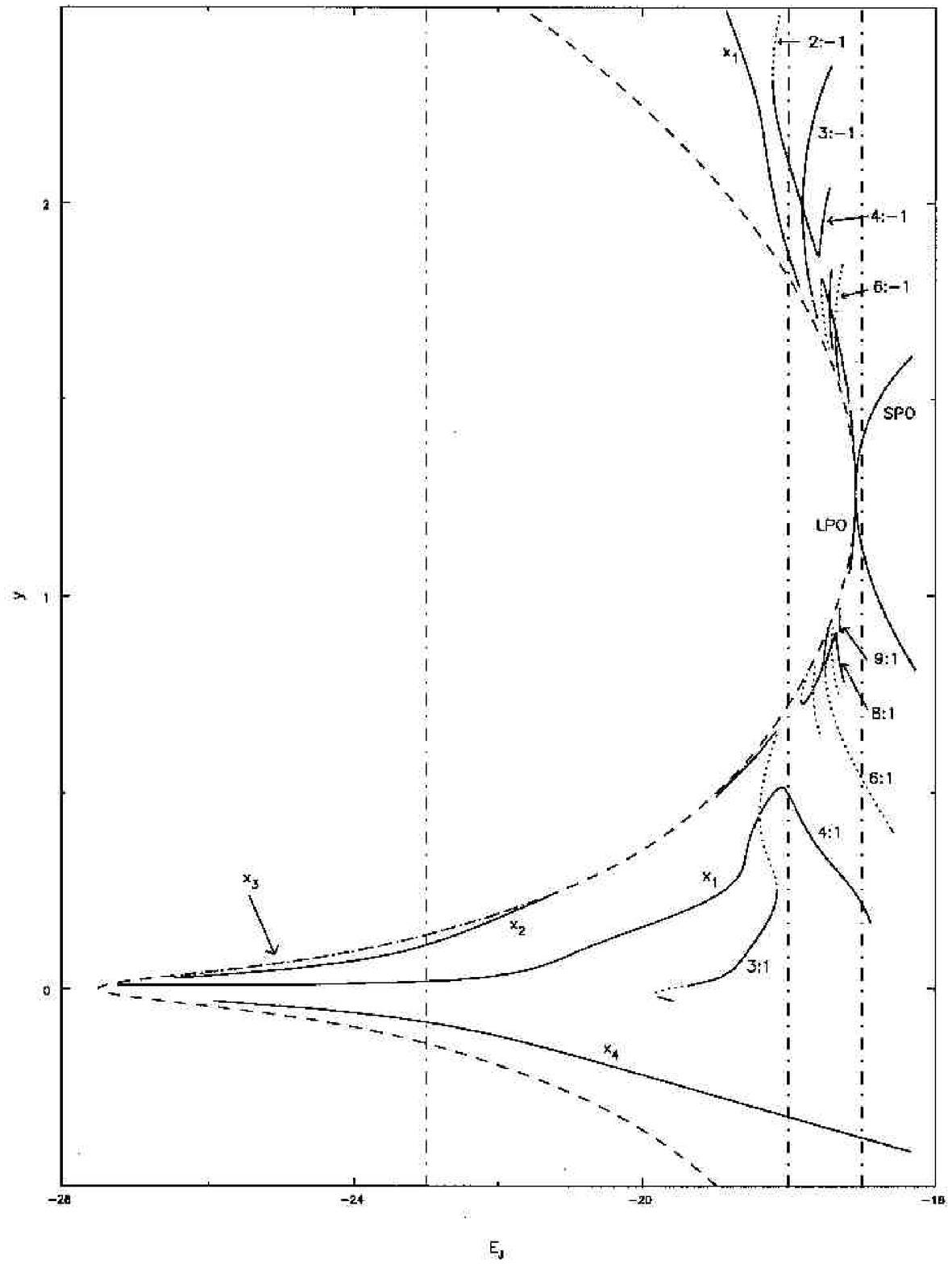,width=.7\hsize,clip=}}
\caption{12}
Characteristic curves for orbits in our bar model.  The point at which
the orbit crosses the bar minor ($y-$)axis is plotted as a function of
$\EJ$ for each family of periodic orbits.  Full-drawn segments of
these curves indicate the orbit family is stable, and unstable parts
of the sequence are marked as dotted lines.  The dashed curve shows
the \ZVC\ and the three vertical dot-dashed lines mark the energies at
which the surfaces of section are drawn in Figure 13.  This diagram
should be compared to figure 7. \par
}\endinsert

The correspondence with \figno{-5}\ is less apparent inside
co-rotation, because what was the circular orbit sequence has been
broken by two yawning gaps.  For $\EJ<-18$, the nearly circular orbits
have become very elongated and the characteristic curve is therefore
much further from the \ZVC.  At the 2:1 resonant gap, the nearly
circular sequence joins to the lower branch of the bubble in
\figno{-5}\ while the continuation of the circular sequence has broken
away and lies much closer to the \ZVC\ to form the $x_2$ sequence.  The
third branch, that which lies very close to the \ZVC\ in \figno{-5},
becomes the highly eccentric and unstable $x_3$ family.  Thus the
$x_1$ family derives from the circular orbits only very near the
centre and near $\EJ=-20$ while at intermediate energies it derives
from the eccentric 2:1 family.  Contopoulos (1983a) calls the closed
loop formed by the $x_2$ and $x_3$ families a floating bubble.  The
appearance of \figno0\ is typical of a strong bar; the resemblance to
\figno{-5}\ is much closer for weaker bars.

The existence of an \ILR\ in the axially symmetrized model is a
necessary, but {\it not sufficient}, condition for the perpendicular
$x_2$ and $x_3$ families to exist.  They are therefore absent when the
central density is less high (Teuben and Sanders 1985), or when the
pattern speed is higher.  However, studies in which the bar strength
is varied show that the gaps at the two 2:1 bifurcations in the
axisymmetric case widen, and the radial extent of the perpendicular
families shrinks, as the bar strengthens.  Eventually, the $x_2$ and
$x_3$ families completely disappear for very strong bars (Contopoulos
and Papayannopoulos 1980).  Van Albada and Sanders (1983) suggest that
the existence and extent of the $x_2$ family can be used to generalize
the concept of an \ILR\ to finite amplitude perturbations.

As also found by Athanassoula \etal\ (1983), the $x_1$ family in our
model does not have loops at higher energies [\figno{-1}(a)].
However, many other papers (Contopoulos 1978, Papayannopoulos and
Petrou 1983, Teuben and Sanders 1985) have reported $x_1$ orbits with
loops, which are sometimes very large.  Athanassoula (1992a) shows
that such loops appear when the bar either rotates more slowly, or has
a high axial ratio, or when the mass distribution is more centrally
concentrated.

\subsubsect{Higher order resonance families}
Because the potential in our model is bi-symmetric, the $m$:$\pm1$
bifurcations in the axisymmetric case become gaps only for even $m$.
Although it is hardly noticeable in \figno0, the $x_1$ family becomes
unstable close to the odd bifurcations; a discussion of this
phenomenon is given by Contopoulos (1983a).

Extensive 3:1 families exist both inside and outside co-rotation, 
\figno{-1}(d).  It is very common for the inner 3:1 family to be unstable, 
but few authors report that the sequence becomes stable again for very
high eccentricities.  Papayannopoulos and Petrou (1983) found a family
of orbits having a similar appearance, but which in their case
resulted from a 1:1 bifurcation in a rather slowly rotating bar.  Our
model rotates too rapidly to exhibit a 1:1 bifurcation; the conditions
under which this could be present are discussed by Martinet (1984).

The gap in the characteristic curves at the inner 4:1 resonance is
particularly wide even for our ellipsoidal bar; square ended bars in
real galaxies can be expected to have still stronger $m=4$ components
to the potential and therefore stronger resonances and yet bigger
gaps.  Notice that the gap in this case is of type 2, which means that
the low energy $x_1$ family joins the more nearly rectangular orbits
of the 4:1 family.  As orbits on this branch remain aligned with the
bar, while orbits on the other branch are roughly diamond shape and
anti-aligned, the gap type at this resonance may be important for
self-consistency.  The type of gap to be expected is more difficult to
predict at these higher order resonances as it depends upon the
relative amplitude and radial variation of more than a single Fourier
component (Contopoulos 1988).  Type 2 gaps seem to be more common
(\eg\ Athanassoula 1992a).  Despite the strong resonance, the
qualitative appearance of the 4:1 orbits, \figno{-1}(e), particularly
further out, is not greatly affected by the presence of the bar (\cf\
\figno{-6}c).

As we move closer towards the Lagrange points, the pattern of resonant
gaps for even families and pitchfork bifurcations for odd, recurs
repeatedly.  The additional families are ever more closely confined to
the vicinity of co-rotation and of rapidly vanishing importance to a
self-consistent bar.

\subsubsect{Orbits around the Lagrange points}
The two families of orbits shown in \figno{-1}(f) are present whenever
the Lagrange points L$_4$ and L$_5$ are stable; note that both
families exist about both Lagrange points.  The more energetic short
period orbits (SPO) derive simply from Lindblad epicycles (\S4.2.1) in
the unperturbed potential.

The banana shaped long period orbits (LPO), are the extension of the
nearly circular sequence, $x_1$ which continues through all the
bifurcations as the Lagrange point is approached (Papayannopoulos
1979).  Once the energy exceeds the effective potential at the major
axis Lagrange points L$_1$ and L$_2$, a nearly circular orbit crossing
the minor axis cannot complete an orbit around the entire galaxy, but
will be turned back as it approaches the bar major axis.  The LPOs are
of exactly the same type as the horseshoe orbits discussed in the
field of planetary ring dynamics (\eg\ Goldreich and Tremaine 1982).

Because they lie at large distances on the minor axis, neither of
these families has any relevance to self-consistent bar models, though
the LPO family may be important for rings (\S7).

\subsubsect{Other types of periodic orbit}
All orbits shown in \figno{-1}\ are reflection symmetric about the
minor axis, and cross this axis with $\dot y = 0$.  Yet more families
of asymmetric periodic orbits exist, some of which are stable.  They
are generally ignored in the majority of papers since stars following
asymmetric families are not expected to be present in large numbers in
symmetric bar models.

Moreover, we have described only simple periodic orbits which return
to the same point on the bar minor axis at each crossing.  Many more
periodic orbits exist for which the star returns to the same point
only after more than one orbit around the galaxy.  Again these are not
thought to be present in great numbers in any bar model, but the
existence of these additional families is very important for the onset
of chaos (see \S4.7).

\pageinsert{
\centerline{\psfig{file=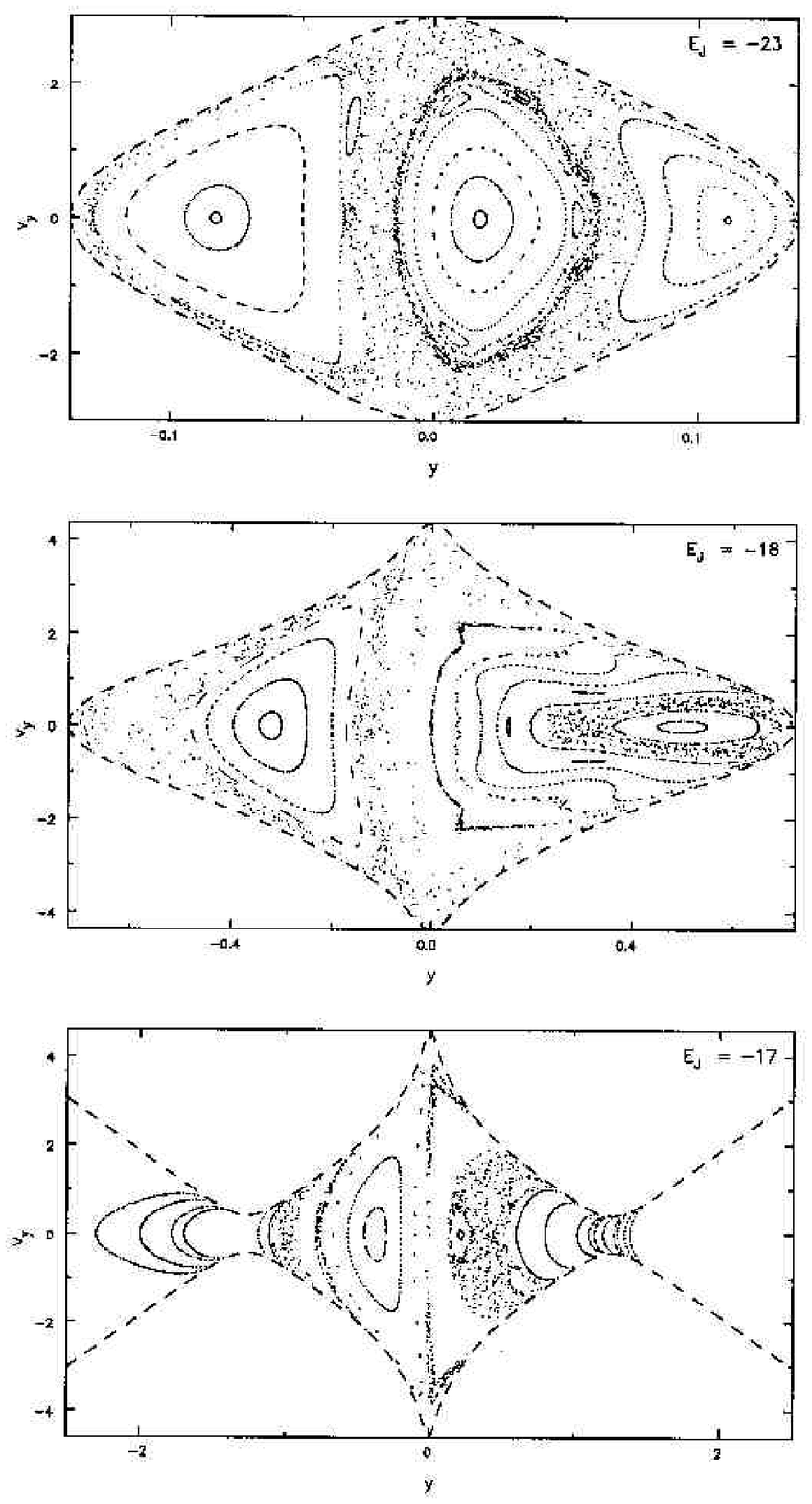,width=.65\hsize,clip=}}
\caption{13}
Surfaces of section at three energies in our model; many different
orbits of the same energy contribute to each plot.  The points mark
the $(y, \dot y)$ values at which an orbit crosses the $y$-axis with
$\dot x < 0$ and more such crossings are marked for complex orbits.
The dashed curves surround the region to which all particles are
confined energetically; at lower energies (top two plots) we show only
the region inside co-rotation, but the energy of particles in the
lower plot is sufficient for them to cross the Lagrange points. \par
\vfill
}\endinsert

\subsect{Non-periodic orbits}
The orbits of most stars in a bar are not periodic, but when most are
trapped to librate about a {\it parent\/} periodic orbit, the
structure of the bar is largely determined by the shapes of the parent
orbits.  It is important therefore to determine the extent to which
orbits are trapped.

A stable periodic orbit must support a trapped region, but the extent
of this region cannot be determined from the stability test (\S4.4.2);
conversely, an unstable orbit has no orbits trapped about it but could
lie in a region where all nearby orbits are trapped about stable
periodic orbits.  The stability test alone therefore does not tell us
the extent of the trapped region and we need the more powerful tool of
the surface of section (SOS) to determine whether one trajectory
through 4-D phase space oscillates about another.  We illustrate this
tool with three examples in \nextfig, again drawn for our model.  Note
that the scales differ substantially between the three panels and that
for the two lower energies, we have not drawn what happens outside
co-rotation.

To construct these diagrams, we integrate the trajectory of a test
particle numerically, and mark the point in the $(y,\dot y)$ plane
each time it crosses the line $x=0$.  The points created by successive
crossings of the $y$-axis are called {\it consequents}.  Only points
for which $\dot x<0$ are plotted; thus points with $y>0$ are created
by a star whose orbital motion is in the direct sense in the rotating
frame.  This is the same Hamiltonian map discussed in \S4.4.1.

Each panel in \figno0\ contains consequents calculated for many
different orbits all having the same Jacobi constant; the dashed
curves mark the boundaries of the region accessible to any orbit of
that energy.  A number of different types of behaviour can be seen.

\subsubsect{Regular orbits}
The single most striking feature in all three panels of \figno0\ is
that in some parts of the plane the consequents lie on closed curves,
known as {\it invariant curves\/}; an orbit which gives rise to an
invariant curve is known as a {\it regular\/} or {\it
quasi-periodic\/} orbit.  Successive consequents are usually well
separated along each curve, but as the orbit is followed for longer
and longer, the consequents populate the curve more and more densely.
Topologically, the star is confined to a two-dimensional toroidal
surface in the 4-D phase space and the closed invariant curve in the
surface of section is simply a cross-section of this torus.  Invariant
curves from different orbits of the same energy can never
cross,\nextfoot{An invariant curve appears to cross itself at an
unstable periodic orbit (\eg\ H\'enon and Heiles 1964) but the curves
there are the limits of two hyperbolae whose apices just touch at that
point.} and characteristically form a nested sequence generally
centred on the single point representing the parent periodic orbit (or
elliptic fixed point).

It is clear from \figno0\ that the trapped region around some periodic
orbits covers a substantial region of accessible phase space.  At
$\EJ=-23$, the regular region on the right is trapped about the family
$x_2$; that near the centre about $x_1$ and the closed curves on the
left are trapped about a retrograde orbit from family $x_4$.

\subsubsect{Irregular orbits}
Not every orbit gives rise to an invariant curve, however.  Well
beyond these three regular regions for $\EJ=-23$, is an example of an
{\it irregular\/} orbit.\nextfoot{In astronomical papers, these orbits
are also variously denoted as {\it chaotic}, {\it stochastic} or even
{\it semi-ergodic\/}, but these terms are not all considered to be
fully interchangeable in other fields.}  The consequents from an
irregular orbit gradually fill an area of the plane more and more
densely with a random scatter of points as the integration is
continued.  In this case, the area is bounded by the zero velocity
curve and the regions occupied by regular orbits.  As can clearly be
seen for $\EJ = -18$ in \figno0, an irregular orbit may also be
confined between two closed invariant curves.  This is an example of a
{\it semi-trapped\/} orbit.

MacKay \etal\ (1984) note that some irregular orbits appear to be
confined to a part of the surface of section for a very large number
of periods, before suddenly crossing to spend a long time in another
part.  The boundary, or {\it cantorus} (Percival 1979), which appears
to separate these regions may be likened to an invariant curve that is
not ``water-tight''.  We have not found a very clear example of this
behaviour in our model.

\subsubsect{Other features}
In the surface at $\EJ=-23$, just outside the fourth invariant curve
around $x_1$, there are two examples of a more complicated invariant
curve: one has three distinct loops, the other seven.  This type of
invariant curve is created by a regular orbit trapped about a periodic
orbit which closes only after several crossings of this plane.  A
further example occurs around the retrograde family, but in that case
the orbit closes after two passages and is also asymmetric.  We have
not drawn the twin of this orbit, its reflection about the $y$-axis,
since it would no longer have been clear that these were a pair of
distinct periodic orbits.

The final type of behaviour of note is to be seen surrounding the
seven-fold regular orbit.  This orbit illustrates a dissolving
invariant curve; the consequents surround the chain of islands, and
although the orbit is clearly not regular, it does not escape to the
stochastic region just outside it for many crossings.  This is a
characteristic feature in the SOS when the degree of regularity is
changing with energy (H\'enon and Heiles 1964); at energies just a
little lower ($\EJ = -24$), phase space is entirely regular, while at
slightly higher energies ($\EJ = -22$), the trapped region around
$x_1$ is very small.

\subsubsect{Higher energies}
In the SOS at $\EJ=-18$, the regular region on the directly rotating
side has recovered somewhat, but it is centred around the 4:1 orbit
near $y=0.5$, which is well outside the bar.  (The bar semi-minor axis
is $1\over3$.)  The regular region around the retrograde $x_4$ family
is still strongly evident, but the energy is now too great for the
$x_2$ family to be present.

At $\EJ=-17$, the orbits are unconstrained energetically.  Phase space
near the Lagrange points is still regular, but some invariant curves
appear as segments at both positive and negative $y$ because, as it
oscillates about the parent periodic orbit, the orbit may cross the
minor axis going in either direction on either side of the galaxy.  As
such orbits spend most of their time on the minor axis well outside
the bar they cannot be of any real importance to the maintenance of
the bar density.

\subsect{Onset of chaos}
The reason for the dissolution of invariant curves is one of the
principal concerns of non-linear dynamics.  Lichtenberg \& Lieberman
(1983) and H\'enon (1983) give accounts of this complex issue, but we
have also found the article by Dragt and Finn (1976) and the informal
review by Berry (1978) especially helpful.

Unstable periodic orbits provide the key.  Although most lie in
irregular regions of phase space, not all of them do -- examples are
in the original H\'enon and Heiles experiment and our model discussed
here.  Unstable periodic orbits can be found in regular regions when
the two branches of the invariant curve from that point meet again
only at another unstable periodic orbit, forming a {\it separatrix}.
The region around the fixed point suddenly becomes ergodic as soon as
two branches from hyperbolic fixed points fail to join smoothly, but
cross at some point other than a fixed point.  The breakdown of
large-scale regularity seems to take place in stages, however, as
chains of islands of stability survive for short energy ranges around
multiply periodic stable orbits.  These chains are often surrounded by
dissolving orbits (as in \figno0) signifying the breakdown of the
corresponding separatrix.

The KAM (Kolmogorov, Arnol'd and Moser, \eg\ Moser 1973) theorem
concerns the survival of invariant tori when an integrable system,
such as an axisymmetric disc, is subjected to a perturbation, such as
a bar.  The theorem states that those tori ``sufficiently far from
resonance'' survive, in a deformed state, when a ``sufficiently
small'' perturbation is imposed.  It does not predict the strength of
the perturbation sufficient to destroy a torus, but it does give a
scaling law related to the order of the resonance concerned and the
associated frequencies.  Thus we should expect, as we have found, that
trapped orbits exist in non-axisymmetric potentials.  We also observe
that the extent of the regular region of phase space generally
diminishes as the strength of the bar perturbation rises.

This last observation is particularly true near co-rotation where ever
more resonances occur as co-rotation is approached.  As the strength
of the perturbation rises, more and more of these resonances overlap,
which is a condition for the destruction of regularity (Chirikov
1979).  Athanassoula (1990) has argued that the square-ended nature of
the bar density distribution may also hasten the onset of chaos in
this region, because the $m=4$ component of the potential has greater
strength relative to the $m=2$ than would a more elliptical potential,
causing greater overlap of resonances for a fixed bar strength.

Further, Martinet and Udry (1990) note that interactions of higher
order resonances in the vicinity of the unstable $x_3$ family seem to
be a particularly effective generator of chaos.  They argue that the
contraction of this family as the angular speed of the bar is raised
may account for the apparent reduction in the chaotic fraction of
phase space found in faster bars.

Contopoulos (1983a) noted that an infinite sequence of period-doubling
bifurcations in the characteristic diagram is a second factor which
gives rise to ergodic behaviour.  This seemed to occur over a wide
region around co-rotation in his strongly barred case (Contopoulos
1983b).

Liapunov exponents (Liapunov 1907), which give a quantitative measure
of the degree of stochasticity in an irregular region, have been
calculated in recent studies (\eg\ Udry and Pfenniger 1988,
Contopoulos and Barbanis 1989).  They are a set of exponents
describing the rate of separation of two nearby orbits in phase space
as the motion proceeds; the region is regular if all the exponents
vanish, irregular otherwise.  They are formally defined for orbits
infinitely extended in time, but in astronomical systems the interest
lies in the behaviour over a Hubble time.

\subsect{Actions}
In steady potentials, one quantity is conserved for all orbits: the
energy, if the potential is non-rotating, or Jacobi's integral in a
non-axisymmetric rotating potential.  If this were the only conserved
quantity, the particle would be able to explore all parts of phase
space accessible with this energy and the entire surface of section
would be filled by one single irregular orbit.

We already noted that regular orbits are, however, confined to a
two-dimensional toroidal surface, and therefore respect an additional
integral, other than the energy.  In an axisymmetric potential, this
additional conserved quantity is obviously the angular momentum, but
no such simple physical quantity can be identified in non-axisymmetric
potentials, whether stationary or rotating.

Binney and Spergel (1982, 1984) give a vivid illustration that regular
orbits have just two independent oscillation frequencies and can
therefore be described by action-angle variables.  Irregular orbits on
the other hand, are not quasi-periodic and cannot be described by such
variables.  Their first application was for a planar non-rotating
potential but the technique is not restricted to this simple case.

Since the actions {\it are\/} a set of integrals, they would furnish
the ideal variables with which to describe regular regions, in as much
detail as for integrable systems, such as St\"ackel models (\S3.3).
We could write the dynamical equations in a very simple form (\eg\
\S9.2), express the distribution function in terms of the actions
(Jeans theorem), and take advantage of their adiabatic invariance
(away from resonances) for the study of slowly evolving models.

Their principal drawback, however, is that we have no analytic
expressions for them, though they can be determined numerically from
the area bounded by the invariant curve in the appropriate surface of
section (\eg\ Binney \etal\ 1985).  Worse, we cannot easily transform
back to real space coordinates in order, for example, to compute the
shape of an orbit of known actions.  This very serious limitation is a
major handicap to progress in the entire subject.  Ratcliff \etal\
(1984) suggested a general technique based upon Fourier expansion, but
ran into a number of operational difficulties.  The canonical mapping
approach outlined by McGill and Binney (1990) may be more successful.

\subsect{Self-consistency}
It is widely believed that self-consistent bars are largely supported
by stars on orbits trapped or semi-trapped about the $x_1$ family,
which is highly elongated in the direction of the bar.  Most other
families are too round, or elongated in the opposite sense, to make
any useful contribution in a self-consistent model.  Somewhat
surprisingly, the 4:1 family does not appear to be responsible for the
``rectangular'' shapes of real bars (\S2.3.1).

Contopoulos (1980) was the first to argue that the properties of the
$x_1$ family suggest that self-consistent bars were likely to extend
almost as far as co-rotation.  Teuben and Sanders (1985) concluded
that stars in such rapidly rotating bars are likely to move in a well
organized streaming pattern, much as observed (\cf\ \S2.5) whereas
much less coherent streaming would be expected were the bar to rotate
more slowly.  Petrou and Papayannopoulos (1986) argued that
self-consistent models having much lower pattern speeds might also be
possible.  However, the majority of authors favour pattern speeds high
enough that co-rotation is not far beyond the end of the bar.

We have already noted that the nearly round, anti-aligned shapes of
retrograde orbits within the bar implies that they cannot be
significantly populated in any self-consistent model of a bar (see
also Teuben and Sanders 1985).  However, these orbits do play an
important negative r\^ole in self-consistent bars, since the
remarkably large regular regions that surround this family in \figno0\
(and in most other studies) mean that it ``reserves'' a substantial
fraction of the phase space volume which the stochastic orbits cannot
enter.

Irregular orbits raise a major difficulty for exactly self-consistent
models that would survive indefinitely, though for practical purposes
this may be too abstract a theoretical requirement.  As their name
implies, irregular orbits follow chaotic trajectories with no
periodicities whatever.  The density distribution produced by a
population of stars having irregular orbits can never be steady,
therefore.  It is not clear what this means in practice: bars in
galaxies are typically quite young dynamically, perhaps only 50
rotation periods.  We might speculate that a significant fraction of
irregular orbits may not have begun to explore the full space
available to them, or the changes they cause could be sufficiently
small and slow that the bar can adjust continuously without being
weakened or destroyed.  It is far from clear how these ideas can be
tested.

\subsubsect{Tour de force}
Pfenniger (1984b) presents by far the most extensive attempt to show
that some mathematically convenient elliptical bar model could be made
self-consistent.  Adopting Schwarzschild's approach (\S3.5), but with
a non-negative least-squares algorithm instead of linear programming,
he managed to obtain self-consistent models which, to our knowledge,
remain the only solutions in the literature for a rapidly rotating
two-dimensional bar.

In constructing his orbit library, Pfenniger considered that the
density distribution had converged to a steady state when the largest
change in the occupation number of any cell dropped below 0.5\% upon
doubling the integration time.  Most regular orbits converged after a
short integration, but when his criterion proved impractical for some
irregular orbits, he imposed an arbitrary maximum integration time of
about 550 bar rotation periods.

Pfenniger was able to use his approximately self-consistent solutions
to calculate velocity and velocity dispersion fields, and found at
least four different forms.  The simplest flow, for the maximum
angular momentum, consisted of directly rotating elongated flowlines
within the bar and circular flowlines corresponding to the inner ring.
This flow pattern was the one which agreed most closely with
observations of early-type galaxies and $N$-body models.  Other more
complicated eddying flows were also possible.  Some retrograde orbits
were always required, involving typically between 10 and 30\% of the
mass, and populating the lens-like part of his assumed elliptical bar
model.  Models could be constructed without irregular orbits, but
usually about 10\% of irregular orbits were required.  The dispersion
fields showed little anisotropy, with the dispersion decreasing from
the central value by a factor of about two by the co-rotation radius.

\subsubsect{N-body model}
An entirely different approach was adopted by Sparke and Sellwood
(1987), who examined the orbital structure of a bar formed in an
$N$-body simulation.  They were able to find many of the usual orbit
families in the frozen potential of the model, and determined, from
their distribution in the SOS, that the majority of particles making
up the bar were either trapped, or semi-trapped about the $x_1$
family.  Their $N$-body bar had a quite realistically rectangular
appearance, yet they found, somewhat surprisingly, that the 4:1 family
was of little importance.  Instead, the orbits librating about the
$x_1$ family seemed to be responsible for the rectangular shape.

Sparke and Sellwood noted that their two-dimensional bar model was
remarkably robust and could adjust essentially immediately to major
alterations of the global potential.  Unfortunately, their conclusion
applies only to models restricted to two-dimensions; once particles
are allowed to move normal to the symmetry plane, the bar appears to
suffer another type of instability (see \S10.1).

\subsubsect{Photometric models}
There is a recent welcome trend in the literature to adopt mass
distributions which bear some resemblance to the light distributions
of barred galaxies, which nature has constructed self-consistently.
One of the first such attempts was made by Kent and Glaudell (1989)
for the well studied SB0 galaxy, NGC~936.  This galaxy is one of the
most favourable for such a study (Figure~2), being bright and nearby,
largely free from star-formation regions and obscuring dust and viewed
from an ideal angle.  Unfortunately, even for this galaxy, major free
parameters remain essentially undetermined by the observational data,
the most important being the pattern speed of the bar and the
mass-to-light ratio of the stellar populations; the latter is most
unlikely to be a universal constant throughout one galaxy.  Kent and
Glaudell attempted to constrain the mass-to-light ratio from the
observed velocity field and experimented with two pattern speeds.
Their results were rather weakly in favour of the higher of the two
pattern speeds, which places co-rotation a little beyond the end of
the bar.

A more recent study of the same galaxy, based on new photometric data,
has been undertaken by Wozniak and Athanassoula (1992).  Instead of
trying to decompose the model into components, they adopt a mass model
based directly on the observed light distribution.  They concur with
Sparke and Sellwood (1987) that the rectangular bar shape is supported
by orbits trapped about $x_1$, and owes little to the 4:1 family.

\sect{Three-dimensional bar models}
The $N$-body experiments of Combes and Sanders (1981) provided the
first indication that an exclusively two-dimensional treatment of
barred galaxies is inadequate, although their result was not
understood at the time.  We now believe that the thin disc
approximation cannot be invoked for bars for two distinct reasons:
firstly, vertical resonances occur in the bar which couple horizontal
to vertical motions and secondly, thin bars are subject to
out-of-plane buckling instabilities.  (We describe buckling modes
together with other forms of bar evolution in \S10, after discussing
bar formation.)  Thus, a simple addition of small vertical
oscillations to the orbits discussed in the previous section would
give a seriously incomplete description of the three-dimensional
dynamics of bars.  Work on fully three-dimensional models of rapidly
rotating bars is still in its infancy, however.

\subsect{Vertical resonances}
For a nearly circular orbit in a weakly perturbed potential, we expect
resonances between the vertical oscillation and a rotating, $m$-fold
symmetric perturbation wherever $m[\Omegac(r) - \Omegap] =
n\kappaz(r)$; here $\kappaz$ is the frequency of oscillation normal to
the symmetry plane and $n$ is an integer.  The vertical co-rotation
($n=0$) resonance is of no dynamical importance, but the $n \ne 0$
resonances could couple motion in the plane to vertical excursions
(Binney 1981).  The $n=\pm1$ resonances are sometimes known as
``vertical Lindblad resonances'' and the $n=\pm2$ resonances give rise
to the much discussed ``Binney instability strips'' which occur only
for retrograde orbits in the inner galaxy or far outside co-rotation
for direct orbits.  (\nb\ $m/n$ in our notation $=n$ in Binney's.)

It is useful to compare $\kappaz$ to $\kappa$.  We write Poisson's
equation for an {\it axisymmetric\/} mass distribution as $$
{1\over r}{\partial \over \partial r}\left( r{\partial \Phi \over
\partial r}\right) + {\partial^2 \Phi \over \partial z^2} = 4\pi
G\rho. \eqno(\equno)
$$ Recognizing that the combination $r{\partial \Phi \over \partial
r}$ is the square of the circular velocity, the first term vanishes in
the mid-plane of a galaxy having a flat rotation curve.  The remainder
of the equation then gives us the standard result $$
\kappaz = \sqrt{4\pi G\rho_0}, \eqno(\equno)
$$ where $\rho_0$ is the density in the mid-plane.  To estimate
$\kappa$, we note that $\kappa^2 \sim 2\Omegac^2$ (exact for a flat
rotation curve), and that $\Omegac^2 \sim {4\pi \over 3}G \bar\rho$,
where $\bar\rho$ is the mean, spherically distributed, density of
matter interior to the point in question.  Thus the ratio $\kappaz /
\kappa \sim \sqrt{\rho_0 / \bar\rho}$ which is large wherever the
self-gravity of the disc is important (Tremaine 1989).

The inequality $\kappaz\gg\kappa$ holds for the majority of stars
which remain close to the plane and implies that the $n=\pm1$
resonances lie further from co-rotation than do the horizontal
Lindblad resonances, which delimit the spiral pattern.  The perturbing
potential for spiral waves in a cool disc can therefore be expected to
be negligible at the first, and all subsequent, vertical resonances.
Thus it is legitimate to ignore vertical resonant coupling when the
orbits of stars are nearly circular, the disc is thin and the
non-axisymmetric component of the potential weak.

Obviously, expression (\equat0) for $\kappaz$ fails in a strong bar.
Not only does a strong bar add a large non-axisymmetric term to
(\equat{-1}), but stars also move on highly eccentric orbits.
Therefore local estimates of the vertical oscillation frequency cease
to be meaningful for an orbit and the existence of resonances can be
determined only from orbit integrations.  Pfenniger (1984a) was the
first to show that vertical resonances were important within the bar
of a reasonably realistic three-dimensional model.

\subsect{Periodic orbits in three-dimensions}
Unfortunately, very few studies of three-dimensional periodic orbits
have been published for models which bear much resemblance to barred
galaxies; the large majority are concerned with slowly rotating,
tri-axial ellipsoids.  We mention this work insofar as it seems
relevant, but space considerations preclude a thorough review of the
literature on the orbital structure of slowly rotating ellipsoidal
models (see \eg\ de Zeeuw and Franx 1991).

\subsubsect{Orbital stability}
As for rotating two-dimensional potentials, Jacobi's integral is
conserved for all orbits (\S4.3.2) in a potential which is steady in
rotating axes.  There are therefore four adjustable coordinates for
the shooting method (\S4.4), which are usually $(y, \dot y, z, \dot
z)$, and again we seek solutions such that $T(\bx_0) = \bx_0$.
(For consistency with the previous section, we use $y$ for the bar
minor axis.)

The stability test for a periodic orbit now requires the determination
of the eigenvalues of a $4 \times 4$ Jacobian matrix of partial
derivatives (equation \equat{-2}).  As for two-dimensions, the orbit
is stable when all eigenvalues lie on the unit circle.  However, there
are six physically distinct types of instability which have been
discussed for a very similar celestial mechanics problem by Broucke
(1969), Hadjidemetriou (1975) and H\'enon (1976), and for bars by
Pfenniger (1984a) and Contopoulos and Magnenat (1985).  The phenomenon
of {\it complex instability}, which appears only in systems with more
than two degrees of freedom, has been discussed extensively; in this
case, all four eigenvalues are complex and not located on the unit
circle.  What causes an orbit sequence to become complex unstable is
not well understood, but the transition frequently occurs at
boundaries of stochastic regions (Magnenat 1982b).

\subsubsect{Notation}
A very substantial fraction of the work in this area has been carried
out by the group at the Geneva Observatory.  Notwithstanding the
importance of their work, we find the notation they adopt, which is an
attempt to generalize from two-dimensions, so cumbersome that we feel
unable to use it here.  To our knowledge, there are at least four
other competing systems in the literature -- an indication perhaps of
the new and unsettled nature of the field, and a source of confusion
to all but the expert.  Since no consensus on nomenclature has been
reached, and we do not find any suggested systems attractive, we have
developed our own!  It does, however, share features of some of the
other systems.

We extend the $m$:$l$ notation in the plane to become $m$:$n$:$l$,
where $m$ is the number of radial oscillations, when viewed from the
rotation axis (as for two-dimensions) and $n$ the number of vertical
oscillations before the orbit closes after $l$ rotations about the
centre.\nextfoot{This differs from the superficially similar notation
used by Schwarzschild, which counts oscillations parallel to the
coordinate axes.  His notation is well suited to non-rotating bars.}
Unfortunately, this nomenclature needs to be supplemented to indicate
whether the orbit is symmetric or anti-symmetric about the $(x,z)$
plane, \ie\ the plane through the bar centre normal to its major axis.
When there is a need to distinguish these we add a subscript; thus,
$m$:$n_{\rm s}$:$l$ and $m$:$n_{\rm a}$:$l$ refer to the symmetric and
anti-symmetric families respectively.  We illustrate this nomenclature
with the orbits discussed in this section.

Our system, like most others, is too rigid to describe transition
regions between families.  A glance at \figno{-1}\ shows that
distinctly named families frequently join smoothly into a single orbit
sequence in two-dimensions; any notation system based upon orbital
shapes will also ascribe different names to parts of a sequence in
three-dimensions.  The gradual change in the properties of orbits over
the transition region is not allowed for in notation systems which
require an abrupt shift to a new designation at some arbitrary point
along a sequence.  We regard this weakness as more than outweighed by
the advantage of a notation which clearly indicates the shape of the
orbit away from transition regions.

\topinsert{
\centerline{\psfig{file=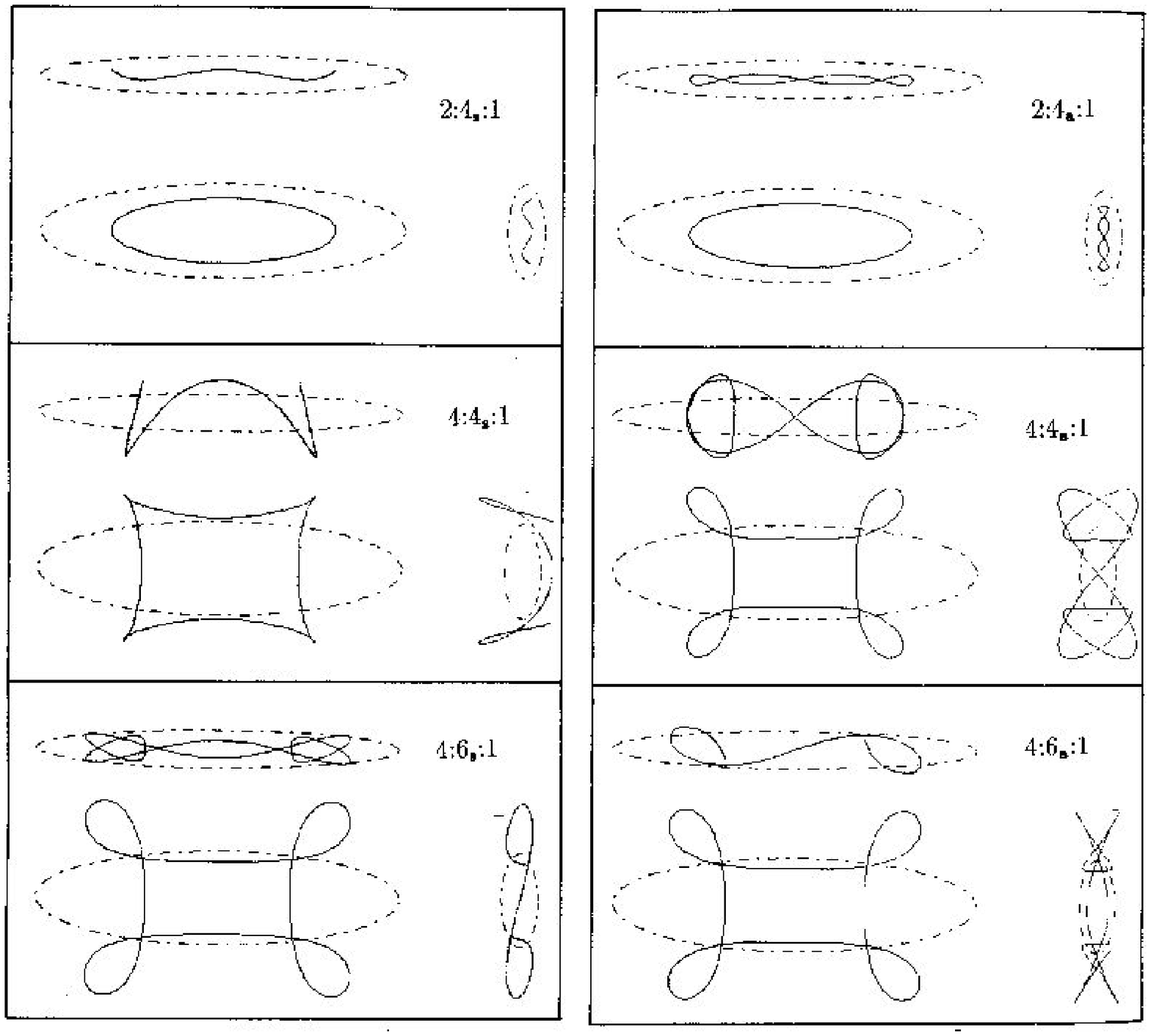,width=.7\hsize,clip=}}
\caption{14}
Six three-dimensional periodic orbits found by Pfenniger (1984a); the
three on the left are symmetric and those on the right are their
anti-symmetric counterparts. We have re-labelled these orbits using
our notation. \par
}\endinsert

\subsubsect{Periodic orbit families}
Pfenniger (1984a) investigated orbits in the combined potential of a
rotating, tri-axial Ferrers bar with axes in the ratio 1:4:10, and a
thickened Kuz'min disc (Miyamoto and Nagai 1975).  He made the bar
rotate about the short axis at a rate which placed co-rotation at the
end of the long axis.  First he examined the periodic orbits in the
equatorial plane and found that the main $x_1$ family had something
like the usual form.  As his model was insufficiently centrally
condensed to contain \ILR s when axially symmetrized, it could not
support the perpendicular families $x_2$ and $x_3$ of periodic orbits
in the plane.

The $x_1$ family was vertically stable throughout most of the inner
part of the bar, but was vertically unstable over several short
stretches at higher energies.  These instability strips lay between
pairs of bifurcations from which new periodic orbit families with
non-zero vertical excursions branched off.  The first pair of
bifurcations gave rise to families of orbits which retained an oval
appearance from above, while developing small wrinkles when viewed
from the side (\nextfig\ top).  These were symmetric and
anti-symmetric 2:4:1 families.  As both formed short sequences in the
characteristic diagram and soon rejoined the main $x_1$ family, they
were probably of little importance.  At higher energies, the next two
vertical families to appear were 4:4:1 orbits (again symmetric and
anti-symmetric, \figno0\ middle) having much more extensive
characteristic curves.  They had the distinctive shape of the 4:1
orbits in the plane as well as a 4:1 oscillation normal to the plane.
A third pair of bifurcations led to the 4:6:1 orbit families (\figno0\
bottom), which Pfenniger again viewed as less important.  He found
that all three anti-symmetric families were generally unstable over
most of their length, but the symmetric sequences were stable over
long regions.

Only the 4:4:1 families remained when he reduced the bar mass
substantially.  It seems likely that the low central density of his
model prevented him from finding any $m$:2:1 families.

Pfenniger also tested the effect of changing the axial ratios of the
bar.  He found that both the horizontal and vertical stability of the
$x_1$ family of orbits decreased as the bar became thinner in the
plane of the disc; when the axis ratio in the plane exceeded $\sim 5$
-- 7 most of the $x_1$ sequence became unstable.  Increasing the bar
thickness perpendicular to the plane (as far as making it prolate)
also improved both the vertical and horizontal stability, probably
because the non-axisymmetric component of the potential weakens as the
bar is made thicker.

Most subsequent studies have been concerned with slowly rotating
tri-axial ellipsoids, though some of the results are relevant to
barred galaxies.  In particular, directly rotating 2:2:1 resonant
families seem to be most relevant to galactic bars.  The 2:2$_{\rm
s}$:1 orbits are commonly known as ``bananas'', and the Geneva group
uses the deplorable term ``anti-bananas'' to describe the 2:2$_{\rm
a}$:1 orbits!\nextfoot{It should be noted that these orbits in a
rapidly rotating bar differ from those also termed anti-bananas in
non-rotating bars (Miralda-Escud\'e and Schwarzschild 1989) which pass
through the exact centre of the bar.}  Both these families bifurcate
from the main $x_1$ family in most models (Mulder and Hooimeyer 1984,
Cleary 1989) but they can also be found as bifurcations from $x_2$
when the bar rotates sufficiently slowly (Udry 1991). The family in
the plane seems always to be vertically unstable between the two
bifurcation points, which are generally quite close together.  Most
authors find that the symmetric family is stable and the
anti-symmetric unstable from their bifurcation points, but that
stability is exchanged between them at a higher energy.  Mulder and
Hooimeyer (1984) found an additional intermediate family connecting
the two at the energy where stability was exchanged, but Udry (1991)
could not, unless the potential was perturbed.

Udry also maps out the limiting bar axis ratios for which the 2:2:1
families can be found in a model having a tri-axial modified Hubble
density profile.  He finds that the short axis should not be more than
30\% to 40\% of the long axis, with only a slight variation over the
entire range possible for the intermediate axis.  He also notes that
these values are hardly affected by rotation, although he did not
investigate very rapidly rotating models.  The addition of a small
Plummer sphere at the centre of the mass model further confined the
existence of these families to much flatter bars, however.

\subsect{Structure of a three-dimensional $N$-body model}
Pfenniger \& Friedli (1991) have carried out a detailed study of the
orbital structure of one of their three-dimensional $N$-body
simulations.  As the bar in their model thickened in the $z$-direction
quite markedly during its first few tumbling periods (see \S10.1),
they chose two different moments for their study: one soon after the
bar formed as a thin structure, and the second much later when
evolution seemed to have ceased.  They first searched for periodic
orbits in the frozen potentials at the two instants, both in the raw
potential of the model and when they imposed reflection symmetry about
the three principal planes.  The pattern speed of the bar dropped by
some 20\% and the approximate axis ratios of the bar shape rose from
$\sim 1:0.42:0.33$ to $\sim1:0.51:0.40$ between the two moments
analysed.

\topinsert{
\centerline{\psfig{file=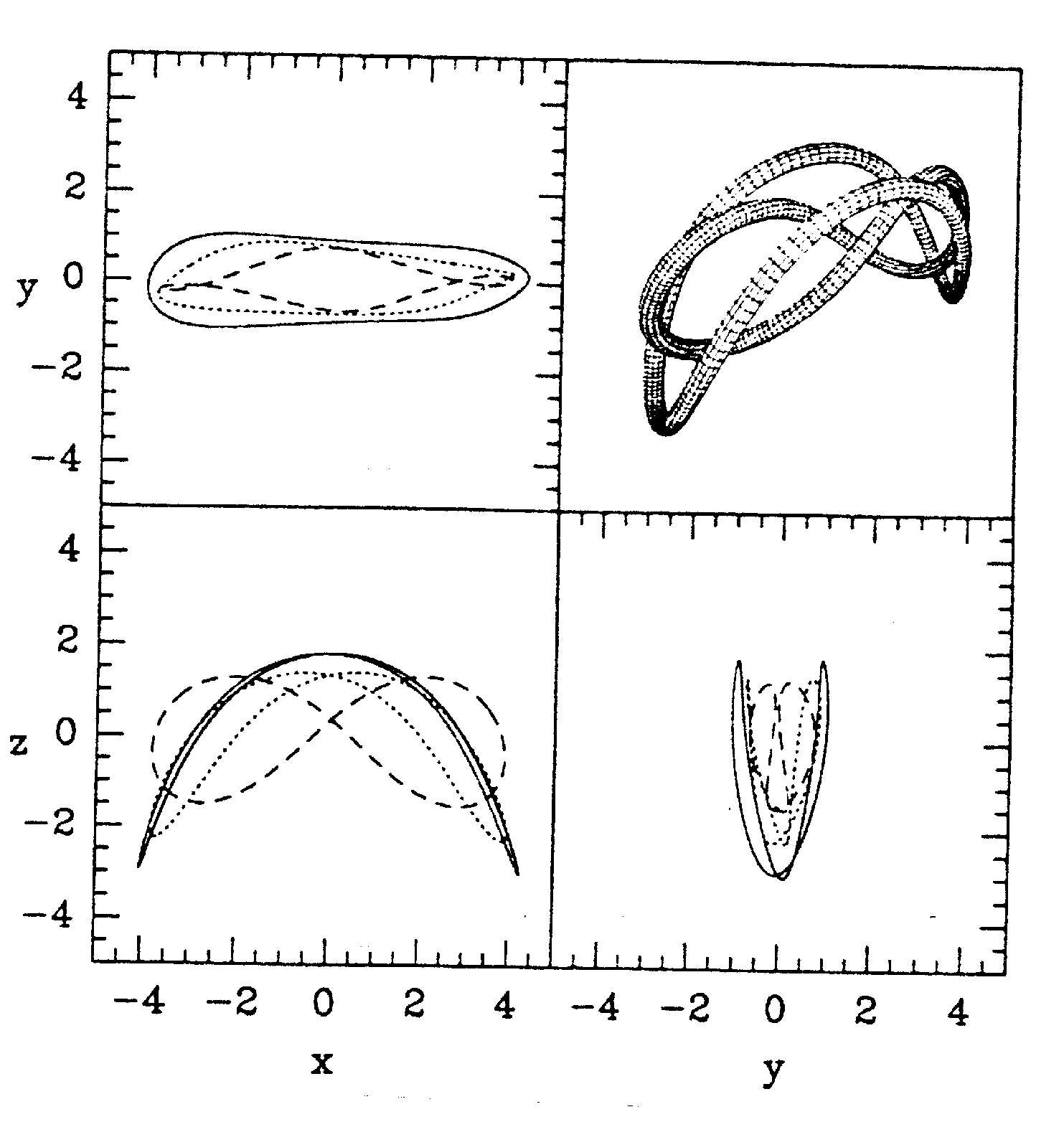,width=.5\hsize,clip=}}
\caption{15}
Three orthogonal projections and a ``tube view'' showing the
three-dimensional shapes of both the symmetric and anti-symmetric
2:2:1 orbits.  The symmetric orbit is drawn as a continuous line and
the anti-symmetric is dashed; the dotted lines show an intermediate
case!  [Reproduced from Pfenniger and Friedli (1991).] \par
}\endinsert

In the symmetrized cases, they found that the planar $x_1$ family was
stable, both horizontally and vertically, from the centre almost to
the 3:1 bifurcation, except for a short vertical instability strip.
The perpendicular families in the plane ($x_2$ and $x_3$) were absent
from their model, but only marginally so.  The bifurcations at either
end of the short vertical instability strip in the main $x_1$ family
lead to the symmetric and anti-symmetric 2:2:1 families, illustrated
in \nextfig.  As usual, the anti-symmetric family was unstable and the
symmetric family stable near the plane, but they exchanged stability
at a higher energy.  The vertical extent of the stable part of the
symmetric family had decreased quite markedly, and the bifurcation
points moved out along the bar, by the later time.

There were no simple periodic orbits which remained precisely in the
plane in the raw (unsymmetrized) potential of the $N$-body model,
because the potential was not exactly symmetric in $z$ at either time.
Nevertheless, they could trace families generally resembling those in
the symmetrized model, but with two important differences: first, the
vertical bifurcations, which were of the pitchfork type in the
symmetrized model, became resonant gaps in the raw potential, and
second, the 2:2$_{\rm a}$:1 families became much more stable while the
2:2$_{\rm s}$:1 families lost still more of their stability.  These
stability differences were particularly marked at the later time.

Pfenniger and Friedli noted that the majority of particles in the bar
seemed to follow quasi-periodic orbits trapped about the $x_1$ and the
2:2:1 families.  Very few orbits were retrograde or belonged to any of
the other families they found from their periodic orbit analysis.  For
energies approaching the Lagrange point, they found larger and larger
fractions of irregular orbits.

Given the stability properties of the 2:2:1 families, it seems likely
that stars in the box-shaped final model are trapped about the
2:2$_{\rm a}$:1 family; a conclusion supported by the preliminary
orbit analysis of his own models performed by Raha (private
communication).  These anti-symmetric orbits were first reported by
Miller and Smith (1979) who found them in large numbers in their
slowly rotating ellipsoidal model.

Pfenniger and Friedli note that their model also supported vertical
bifurcations along the retrograde $x_4$ family, which gave rise to
2:1:1 orbits, also known as anomalous retrograde orbits.  Their study
of the orbital make-up of the bar makes it clear that these are of
little importance in fast bars, though they have been of major
interest for slowly rotating tri-axial systems (Heisler \etal\ 1982,
Magnenat 1982a, Mulder and Hooimeyer 1984, Martinet and Pfenniger
1987, Martinet and de Zeeuw 1988, Cleary 1989, Martinet and Udry 1990,
\etc).

\subsect{Stochasticity in three-dimensions}
The most important difference between two and three degrees of freedom
is that the phenomenon of {\it Arnol'd diffusion\/} appears.  Regular
orbits form boundaries which irregular orbits cannot cross; for test
particles in two-dimensions these invariant tori divide phase space
into separate volumes so that an irregular orbit can be semi-trapped
``inside'' an invariant curve.  The extra degree of freedom in
three-dimensions means the chaotic regions are no longer isolated from
each other by the invariant surfaces, which have too few dimensions,
and are thought to be connected into a single network known as the
Arnol'd web.

Self-consistent models containing a significant fraction of chaotic
orbits would be difficult to construct, since any chaotic orbit fills
a volume bounded only by its energy surface which is always more
nearly spherical than the density distribution giving rise to the
potential.  Models containing some stochastic orbits are therefore
likely to be only quasi-stationary.  Since bars have existed for some
50 orbital periods only, the fact that some stochastic orbits will
cause them to evolve on a longer timescale may not be of great
concern, as the essentially stationary three-dimensional models of
Pfenniger and Friedli (1991) show.

Most orbits in non-rotating three-dimensional ellipsoids are of the
box or short- and long-axis tube types (Schwarzschild 1979, Binney and
Tremaine 1987), which are regular orbits trapped about the simplest
periodic orbit families.  However, the introduction of even slow
rotation appears to change the orbital structure drastically -- many
orbits which were regular in stationary bars become stochastic.  There
seem to be a number of additional ways, apart from slow rotation, to
foment stochasticity, but only rapid rotation is known to reduce it
(\S4.7).

Martinet and Pfenniger (1987) investigated the effect of a mass
concentration in the galaxy core, and showed that for even a small
mass, the $z$-motion close to the centre was quickly destabilized.
Hasan and Norman (1990) confirmed that a central mass concentration is
extremely effective at causing orbits close to the mass to become
chaotic, particularly once the central condensation contains more than
5\% of the total mass.

Udry and Pfenniger (1988) found that stochasticity rises when the bar
is strengthened, \eg\ by making it narrower or squaring off its ends,
and again when the central concentration was raised.  They also
examined the effects of graininess in the potential; a very reasonable
degree of granularity destabilized yet more regular orbits.

Though many questions remain unanswered, collectively, these results
suggest that weak, rapidly-rotating, bar models should have the fewest
chaotic orbits, and strong bars which end well before co-rotation
would have the most.

\sect{Gas and dust}
Although a small fraction of the total mass, at least in early type
galaxies, the gas component is of considerable interest to the
dynamicist mainly because it is an excellent tracer material.  We have
much more detailed knowledge of the flow patterns of gas in galaxies
than we do of the stars because the Doppler shifts of the emission
lines from excited gas are easier to measure than for the broader,
weaker absorption lines seen in the composite spectra of a stellar
population.  Comparison between the observed flow pattern and the
calculated gas behaviour in a number of realistic potentials can be
used as a means to estimate such uncertain quantities as the pattern
speed and mass-to-light ratio of the bar.

The dust lanes, which are dark narrow features along spiral arms and
bars where the gas and dust density may be several times higher than
normal, also demand an explanation.  The widely accepted view that
these delineate shocks in the inter-stellar gas seems to have been
first proposed by Prendergast (unpublished c1962).

Finally, the specific angular momentum of gas elements changes with
time, causing significant radial flows of material.  These are
important for evolution of the metal content of the galaxy and can
cause a build-up of gaseous material in rings where the flow stops.

Almost all the theoretical work and simulations have neglected motion
in the third dimension.  This approximation may still be adequate,
notwithstanding the existence of transient bending instabilities
(\S10.1) and vertical instability strips within the bar since
dissipation must ensure that the gas clouds remain in a thin layer.
However, the work of Pfenniger and Norman (1990) may indicate that the
radial flow of gas is accelerated as the material passes through
vertically unstable regions.

\subsect{Observations of gas in barred galaxies}
Most data come from optical or 21cm HI observations and much less is
known of the distribution and kinematics of the possibly dominant
molecular gas component.  This is because molecular hydrogen must be
traced indirectly through mm-wave emission of CO and other species;
the resolution of single dish antennae is low and only small portions
of galaxies can be mapped with the current mm interferometers; for a
recent review of available CO data see Combes (1992).

\subsubsect{Gas distribution}
The distribution of neutral hydrogen within each galaxy shows
considerable variation.  Neutral hydrogen appears to be deficient
within the bar in a number of galaxies: \eg\ NGC~1365 (Ondrechen and
van der Hulst 1989) and NGC~3992 (Gottesman \etal\ 1984).  On the
other hand, counter-examples with significant HI in the bar are NGC
5383 (Sancisi \etal\ 1979), NGC~3359 (Ball 1986), NGC~4731 (Gottesman
\etal\ 1984), NGC~1073 (England \etal\ 1990) and NGC~1097 (Ondrechen
\etal\ 1989).  The CO is sometimes distributed in a ring around the
bar (\eg\ Planesas \etal\ 1991) and sometimes concentrated towards the
nucleus (\eg\ Sandqvist \etal\ 1988).

Early type barred galaxies contain little gas, in common with their
unbarred counterparts (\eg\ Eder \etal\ 1991).  Moreover, van Driel
\etal\ (1988), who had two strongly barred galaxies (NGC~1291 and
NGC~5101) in their sample having sufficient HI to be mapped, found
that in both cases the gas was concentrated in an outer ring.

\subsubsect{Kinematics}
The position-velocity maps of gas in a barred galaxy indicate that the
flow pattern is more complicated than the simple circular streaming
(sometimes) seen in an approximately axisymmetric galaxy.  In general,
systematic variations in the observed velocity field produce
characteristic $\cal S$-shaped velocity contours and non-zero
velocities on the minor axis, which are indicative of radial
streaming.  However, these features of the velocity field are seen
only in those galaxies for which the viewing geometry is favourable,
as emphasized by Pence and Blackman (1984b).  The general morphology
of the pattern is consistent with the gas following elliptical
streamlines within the bar, but high resolution data sometimes show
very abrupt changes in the observed velocity across a dust lane.

Optical and radio observations of the same galaxy are generally
complementary.  Although the HI gas is quite widely distributed, data
from the high resolution aperture synthesis arrays has to be smoothed
to a large beam (to improve the signal to noise) which blurs the maps,
particularly near the centre where the velocity gradients are steep.
Higher spatial resolution optical measurements of excited gas in the
bright inner parts can overcome this inadequacy to some extent,
especially from Fabry-Perot interferograms (\eg\ Buta 1986b, Schommer
\etal\ 1988, Duval \etal\ 1991), but strong optical emission tends to
be very patchy, and is rarely found near the bar minor axis.  mm data
on molecular gas is also helpful in localized regions (\eg\ Handa
\etal\ 1990, Lord and Kenney 1991).

An excellent example is NGC~5383, one of the best studied galaxies;
the Westerbork data of Sancisi \etal\ (1979) taken together with the
optical slit data from Peterson \etal\ (1978), later supplemented by
Duval and Athanassoula (1983), provided the sole challenge to
theoretical models for many years.  Fortunately, HI data from the Very
Large Array (VLA) has become available in recent years, and the number
of barred galaxies with well determined velocity fields is rising,
albeit slowly.  We mention individual papers in \S6.7.

In the majority of galaxies, the gas rotates in the same sense as the
stars.  However, exceptions have been found: NGC~2217 (Bettoni \etal\
1990) and NGC~4546 (Bettoni \etal\ 1991), in which the gas in the
plane can be seen to rotate in a sense counter to that of the stars.
This most surprising aspect strongly suggests an external origin for
the gas in these two early type galaxies and such cases are believed
to be rare.

\subsubsect{Dust lanes}
Dust lanes occur more commonly in types SBb and later.  Those along
the bar are offset from the major axis towards the leading side
(assuming the spiral to be trailing).  Athanassoula (1984)
distinguished two types: straight, lying at an angle to the bar as in
NGC~1300, or curved as in NGC~6782 and NGC~1433.  Sometimes
predominantly straight lanes curve around the centre to form a
circum-nuclear ring.  Dust can also be distributed in arcs and patches
across the bar: NGC~1365 (Figure~1) is a good example.

\subsubsect{Evidence for shocks}
Direct observations of steep velocity gradients across dust lanes,
which would be the most compelling reason to believe these are shocks,
have been hard to obtain.  The two best examples are for NGC~6221
(Pence and Blackman 1984a) and for NGC~1365 (Lindblad and J\"ors\"ater
1987).

Evidence for gas compression also comes from the distribution of
molecular gas, through CO emission, which appears to be concentrated
in dust lanes (\eg\ Handa \etal\ 1990, but see also Lord and Kenney
1991).  Less direct evidence comes from the non-thermal radio
continuum emission which is frequently strongly peaked along the dust
lanes (\eg\ Ondrechen 1985, Hummel \etal\ 1987a, Tilanus 1990) -- the
enhanced emission is consistent with gas compression, but could also
have other causes.

\subsubsect{Star formation and other activity}
It has been noted frequently (\eg\ Tubbs 1982, and references therein)
that the distribution of young stars and HII regions is not uniform in
barred galaxies.  Stars appear to be forming prolifically near the
centres (Hawarden \etal\ 1986, Sandqvist \etal\ 1988, Hummel \etal\
1990) and at the ends of the bar, but not at intermediate points along
the bar.  This situation in some galaxies is so extreme as to have
been interpreted as a star forming burst either in the nucleus or at
the ends of the bar, \eg\ NGC~4321 (Arsenault \etal\ 1988, Arsenault
1989).  Dense concentrations of molecular gas are also sometimes found
near the centres of barred galaxies (\eg\ Gerin \etal\ 1988).  It has
also been noted by several authors (\eg\ Simkin \etal\ 1980, Arsenault
1989) that active galactic nuclei are somewhat more likely to occur in
galaxies having bars, than in those without.

\subsect{Modelling the ISM}
When speaking of shocks \etc\ in a gas flow, it is customary to think
of a continuous fluid having a well defined sound speed.
Unfortunately, the ISM (inter-stellar medium) is not that simple,
which prompted Prendergast (1962) to muse that ``it is unclear what to
assume for the equation of state''.

It might seem natural to assume that the inter-stellar gas in the
neighbourhood of the Sun, which is the best studied portion of the
ISM, has properties typical of that throughout all galaxies.  Here,
the bulk of the gas mass is contained in cool, dense clouds which
orbit ballistically, virtually unaffected by external pressure forces,
except when in collision with another dense cloud.  By contrast, the
bulk of the volume is filled with high temperature gas at a much lower
density, which is in rough pressure balance with the dense material --
the so called hot phase.  The balance between the components is
regulated by star formation and supernovae; for more information, see
reviews by Cox and Reynolds (1987) or Spitzer (1990).  Unfortunately,
the star formation rate (and probably also the supernova rate, though
no data are available) within bars seems to differ from that near the
Sun, and the ISM in bars may have somewhat different properties.

This ``variously damped and heated multi-phase stew'' (Toomre 1977) is
believed to experience shocks of some form or other in the
intriguingly located dust lanes.  We cannot realistically hope to
understand the large-scale behaviour of the ISM if we include the
intricate small-scale dynamics of each fluid element, and most
calculations assume some gross physical properties for the medium.
The pervasive hot phase is the only component which can reasonably be
described as a smooth fluid on galactic scales, but both the sound
speed and the Alfv\'en speed are likely to be well in excess of 100 km
sec\per, implying that there is little chance of it being shocked by a
potential perturbation with relative motion a fraction of the orbital
velocity.

One approach has been to treat the dense material as a collection of
ballistic particles having a finite cross-section for collision
(Miller \etal\ 1970, Schwarz 1979, Matsuda and Isaka 1980, Combes and
Gerin 1985).  There is some disagreement over whether to dissipate all
or just some fraction of the energy in the collision, whether to merge
the colliding particles and what to assume for the collision
cross-section.

It is not unreasonable, however, to view the collection of cool clouds
as a fluid with a sound speed of the order of the velocity dispersion
of the clouds, which is typically 5 -- 10 km sec\per.  The conditions
under which this might be valid were examined by Cowie (1980), who
attempted to calculate an equation of state for the cloud ensemble.
This vastly simplifying assumption has led to a rival group of papers
which calculates the gas flow as a continuous fluid using conventional
two-dimensional fluid dynamical codes.  A number of different codes
have been tried; van Albada \etal\ (1982) assessed the relative
performance of many common techniques and a multi-grid method was
later introduced by Mulder (1986).  Some authors (\eg\ Roberts \etal\
1979), with an understandable desire for yet higher spatial
resolution, advocate 1-D codes which neglect pressure forces normal to
the flow lines.

Some techniques are of intermediate type, such as the {\it beam
scheme\/} (Sanders and Prendergast 1974), which has proved very
popular, and {\it smooth particle hydrodynamics\/} or SPH (Lucy 1977,
Gingold and Monaghan 1977).  Both codes combine aspects of the
previous two distinct approaches.  Hernquist and Katz (1989) describe
a three-dimensional fluid-dynamical scheme which uses SPH with
self-gravity.

As the ``gas'' in all three types of code obeys equations which are at
best very crude approximations to the real dynamics of the ISM, a
discussion of which is intrinsically ``the best'' misses the point.
Several authors argue the virtue of low numerical viscosity in high
quality fluid codes, without pausing to consider the extent to which
the ISM differs from an inviscid gas with a simple equation of state.
In fact, bulk viscosity may be the most important physical property
distinguishing the gas from the stars (\eg\ Sanders 1977).  Numerical
viscosity is, of course, undesirable because its properties and
magnitude are set by the nature of the numerical code, grid cell size,
collision cross-section \etc, whereas it would be preferable to employ
a viscous coefficient related to the properties of the medium!  In
summary, it is very helpful to have tried a variety of codes having
different numerical weaknesses, because we gain confidence in results
which all can reproduce, and learn to be suspicious of those unique to
one type of code.

\vfill\eject
\subsect{Streamlines and periodic orbits}
Because the velocity dispersion of the gas clouds is so much lower
than their orbital speeds, the influence of ``pressure'' (collisions)
on the trajectories will generally be small.  When pressure is
completely negligible, the gas streamlines must coincide with the
periodic orbits in the system.  However, gas streamlines differ from
stellar orbits in one crucial respect: they cannot cross, \ie\ the gas
must have a unique stream velocity at each point in the flow.  This
very obvious fact implies that when periodic orbits cannot be neatly
nested, pressure or viscous forces must always intervene to prevent
gas streamlines from crossing.

Even when the perturbing potential is a weak, rotating oval distortion
and orbits can be computed by linear theory, as in \S4.2, periodic
orbits are destined to intersect at resonances.  Not only do the the
eccentricities of the orbits increase as exact resonance is
approached, but the major axes switch orientation across all three
principal resonances, making the crossing of orbits from opposite
sides of a resonance inevitable.  Sanders and Huntley (1976), using
the beam scheme, showed that the gas response between the inner and
outer Lindblad resonances takes the form of a regular two-arm spiral
pattern in the density distribution.  They argued that the orientation
of the streamlines slews gradually over a wide radial range and the
locus of the density maximum marks the regions where ``orbit
crowding'' is greatest.

Each spiral arm winds through, at most, 90\degrees\ per resonance
crossed.  As Sanders and Huntley's first model had a power-law
rotation curve, only one \ILR\ was present.  In models having two \ILR s,
the orientation must change again through 90\degrees\ at the inner
\ILR; note however, that we should expect a leading spiral arc at this
resonance, because the dynamical properties of the orbits inside the
inner \ILR\ revert to those between the outer \ILR\ and co-rotation.  In a
subsequent paper, Huntley \etal\ (1978) showed that the result in
their case is a density response which leads the major axis of the
potential by a maximum of about 45\degrees.

Where a weak bar potential rotates fast enough for no \ILR s to be
present, orbit crossings might be avoidable everywhere inside
co-rotation.  The flow may then remain aligned with the bar all the
way from the centre to co-rotation (\eg\ Schwarz 1981), changing
abruptly at co-rotation to trailing spiral arcs extending to the \OLR.

It should be noted that all the results mentioned in this sub-section
were obtained from a mild oval distortion of the potential having a
large radial extent.

\subsect{Strong bars}
Streamlines still try to follow periodic orbits even in strongly
non-axisymmetric potentials, though it becomes increasingly difficult
to find circumstances in which the orbits can remain nested; not only
can adjacent orbits cross, but a periodic orbit can also cross itself
(\eg\ \figno{-4}).  Because this greatly complicates the relationship
between periodic orbits and streamlines, we find the alternative
picture described by Prendergast (1983), paraphrased here, easier to
grasp.

As there is a formal analogy between compressible gas dynamics and
shallow-water theory (\eg\ Landau and Lifshitz 1987, \S108), we can
think of the gas flow within a bar as a layer of shallow water
circulating in a rotating non-axisymmetric vessel having the shape of
the effective potential (\eg\ \figno{-5}); for a rapidly rotating bar,
this has the shape of a non-circular volcano crater (see \S4.3.2).  As
the crater rim defines co-rotation, the water within the crater flows
in the same sense as the bar.  It flows along the sides of the crater,
but has too much momentum to be deflected round the end and back along
the far side by the comparatively weak potential gradients.  Instead
it rushes on past the major axis of the potential and on up the sides
of the vessel, finally turning back when the flow stalls.  The
hydraulic jump which must form where fresh material encounters the
stalled flow is the analogue of a shock in gas dynamics.

This analogy provides an intuitive explanation for the location of
shocks on the leading edge of the bar, something which is not so
easily understood from the discussion in terms of periodic orbits
presented by van Albada and Sanders (1983).  As their main conclusion
is that the periodic orbits must loop back on themselves, a condition
implicit in Prendergast's description, the two arguments are
equivalent.

\topinsert{
\centerline{\psfig{file=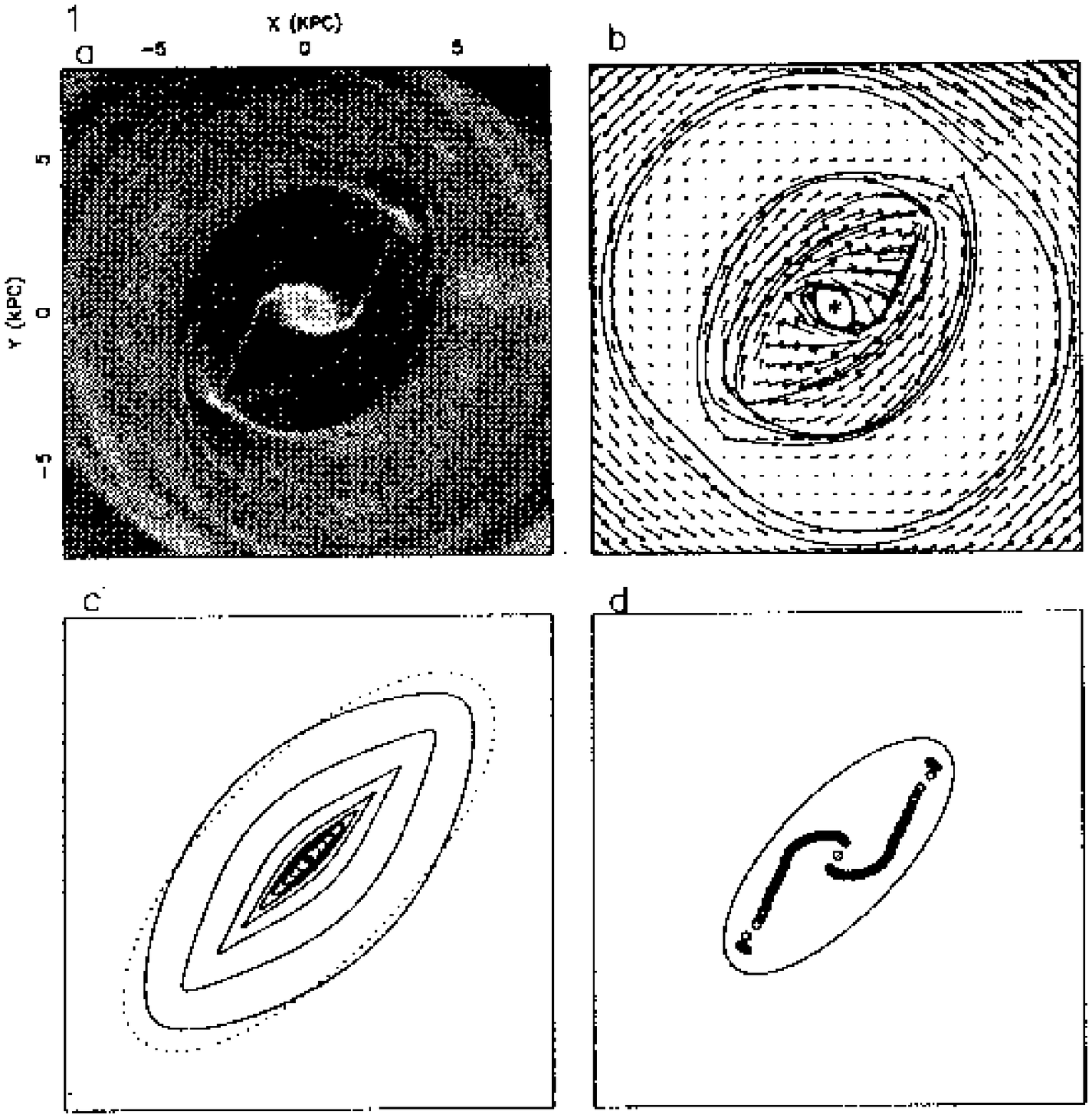,width=.5\hsize,angle=90,clip=}}
\caption{16}
The gas flow pattern in one of Athanassoula's (1992b) simulations in
which the inhomogeneous bar is positioned across the diagonal of each
frame. (a) shows a grey scale representation of the gas density
distribution (highest densities are white), (b) shows the velocity
vectors and a few streamlines in the restframe of the bar, (c) shows
some periodic orbits from the $x_1$ family (on an expanded scale) and
(d) shows the loci of density maxima and the outline of the bar on the
same scale as in (a) and (b). \par
}\endinsert

Strong offset shocks of this type were first revealed in the fluid
dynamical simulations by S\o rensen \etal\ (1976) and have been
reproduced many times (Roberts \etal\ 1979, Sanders and Tubbs 1980,
Schempp 1982, Hunter \etal\ 1988, \etc).  \nextfig\ shows such a
result from the high quality simulations by Athanassoula (1992b).  As
expected, she finds that shocks develop only when the $x_1$ family of
orbits possess loops, or are at least very sharply curved.

Athanassoula also concludes that the range of possible pattern speeds
which gives rise to straight shocks is such that the major axis
Lagrange points should lie between 1.1 and 1.3 times the bar
semi-major axis, for a Ferrers bar model.  If this result proves to be
more general, and if the straight dust lanes are indeed the loci of
shocks, then it supplies the tightest available constraint on the
pattern speeds of bars in galaxies.

An additional result from her study is that the shocks are offset
along the bar only when the potential supports a moderately extensive
$x_2$ (perpendicular) family of orbits.  If the mass distribution is
insufficiently centrally concentrated, then the shocks lie close to
the bar major axis.

\subsect{Driven spiral arms?}
The spirals that Sanders and Huntley (1976) and Schwarz (1981) were
able to produce extended out to, and even a little beyond, the \OLR.
However, many authors have reported that they are unable to reproduce
such extensive spiral responses in passive gas to forcing by an
ellipsoidal bar model instead of an oval distortion to the potential.
This is because the quadrupole field of a realistic bar, which ends
near co-rotation, falls off too rapidly at larger radii to induce a
spiral response in the gas (\figno{-7}).  In order to maintain a
spiral response well beyond the bar, Roberts \etal\ (1979) made the
rather {\it ad hoc\/} assumption that the bar near the centre goes
over to a trailing spiral perturbation beyond co-rotation, the whole
pattern rotating at the same rate.  Hunter \etal\ (1988) added a
co-rotating oval distortion for similar reasons.

Spirals are more likely to be independent patterns, as in normal
galaxies, and probably have a different pattern speed (Sellwood and
Sparke 1988).  Such a suggestion would seem to imply a random
distribution of phase differences between the bar and the start of the
arms which, it is generally felt, conflicts with the observed
situation.  Sellwood and Sparke point out that this is not a valid
objection, for two reasons: Firstly, a phase difference can, in fact,
be seen in a number of barred galaxies -- even where a two arm ``grand
design'' spiral pattern dominates (\eg\ NGC~5383, Figure~2).
Secondly, contour plots of the non-axisymmetric density in their model
show that the spiral arms appear to the eye to be joined to ends of
the bar for most of the beat period.

A review of the problems and ideas of spiral arm generation in disc
galaxies would take us too far from the subject of barred galaxies.

The most convincing evidence for two separate pattern speeds in a
galaxy is that for NGC~1365 (Figure~1).  Ironically, the impressive
grand design spiral arms of this galaxy have frequently been cited as
a clear example of bar forcing, yet the kinematic data, presented in
\S6.7, strongly support two separate co-rotation radii for the bar and
spiral patterns.  Weaker evidence can also be found for other
galaxies.

\subsect{Angular momentum changes}
Whenever gas is distributed asymmetrically about the major axis of the
potential, it will experience a net torque which causes a secular
change in its angular momentum.  Where the density maximum leads the
bar, the gas will systematically lose angular momentum, and conversely
a trailing offset will cause it to gain.  (The angular momentum is
removed from, or given up to, the stellar population creating the
non-axisymmetric potential.)

This process was emphasized by Schwarz (1981), who found that ``gas''
particles between co-rotation and the \OLR\ were swept out to the \OLR\ in
a remarkably short time.  The swept-up material quickly formed a ring,
which was slightly elongated either parallel to the bar if the initial
gas distribution extended to radii beyond the \OLR, or perpendicular to
it if the distribution was not so extensive.  He obtained this result
in an isochrone background potential using a bar pattern speed
sufficiently high to have no \ILR s, but in his thesis Schwarz (1979)
also reports inner ring formation in a different model having an \ILR.

Schwarz finds that the high rate at which gas is swept up into rings
depends only weakly upon his numerical parameters: the collision box
size, coefficient of restitution, \etc~ Since the torque responsible
for these radial flows is proportional to the density contrast in the
arms, as well as the phase lag (or lead) and strength of the
non-axisymmetric potential, any realistic density contrast in the
spiral or bar must give a similar flow rate.  Schwarz's flow rates are
probably too high, however, because the bar-like potential
perturbation he used, which peaks at co-rotation, is unrealistically
strong in the outer parts.

Simkin \etal\ (1980) proposed a causal link between the inflow of gas
within co-rotation and the existence of an active nucleus, which they
suggest occurs somewhat more frequently in barred galaxies.  This
suggestion has been endorsed by Noguchi (1988), Barnes and Hernquist
(1991) and others.  While gas inflow along the bar is expected to
raise the gas density in the central few hundred parsecs, its angular
momentum must be further reduced by many more orders of magnitude
before the material could be used to fuel a central engine.

\subsect{Comparison with observations}
NGC~5383 (Figure~2) is probably the most extensively studied and
modelled barred galaxy.  Sanders and Tubbs (1980) made a systematic
attempt to model the gas flow pattern measured by Peterson \etal\
(1978) and Sancisi \etal\ (1979).  By varying the bar mass, axis
ratio, pattern speed and other parameters they were able to find a
model which broadly succeeded in reproducing the qualitative features
of the observed flow pattern, though discrepancies in detail remained.
An altogether more comprehensive attempt to model this galaxy was made
by Duval and Athanassoula (1983), who used the distribution of surface
brightness to constrain the bar density distribution and added more
high resolution optical observations to map the flow pattern within
the bar in more detail.  They ran simulations to determine the flow
pattern when co-rotation was close to the bar end and experimented
mainly with a range of mass-to-light ratios for the bar; again their
best model resembled the observed flow pattern within the bar, though
still not impressively so.  It is possible that their low resolution
beam scheme code precluded a better fit.

Pence and Blackman (1984b) found that the velocity field of NGC~7496
closely resembled that of NGC~5383.

Following this initial success, a number of attempts have been made to
model other galaxies, notably by the Florida group.  One galaxy in
their sample, NGC~1073 (England \etal\ 1990), is too nearly face on
for the kinematic data to constrain a model.  In both the other two,
NGC~3992 (Hunter \etal\ 1988) and NGC~1300 (England 1989), they
encountered considerable difficulties in modelling the outer spiral.
When a pure bar model failed to produce a sufficiently strong density
contrast in the outer spiral arms, they added a global oval
distortion, but that seemed to produce too open a spiral pattern.  It
seems likely, therefore, that the spiral arms in these galaxies do not
result from forcing by the bar, but are independent dynamical
structures of the type described in \S6.5.

NGC~1365 (Figure 1) has also been observed extensively, though no good
model for the whole galaxy has yet been published.  Teuben \etal\
(1986) find quite convincing evidence for gas streamlines oriented
perpendicularly to the bar near the very centre.  They identify the
location where this is observed with the $x_2$ family of periodic
orbits, giving them a rough indication of the pattern speed; the value
obtained in this manner places co-rotation close to the end of the bar
-- a reassuring circumstance, which is also corroborated by the
presence of the offset dust lanes along the strong bar (see \S6.4).
However, Ondrechen and van der Hulst (1989) note that the inward
direction of the gas flow on the projected minor axis provides an
unambiguous indication that the spiral arms at this point are still
inside co-rotation.  These two conclusions can be reconciled only by
accepting that the bar and spirals are two separate patterns, with the
bar rotating much faster than the spirals.

Separate pattern speeds for the bar and spiral may occur in many
galaxies.  Co-rotation for the spiral pattern in NGC~1097 appears to
lie beyond the bar (Ondrechen \etal\ 1989); Chevalier and Furenlid
(1978) had difficulty in assigning a pattern speed for NGC~7723; the
dust lanes in NGC~1365 (Figure~1) and NGC~1300 (Sandage 1961) cross
the spiral (another, much weaker indication of co-rotation) well
beyond the end of the bar.  The co-rotation resonance for the two
tightly wrapped spiral arms which make up the pseudo inner ring of
NGC~6300 appears to lie inside the point where they cross minor axis
(Buta 1987), yet this is close to the bar end; this may be an example
where the spirals rotate more rapidly than the bar -- the misalignment
between the spirals and the bar also supports the idea of separate
patterns.

The conclusions from all these studies are:
\item{i} Most radio observations need to be supplemented by high resolution 
optical data before modelling of the observed flow pattern provides
useful constraints on the properties of the bar.
\item{ii} The gas flow {\it within the bar\/} can be modelled fairly 
successfully when the bar pattern speed is about that required to
place co-rotation just beyond the end of the bar.
\item{iii} Shocks along the bar also develop under the same conditions for the 
bar pattern speed, but are offset only if the $x_2$ family is present.
\item{iv} The outer spiral arms usually cannot be modelled without assuming 
some additional non-axisymmetric component to the potential.
\item{v} Not one of the well studied cases provides evidence that the spiral 
patterns are driven by the bar, while there is frequently a suggestion
that the spiral arms have a lower pattern speed than does the bar.

\sect{Rings and lenses}
Many galaxies, both barred and unbarred, exhibit rings which are
believed to lie in the disc plane.  These are thought to have an
entirely different origin and properties from the much rarer polar
rings, so named because they lie in a plane almost perpendicular to
that of the disc.  We do not discuss polar rings in this review (see
\eg\ Whitmore \etal\ 1990).

Unlike the spiral patterns in barred galaxies just discussed, there is
considerable evidence that many rings share the same pattern speed
with the bar, and therefore seem very likely to be driven responses to
forcing by the bar.

\subsect{Observed properties of rings}
Statistical properties of some 1200 ringed galaxies selected from the
southern sky survey are presented by Buta (1986a); these appear to be
quite representative of the complete catalogue (Buta 1991), which will
contain about twice this number.  Three major ring types are
distinguished by the radii, relative to the bar major axis, at which
they occur.

\subsubsect{Outer rings}
Outer rings are the largest, relative to the host galaxy, having a
diameter some $2.2\pm0.4$ times the bar major axis (Kormendy 1979).  A
good example is NGC~2217 in Figure~2 and others include NGC~1291,
NGC~2859 (de Vaucouleurs 1975) and NGC~3945 (Kormendy 1981).  The
frequency of outer rings is difficult to estimate because they could
be missed on all but the deepest plates; early estimates (de
Vaucouleurs 1975, Kormendy 1979) suggested they occur in only 4 -- 5\%
of all galaxies, but Buta's survey may well indicate a higher fraction
(Buta, private communication).

The ring is centred on the nuclear bulge and is thought to lie in the
disc plane.  Statistical arguments indicate that outer rings have
intrinsic axis ratios in the range 0.7 to 1.0 (Athanassoula \etal\
1982, Schwarz 1984b, Buta 1986a).  Buta notes that although the longer
axis is usually perpendicular to the bar, there is a significant
sub-population for which the ring is parallel to the bar.

True outer rings are not always easily distinguished from {\it pseudo
outer rings}, which occur when the outer spiral arms almost close.

\subsubsect{Inner rings}
Inner rings are somewhat smaller, and generally have a diameter
similar to the bar major axis -- good examples are NGC~1433 and
NGC~2523 in Figure~2 -- but are sometimes noticeably larger, \eg\
NGC~936 (also in Figure~2).  They are more common than outer rings,
but are found mainly in later types; Kormendy (1979) reports that 76\%
of SBab-SBc galaxies have inner rings, while few early-type galaxies
have them.

Inner rings in barred galaxies are generally more elliptical than
outer rings, having an axis ratio in the range 0.6 -- 0.95 (Buta
1986a) and are elongated parallel to the bar (Schwarz 1984b) with few
exceptions.\nextfoot{Exceptions are NGC~4319, which is a tidally
interacting system, and NGC~6300 in which two tightly wrapped arms
make a pseudo-inner ring (Buta 1987).}  Buta (1988) finds some
evidence for non-circular motion in the rings of a few nearby
galaxies, thereby confirming their intrinsic non-circular shape.  Buta
(1991) notes that some are more rectangular or even hexagonal; the
best example of hexagonal isophotes is for the weakly barred galaxy
NGC~7020 (Buta 1990b).

In galaxies having both types of ring, the ratio of the outer to inner
ring major axis diameters is on average $2.21 \pm 0.02$ with a long
tail to higher values (Buta 1986a).

Both inner and outer rings tend to be bluer than the surrounding disc
and many have HII regions, as do spiral arms (Buta 1988, Buta and
Crocker 1991).

\subsubsect{Nuclear rings}
A third type of ring has so far been found in relatively few systems,
\eg\ NGC~1512 (J\"ors\"ater 1979), NGC~1365 (Teuben \etal\ 1986),
NGC~1097 (Hummel \etal\ 1987b, Gerin \etal\ 1988), NGC~4321 (Arsenault
\etal\ 1988), NGC~5728 (Schommer \etal\ 1988) and NGC~4314
(Garcia-Barreto \etal\ 1991a).  They are usually very small, radius a
few hundred parsecs, nearly round, and not aligned with the bar (Buta
1986a).  An exceptionally large nuclear ring is seen in ESO~565-11
(Buta and Crocker 1991).  As they are hard to find, because of their
small size, their apparent rarity could again simply be a selection
effect.  In some cases, co-incident radio and optical emission from
discrete sources lying in the ring is found, together with significant
quantities of molecular gas and dust (\eg\ Sandqvist \etal\ 1988).
Hawarden \etal\ (1986) also note an excess of 25$\mu$m emission from
barred galaxies, which they interpret as being due to enhanced star
formation in a nuclear ring.  However, Garcia-Barreto \etal\ (1991b)
point out that similar phenomena can also occur without a detectable
ring.

Several galaxies contain short {\it nuclear bars\/} within the nuclear
ring.  We discussed such features in \S2.4.

\subsect{Lenses and oval distortions}
While inner rings are rare in early type galaxies, some $\sim 54$\% of
SB0-SBa galaxies (Kormendy 1981) manifest a flattened ellipsoidal
structure known as a lens.  Clear examples are NGC~5101 (Sandage 1961)
and NGC~3945 and NGC~4596 (Sandage and Brucato 1979).  This feature is
similar in size to the inner rings discussed above; Kormendy (1979)
notes that the bar usually fills the lens in one (frequently the
longest) dimension.  The axial ratio in the disc plane is typically
$\sim0.9\pm0.05$, \ie\ they are slightly rounder than inner rings.

Comparatively few late-type galaxies (SBb to SBm types) are classified
as having a lens, but many have a so-called {\it oval disc\/} or {\it
distortion}.  Kormendy suggests that lenses and oval discs are
distinct phenomena, on the grounds that the kinematic properties
appear to be different: the rotation curves of oval discs are flat
(\eg\ NGC~4736, Kormendy 1979), whereas the rotation curve of a lens
seems to rise with radius (Kormendy 1981).  Clearly more data are
required to establish his case.

\subsect{Formation of rings}
By far the most popular theory is that rings form from radial flows of
gas driven by the bar.  The gas dynamical simulations by Schwarz
(1979) showed that material gathers in rings where the radial flow,
caused by the spiral response to the bar, ends at a major resonance of
the pattern.  In this picture, the outer rings lie at the \OLR\ for the
bar while the inner rings occur either at co-rotation or at the 4:1
resonance.  The nuclear rings are thought to lie at the \ILR\ for the
bar (\eg\ Combes and Gerin 1985, Buta 1986b, Schommer \etal\ 1988),
and the inner bar may be populated by stars on the perpendicular orbit
family $x_2$ (\eg\ Teuben \etal\ 1986).

There are two pieces of evidence in favour of this interpretation.
Firstly, making plausible assumptions about the shapes of rotation
curves, Athanassoula \etal\ (1982) concluded that radii were
consistent with the hypothesis that the outer rings lay at the \OLR\
while the inner were located at the 4:1 resonance.  Secondly, and more
convincingly, the ring orientations fit extremely well with the
results from the simulations and suggest that rings trace the major
periodic orbits.  In particular, Schwarz (1979, 1981) found two
orientations for the outer ring, depending upon the extent of the
original gas disc; both cases seem to occur abundantly in nature, and
there are even a few galaxies which seem to possess both
simultaneously (Buta 1986a, 1991).

However, the theory does not offer convincing reasons for the
existence of rings in unbarred galaxies -- less common, but by no
means rare (Buta 1991) -- and the absence of rings in other barred
galaxies.  The first might be explained by arguing that the bars are
unseen in the optical (\S1) or that they have dissolved after first
forming a ring (\S10.3).  The second may indicate a finite lifetime
for these features.

The theory loses more of its appeal if, as we be have argued, spiral
patterns in barred galaxies rotate at a rate different from the bar.
The spirals themselves may also drive gas radially, but the radii of
rings formed by the spirals cannot be expected to correlate with the
resonances of the bar.\nextfoot{Unless the pattern speeds are related
in some special way (\eg\ Sellwood 1991)} A way to salvage the theory,
is to suppose that outer rings are formed at the same time as the bar.
The bar formation process causes a substantial re-arrangement of
angular momentum in the disc, and frequently forms a transient ring in
the stellar component near the \OLR\ for the bar (\S9.2, a weak example
is shown in \figno1).  (This is a much more efficient way to
accumulate material at the \OLR\ than through slow forcing by the weak
quadrupole field of a steady bar.)  The lifetime of the gaseous ring,
which must be formed at the same time as the stellar ring, is likely
to be much greater because the random motions, which cause the stellar
ring to dissolve, can be dissipated in the gas.  The comparatively
small fraction of outer rings observed may indicate that these are
still rather short lived features, and therefore that the bar may have
formed recently in galaxies where they are seen.

The theory that bar-driven radial flows form rings may account for the
outer and nuclear rings, but it is less clear that it can account for
inner rings.  It has been suggested (\eg\ Schwarz 1984a) that inner
rings contain material trapped about the stable Lagrange points on the
bar minor axis, or on higher resonant orbit families.  This last
suggestion is an attractive one to account for the rectangular, or
even hexagonal, shapes of some inner rings (Buta 1990a, b).

Buta (1988) and Buta and Crocker (1991) have started to acquire more
data on a few good candidates, but detailed photometric and optical
and radio kinematic data are required for several more cases in order
to provide a serious test of the bar driven ring theory.

\subsect{Formation of lenses}
At present there is no good theoretical interpretation of lenses and
oval distortions.  A number of rather speculative ideas have been
discussed in the literature, none of which we find entirely
convincing.  Kormendy (1979) speculated that a lens is formed by the
dissolution of a bar; since current ideas suggest that bars formed
from instabilities in the disc, this hypothesis does not seem to
account for the substantially higher surface brightness of the lens.
Bosma (1983) proposed that the lens is an inner disc formed earlier
than the faint outer disc and Athanassoula (1983) suggested that a
lens results from a bar-instability in a high velocity dispersion
disc.  The latter idea suffers from the same flaw as does Kormendy's,
and requires a cool population of stars to be also present to form the
narrow bar.  Teuben and Sanders (1985) weigh in with a similar
suggestion that the lens is made up of chaotic orbits while Buta
(1990a) suggests that lenses are no more than aging inner rings.

\sect{Asymmetries}
Unfortunately, many barred galaxies are still more complicated, in
that they depart strongly from the perfect bi-symmetry which has been
implicit in all our theoretical discussion so far.  Mild asymmetries
are seen in virtually all spiral galaxies, but gross asymmetries occur
much more frequently in late type galaxies.  They are not confined to
barred galaxies, as emphasized by Baldwin \etal\ (1980).

\subsect{Observed properties}
In some cases, the disc and spiral pattern is simply much more extensive on 
one side of the galaxy.  An example of this type is NGC~4027 (de Vaucouleurs 
\etal\ 1968), in which the velocity field of the emission line gas remains 
centred on the bar (Pence \etal\ 1988) even though the bar is far from
the centre of the outer isophotes.  Other examples, for which there is
no kinematic data available, are NGC~4618 and NGC~4625 (Odewahn 1991).

Another type of asymmetry is for the velocity field to be noticeably
asymmetric while the light distribution is only mildly so.  Examples
are NGC~2525, which has a 50 km s\per\ asymmetry between ends of bar
(Blackman and Pence 1982), NGC~3359 (Duval and Monnet 1985), NGC~55
(Hummel \etal\ 1986) and NGC~7741 (Duval \etal\ 1991).  Bettoni and
Galletta (1988) noted a slight displacement in the centre of symmetry
of the velocity pattern for 6 out of the 15 galaxies in their sample.

In other cases, the bar is displaced from both the optical and
kinematic centre. Two of the most extreme examples are the LMC where
the rotation centre is nowhere near the bar (de Vaucouleurs and
Freeman 1972) and NGC~1313 (Figure~2) in which the rotation centre is
at one end of the bar (Marcelin and Athanassoula 1982).

\subsect{Models}
There have been very few attempts to model such asymmetries.  Marcelin
and Athanassoula (1982) built a mass model based on photometric
measurements of the galaxy NGC~1313, and solved for the gas flow
assuming a uniform rotation of the eccentric mass distribution about
the rotation centre.  They obtained quite a good match between their
heuristic model and the observed velocity field, but declined to
speculate as to why the mass distribution was so asymmetric or how it
could remain so.  Colin and Athanassoula (1989) followed up this study
with a similar treatment of other offset bar geometries.

There are no really convincing theories for the origin or persistence
of these asymmetries.  It is probably significant that these late type
galaxies have almost solid body rotation curves, so it would take a
very large number of orbital periods for asymmetric structures to be
torn apart by the shear.  The rate at which asymmetric patterns are
sheared can be further slowed if they are kinematic density waves and
not material features (Baldwin \etal\ 1980).

It seems likely that they originate through interactions with other
galaxies or accretion of dwarf companions, but there is also a
possibility that they arise from $m=1$ type instabilities.  No
detailed work appears to have been done to develop either hypothesis.
A further possibility is that they are primordial, and that such
galaxies are simply dynamically very young.

\pageinsert{
\centerline{\psfig{file=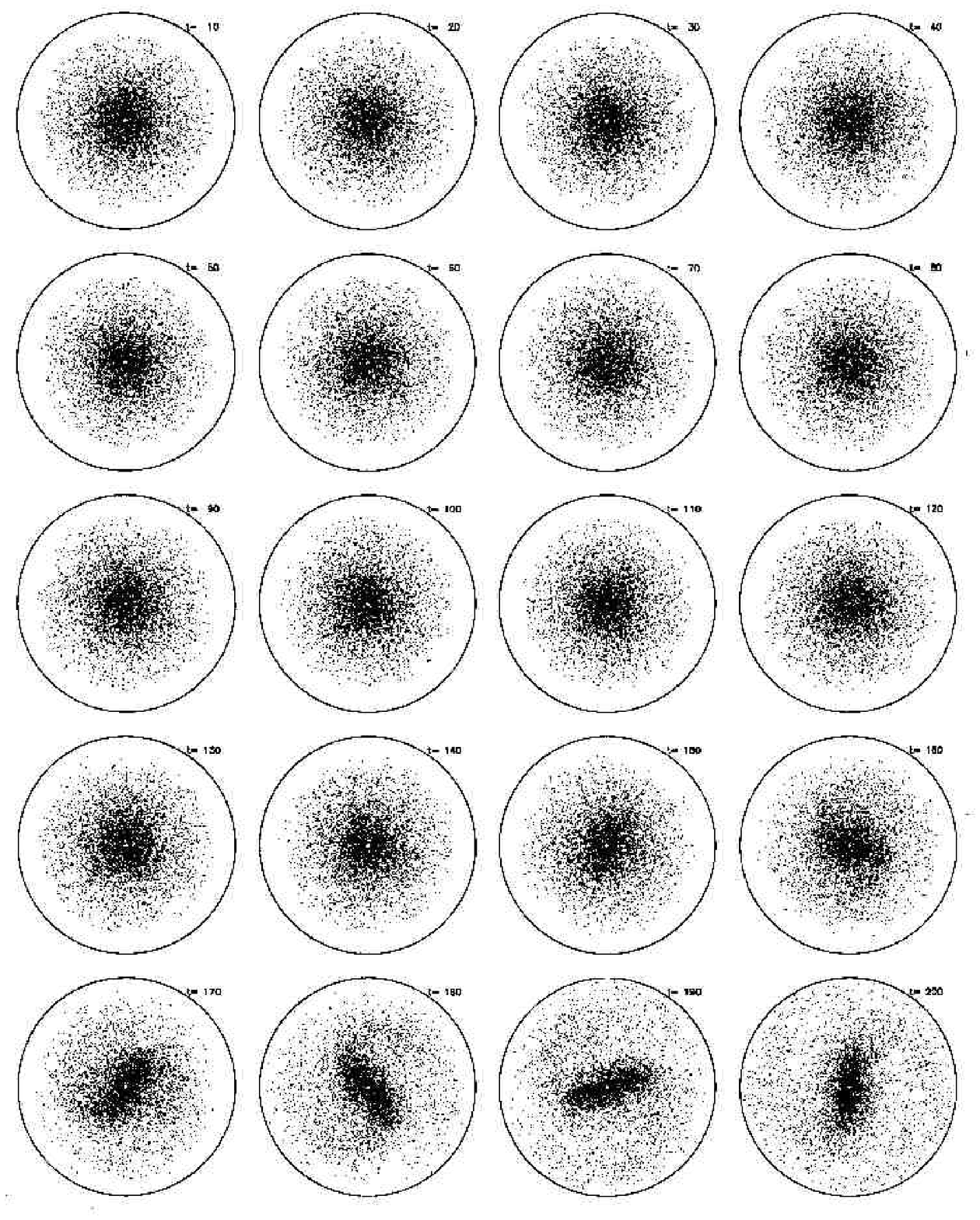,width=.9\hsize,clip=}}
\caption{17}
The formation of a bar through dynamical instability in a largely
rotationally supported disc.  The model is a completely
self-gravitating, two-dimensional isochrone/6 model, whose dominant
unstable mode was calculated by Kalnajs (1978).  100K particles are
employed, but only 5K are shown in each frame, the times are marked in
units of $\sqrt{a^3/GM}$ and the circle is drawn at a radius of
$6.18a$, where $a$ is the isochrone scale length and $M$ is the mass
of the untruncated disc.  The linear growth of the mode can be
detected by Fourier analysis before time 150, and the estimated growth
rate is some 16\% below the predicted value.  The discrepancy is
probably caused by the gravity softening (scale $=0.05a$) which was
introduced in the simulation. \par
\vfill
}\endinsert

\sect{Origin of bars}
The first $N$-body simulations of collisionless stellar discs (Miller
and Prendergast 1968, Hockney and Hohl 1969) revealed that it is easy
to construct a rotationally supported stellar disc which is globally
unstable and forms a large-amplitude bar on a dynamical timescale.
\nextfig\ shows a recent high quality simulation illustrating this
behaviour.  While this instability offers a natural explanation for
the existence of bars in some galaxies, it has proved surprisingly
difficult to construct stable models for unbarred galaxies.  The
problem of the origin of bars in galaxies was therefore quickly
superseded by that of how a fraction of disc galaxies could have
avoided such an instability.

Though this last question is of only peripheral interest to a review
of barred galaxies, it is hard to overstate its importance for disc
galaxy dynamics.  The bar instability has therefore been repeatedly
re-examined from a number of directions, which have all tended to
confirm that rotationally supported discs suffer from vigorous,
global, bi-symmetric instabilities.  It would take us too far from the
subject of this review to discuss the bar instability in great depth,
and we confine ourselves to a description of two aspects: the
formulation of a global stability analysis for discs as an eigenvalue
problem and the mechanism for the instability.

\subsect{Global analysis}
A full stability analysis which leads to an eigenvalue problem for
normal modes of an axisymmetric stellar disc was first formulated by
Shu (1970) and by Kalnajs (1971).  Though much can be formally deduced
from this approach (Kalnajs 1971, 1977), it has proved very difficult
to find eigenvalues in practice.  We describe Kalnajs's method and its
practical difficulties of implementation here not only because it
gives a vivid illustration of the analytical problems encountered in
galactic dynamics, but also because it reveals the power and
limitations of action-angle variables.  The limited results that have
emerged from this work are, however, of immense value as they provide
predictions against which we can test the $N$-body results.  Moreover,
the dominant linear modes found are almost always those which seem
likely to form bars.

\subsubsect{Formulation of the eigenvalue problem}
Kalnajs works in action-angle variables, introduced in \S4.2.3.  As
these are canonically conjugate coordinates, he is able describe the
dynamics using the elegant formulae of classical Hamiltonian mechanics
(\eg\ Goldstein 1980).  In particular, the collisionless Boltzmann
equation (1) can be written very compactly in Poisson bracket form $$
{\partial F \over \partial t} + \left[ F, H \right] = 0, \eqno(\equno)
$$ where $F$ is the distribution function and $H$ is the Hamiltonian.

In order to calculate small departures from equilibrium, we divide the
Hamiltonian into a sum of the unperturbed axisymmetric part $H_0$ plus
a small perturbing potential, $h$.  Similarly, we divide the
distribution function: $F = F_0 + f$.  Substituting into (\equat 0),
expanding the Poisson bracket, making use of the equations of
unperturbed motion $$
\dot J_i = - {\partial H_0 \over \partial w_i} \equiv 0 \qquad {\rm and} 
\qquad \dot w_i = {\partial H_0 \over \partial J_i} = \Omega_i, \eqno(\equno)
$$ and neglecting $[f, h]$ as being second order, we obtain $$
{\partial f \over \partial t} + \bOmega\cdot {\partial f \over 
\partial\bw} = {\partial F_0 \over \partial \bJ } \cdot {\partial h 
\over \partial \bw } \eqno(\equno)
$$ to first order.  We have adopted vector notation for the quantities
$\bw = (\wr,\wa)$, $\bJ = (\Jr,\Ja)$ and $\bOmega =
(\Omegar,\Omegaa)$.

Since all unperturbed orbits are quasi-periodic, we can expand the
perturbation potential, $h$, as a Fourier series in the angle
variables: $$
h(\bJ,\bw,t) = {1 \over (2\pi)^2} \sum_{\sbm} h_{\sbm} (\bJ,t)
e^{i\sbm \cdot \sbw}. \eqno(\equno)
$$ where $\bm = (l,m)$.  The coefficients are given by $$
h_{\sbm}(\bJ,t) = \int_0^{2\pi} \int_0^{2\pi} h(\bJ,\bw,t) e^{-i\sbm 
\cdot \sbw} d^2\bw, \eqno(\equno)
$$ We also transform $f$ in the same way.  Introducing these
decompositions into (\equat{-2}), we see it must be satisfied by each
Fourier component separately, thereby converting the PDE (\equat{-2})
into a set of first order ODEs: $$
{\partial f_{\sbm} \over \partial t} + i\bm \cdot \bOmega f_{\sbm} =
h_{\sbm} i\bm \cdot {\partial F_0 \over \partial \bJ }, \eqno(\equno)
$$ which has the solution $$
f_{\sbm}(\bJ,t) = i \bm \cdot {\partial F_0 \over \partial \bJ} 
e^{-i\sbm \cdot \bOmega t} \int_{-\infty}^t h_{\sbm}(\bJ,t) e^{i\sbm
\cdot \bOmega t^\prime} dt^\prime. \eqno(\equno) $$

Assuming $|f(\bJ, \bw, t)|$ does not grow any faster than
$e^{\alpha t}$, ($\alpha$ real and $>0$), we may write as a Laplace
transform $$ 
f(\bJ, \bw, t) = {1\over 2\pi}
\int_{-\infty+i\alpha}^{\infty+i\alpha} \tilde f(\bJ, \bw,
\omega) e^{-i\omega t}d\omega. \eqno(\equno)
$$ With this substitution into (\equat{-1}), we can evaluate the
$t^\prime$ integral to obtain $$
\tilde f_{\sbm}(\bJ, \omega) = {\tilde h_{\sbm}(\bJ, \omega) \over 
\bm \cdot \bOmega - \omega } \bm \cdot {\partial F_0 \over 
\partial\bJ }. \eqno(\equno)
$$ This is an equation for one Fourier component of the perturbation
to the \DF\ caused by a single Fourier component of a disturbance
potential having frequency $\omega$.  Notice that the denominator
passes through zero for purely real $\omega$ and certain values of
$\bOmega(\bJ)$; these are the familiar resonances where linear
theory breaks down for steady waves, as we saw in \S4.2.  However,
this problem does not arise for growing or decaying disturbance
(complex $\omega$).

The total response is the sum of these components $$
\tilde f(\bJ, \bw, \omega) = {1 \over (2\pi)^2} \sum_{\sbm} \tilde 
f_{\sbm}(\bJ, \omega) e^{i\sbm \cdot \sbw} = {1 \over (2\pi)^2} 
\sum_{\sbm}{\tilde h_{\sbm}(\bJ, \omega) \over \bm \cdot \bOmega - 
\omega } e^{i\sbm \cdot \sbw} \bm \cdot {\partial F_0 \over
\partial\bJ }. \eqno(\equno)
$$ We obtain an eigenvalue equation for the normal mode frequencies, 
$\omega$, by requiring that the perturbation potential arises from the 
disturbance density; \ie $$
\nabla^2 \tilde h(\bJ, \bw, \omega) = 4\pi G \tilde \Sigma_p \delta_z 
= 4\pi G \delta_z \int\int \tilde f(\bJ, \bw, \omega) d^2\bv. 
\eqno(\equno)$$

Two practical difficulties are posed by this eigenvalue problem.
Firstly, the Fourier components of the disturbance potential, $\tilde
h_{\sbm}(\bJ, \omega)$ are coupled by Poisson's equation.
Because of the rotational invariance of the Laplacian operator, the
different angular harmonics of the potential can be treated separately
(for infinitesimal perturbations), but the entire set of $l$
components for a given $m$ remain coupled.  We therefore have to solve
for eigenvalues of an infinite set of coupled equations.

Secondly, it may seem that the use of action-angle variables
aggravates the difficulty of finding solutions to (\equat0), since we
have to evaluate the perturbed surface density, $\tilde \Sigma_p$, by
integration over the velocities.  The transformation needed to express
$\tilde f(\bJ, \bw, t)$ in $(\bx, \bv)$ coordinates cannot be written
in closed form, except for a few special potentials.  Were we to
eschew these variables, and approximate the orbits as Lindblad
epicycles, we would avoid this problem, but we would still be faced
with the difficulty of calculating the disturbance potential through
Poisson's equation.  Since the solution of Poisson's equation for an
arbitrary mass distribution in a disc cannot be written down directly,
the eigenvalue problem is not much easier in $(\bx, \bv)$ coordinates!

Kalnajs (1977) proceeds by introducing a bi-orthonormal set of basis
functions; the surface densities $\{\Sigma_i\}$ and corresponding
potentials $\{h_i\}$ are related through Poisson's equation and
normalized such that $$
-{1 \over 2\pi G} \int h_i^*\Sigma_j d^2\br = \cases{ 1 & if
$i=j$, \cr 0 & otherwise. \cr} \eqno(\equno)
$$ He then writes the (unknown) density perturbation, $\tilde
\Sigma_p$, as a sum over the set $$
\tilde \Sigma_p = \sum_{j=0}^\infty a_j\Sigma_j \qquad {\rm where} \qquad a_j 
= -{1 \over 2\pi G} \int h_j^*\tilde \Sigma_p d^2\br. \eqno(\equno)
$$ The self-consistency requirement for modes is now that both $\tilde
\Sigma_p$ and the perturbed potential, $\tilde h$, have the same expansion in 
the chosen basis which enables him to avoid calculating the perturbed density 
altogether.  With this trick, he is able to rewrite the mode equation 
(\equat{-2}) in matrix form $$
\sum_{j=0}^\infty {\cal M}_{ij}(\omega)a_j = a_i, \eqno(\equno)
$$ where the matrix coefficients are $$
{\cal M}_{ij}(\omega) = -{1 \over 8\pi^3} \sum_{\sbm} \int\int F(\bJ) \; 
\bm \cdot {\partial \over \partial \bJ} \left[ {(\tilde h_i)_{\sbm}^*
(\tilde h_j)_{\sbm} \over \bm \cdot \bOmega - \omega} \right] 
d^2\bJ; \eqno(\equno)
$$ they are more readily evaluated after integration by parts.  The
overall technique involves a laborious calculation of all the
coefficients $(\tilde h_i)_{\sbm}$ (which fortunately are independent
of $\omega$) and a non-linear search for the eigenvalue, $\omega$;
results have so far been obtained for three disc models (Kalnajs 1978,
Zang 1976, Hunter 1992).\nextfoot{The method has also been
successfully used for spherical stellar systems (Polyachenko and
Shukhman 1981, Saha 1991, Weinberg 1991).}

A number of short-cuts have been devised to simplify the method, such
as using cold discs with softened gravity (Erickson 1974, Toomre
1981), gaseous approximations (Bardeen 1975, Aoki \etal\ 1979,
Pannatoni 1983, Lin and Bertin 1985) and moment methods (Hunter 1970,
1979).

\subsect{Bar-forming modes}
The dominant modes found in many of these analyses have very high
growth rates and an open bi-symmetric spiral form.  In several cases
the equilibrium model has also been studied in high quality $N$-body
simulations which reveal a dominant mode with a shape and
eigenfrequency in close agreement with the analytic prediction (Zang
and Hohl 1978, Sellwood 1983, Sellwood and Athanassoula 1986).  A
further example of such a comparison has been made in the simulation
shown in \figno0; as usual, the linear instability leads to a
large-amplitude bar.  The possibility that $N$-body simulations
somehow exaggerate the saturation amplitude, and hence the
significance of the instability, was eliminated by Inagaki \etal\
(1984), who compared an $N$-body simulation with a direct integration
of the collisionless Boltzmann equation for the same problem.  The
bars which form appear to be robust structures that survive for as
long as the simulations are continued (see
\S10 for caveats).

It seems likely that bars in real galaxies were created in this way,
since many of their observed properties are similar to those of bars
in the simulations (\eg\ Sparke and Sellwood 1987), yet the majority
of galaxies do not possess a {\it strong\/} bar (Figure 3).  To
account for this, we must understand how most galaxies could have
either avoided this instability or subsequently regained axial
symmetry.  We discuss several ways in which the instability might be
averted in \S9.5, but we cannot claim to have a completely
satisfactory theory for the formation of bars in some galaxies unless
we can also account for the fraction of galaxies that contain strong
bars.

\subsect{Properties of the resulting bars}
The type of behaviour illustrated in \figno0\ is typical of almost
every two-dimensional simulation for which the underlying model is
unstable to a global bi-symmetric distortion.  As the instability runs
its course, the transient features in the surrounding disc fade
quickly and the only non-axisymmetric feature to survive is the
steadily tumbling bar.

Many authors (\eg\ Sellwood 1981, Combes and Sanders 1981, \etc) have
observed that the bar ends at, or usually just inside, co-rotation (or
more correctly the Lagrange point -- see \S4.3.2).  Thus, as the
linear global mode saturates, its spiral shape straightens within
co-rotation while the trailing arms outside co-rotation become more
tightly wrapped and fade.  Also the pattern speed of the bar {\it
immediately as it forms\/} is generally close to that of the original
global mode, in most cases slightly lower but sometimes higher (\eg\
Figure~8 of Sellwood 1983).  The figure rotation rate may quickly
start to change as the model evolves further (see \S10.3).

The rule that the initial bar nearly fills its co-rotation circle
appears to be widely held and no counter examples have been claimed.
Sellwood (1981) also found that the bar length appeared to be related
to the shape of the rotation curve, but his models were all of a
particular type and different results are obtained from other models
(Efstathiou \etal\ 1982, Sellwood 1989).

The axis ratio of the bar largely depends upon the degree of random
motion in the original disc: generally speaking, the cooler the
initial disc the narrower the resulting bar (Athanassoula and Sellwood
1986).\nextfoot{A possible exception occurs for catastrophically
unstable cold discs (\eg\ Hohl 1975), but as these discs are unstable
on all scales down to the smallest that can be resolved, one can argue
that the local instabilities have caused so much heating that this is
not a real exception to the rule.}  The most extreme axial ratios to
survive for at least a few tumbling periods are in the range 4 -- 5:1.

Sparke and Sellwood (1987) give a comprehensive description of one of
these $N$-body bars.  They emphasize that the bar is much more nearly
rectangular than elliptical and matches the observed profiles
remarkably well (\S2.1).  The stars within the bars in these
simulations exhibit a systematic streaming pattern, which again bears
some resemblance to the observed velocity field (\S2.5.1).

\subsect{Mechanism for the mode}
The impressive agreement between the results from global analysis and
the behaviour in $N$-body simulations leaves little room for doubt
that rotationally supported, self-gravitating discs have a strong
desire to form a bar.  Yet these results do not explain why this
instability should be so insistent, nor do they reveal the mechanism
for the mode or give any indication as to how it could be controlled.
Ideas to answer these questions have emerged from local theory,
however.

Toomre (1981) proposed that the bar-forming mode was driven by
positive feedback to an amplifier.  The inner part of a galaxy acts as
a resonant cavity which may be understood as follows: {\parindent 1cm
\item{(i)} the group velocity of trailing spiral waves is inwards while that 
of leading waves is outwards,
\item{(ii)} trailing spiral waves that can reach the centre are reflected as 
leading waves,
\item{(iii)} a second reflection occurs at co-rotation where a leading wave 
rebounds as an amplified trailing disturbance.
\par}
\noindent Standing waves can occur at only those frequencies for which the 
phase of the wave closes, which implies a discrete spectrum.
Super-reflection off the co-rotation resonance causes the standing
wave to grow, however, making the mode unstable.  The
super-reflection, which Toomre (1981) aptly named ``swing
amplification'', was first discussed by Goldreich and Lynden-Bell
(1965) and by Julian and Toomre (1966).  It has been further developed
by Drury (1980) and invoked by Bertin (1983) and Lin and Bertin
(1985).  Wave action at co-rotation is conserved through a third,
transmitted wave which carries away angular momentum to the outer
galaxy.

\subsect{Controlling the bar instability}
The mechanism for the bar instability just outlined indicates three
distinct strategies for controlling it.  These are: {\parindent 1cm
\item{(i)} Raising the velocity dispersion to inhibit collective effects.  
This option is unattractive as it would also inhibit spiral waves,
though it might be possible to construct a model galaxy which could
sustain spirals in the outer disc and inhibit the bar instability by
having an unresponsive region near the centre of the disc only.
\item{(ii)} Immersing the galaxy in a massive halo, as favoured by Ostriker 
and Peebles (1973).  The additional central attraction increases the
stabilizing contribution from rotational forces while keeping the
destabilizing gravitational forces from the density perturbations
unchanged.  There are two reasons why this method seems unattractive
for galaxies: firstly, the halo must dominate the rotation curve even
in the centre (Kalnajs 1987), whereas evidence suggests that halos
dominate only the very outer rotation curve (van Albada and Sancisi
1986), and secondly, galaxies stabilized in this way would exhibit
multi-armed spiral patterns (Sellwood and Carlberg 1984), whereas the
more common visual impression is of a dominant bi-symmetric pattern.
\item{(iii)} Interrupting the feed-back cycle, as favoured by Toomre (1981).  
The feed-back loop will be broken if in-going waves are prevented from
reaching the centre.  This should occur if the wave encounters an \ILR,
which damps the oscillation through wave-particle interactions.  \ILR s
are likely to arise in galaxies having high circular velocities close
to the rotation centre.  Lin (1975) was probably the first to
conjecture that a dense central bulge might be effective in preventing
the bar instability. \par}

\subsect{Meta-stability and tidal triggering}
The simulations reported by Sellwood (1989) were designed to test this
last stabilizing mechanism.  He found that almost fully
self-gravitating discs could indeed be {\it linearly\/} stable when
the centre was made sufficiently hard; \ie\ he found that no bar
formed when initial perturbations are kept to small amplitudes.  When
the same model was subjected to a larger amplitude disturbance,
however, it formed a bar with properties very similar to those in
other linearly unstable models!

The reason the outcome depends upon the amplitude of the initial
disturbance is that the Lindblad resonance can damp only weak
perturbations; particles become trapped in the valleys of a
large-amplitude wave -- \ie\ the resonance saturates.  Sellwood found
that waves arising from noise fluctuations in the random distribution
of $2\times10^4$ particles were of large enough amplitude to alter the
linear behaviour!

Another method of inducing a bar in a meta-stable galaxy model is by
tidal triggering, which was explored by Noguchi (1987) and by Gerin
\etal\ (1990).  This work has been criticised on the grounds that it
is not fully self-consistent (Hernquist 1989), and Noguchi's choice of
$Q=1$ for the initial discs means that his simulations probably
over-estimate the efficiency of bar formation.  Notwithstanding these
criticisms, this idea may offer an explanation for the possible
over-abundance of barred galaxies amongst binary pairs and small
groups of galaxies (Elmegreen \etal\ 1990).

\subsect{An alternative theory of bar formation}
While this section has focused on the idea that bars are formed
quickly as a result of some large-scale collective oscillation of the
disc stars, we should note that an entirely different viewpoint was
proposed by Lynden-Bell (1979).

In a brief, but elegant paper, Lynden-Bell suggested that bars may
grow slowly through the gradual alignment of eccentric orbits.  He
pointed out that a bar-like perturbing potential would exert a torque
on an elongated orbit close to inner Lindblad resonance.  He then
showed that the torque would cause the major axis of the orbit to
oscillate about an axis orthogonal to the density perturbation in
outer parts of galaxies, but near the centre, in regions where double
\ILR s are possible, the orbit would tend to align with the
perturbation.

In the aligning region, orbits having very similar frequencies would
tend to group together at first, creating a bi-symmetric density
perturbation which would then exert sufficiently strong tangential
forces to trap orbits whose frequencies differ somewhat from that of
the bar.  A large perturbation can gradually be assembled in this
manner over many dynamical times.  Significantly, the pattern speed of
the bar would also decline as the orbits become more eccentric,
enabling yet lower frequency orbits to be caught by the potential.
The excess angular momentum which needs to be shed from the barring
region would be carried away by spiral arms.  Such a bar would not
extend as far as its co-rotation point.

Although Lynden-Bell was originally motivated to understand the weak
bar-like features which developed in the bar-stable simulations of
James and Sellwood (1978), he proposed that the mechanism describes
the general evolution of most galaxies.  Given the ease with which
bars are formed by fast collective instabilities, it is unclear
whether this mechanism plays any r\^ole in the formation of the strong
bars we see in galaxies today.  However, the alternative view of
interacting orbits presented in his paper does provide considerable
insight into the structure of a bar.

This mechanism gives rise to the linear bi-symmetric modes of a
radially hot, non-rotating disc calculated by Polyachenko (1989).  In
the appendix of their paper, Athanssoula and Sellwood (1986) described
a few vigorous bi-symmetric modes in their simulations of hot disc
models which, however, saturated at a very low amplitude.  If these
modes were indeed of the type suggested by Polyachenko. based on
Lynden-Bell's mechanism, the instability is probably unrelated to the
formation of strong bars.

\topinsert{
\centerline{\psfig{file=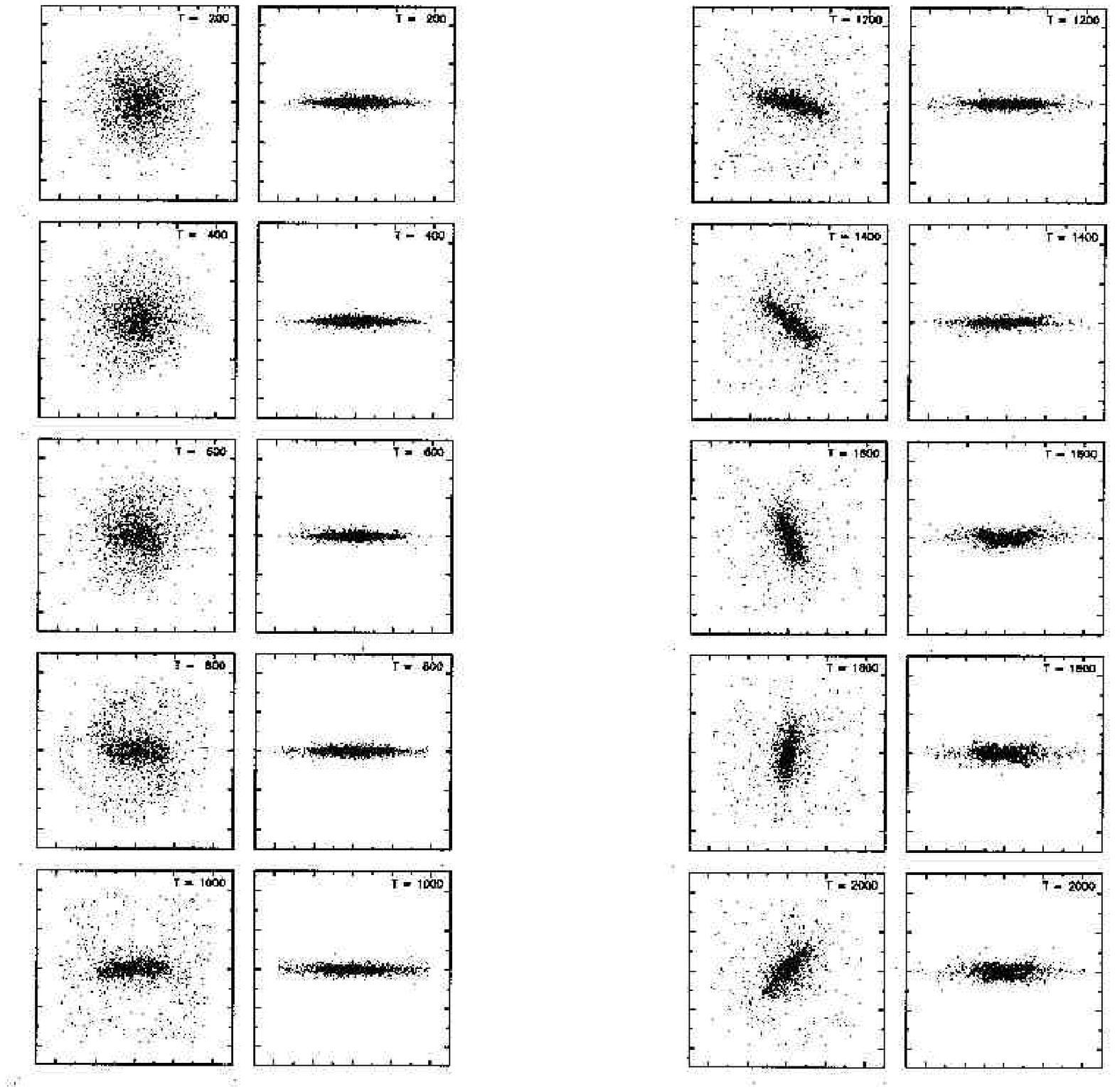,width=.8\hsize,angle=270,clip=}}
\caption{18}
Pole and side views of bar formation and subsequent buckling in the
fully three-dimensional thicker disc experiment reported by Raha
\etal\ (1991).  The side views are always shown from a position
perpendicular to the bar major axis.  Only 2K of the 90K disc
particles are shown in each plot, and none of the 60K spheroid
particles is shown.  The bar rotation period is a little less than 200
time units. \par
}\endinsert

\sect{Evolution of the bar}
Many authors have described the bar as the ``end point'' of the
simulation.  Sparke and Sellwood (1987) presented convincing evidence
that a two-dimensional stellar bar is indeed a steady, robust feature
when all particles except those in the bar are frozen and therefore
prevented from interacting with it.  They knew at the time that the
bar evolves through interactions with other responsive components, but
were unaware that an isolated bar can suffer a further instability
when motion into the third dimension is permitted.

\subsect{Buckling instability}
The ``fire hose'' (buckling or corrugation) instability for a stellar
disc was considered by Toomre (1966), Kulsrud \etal\ (1971), Fridman
and Polyachenko (1984) and Araki (1985) in an idealized, non-rotating,
infinite sheet model.  Subsequent papers have all confirmed Toomre's
original conclusion that buckling modes develop in this simple model
only when the sheet is thin, or more exactly, when the velocity
dispersion in the $z$-direction is less than about one third (Araki
quotes 0.29) that of the velocity dispersions in the plane.  As the
value observed in the solar neighbourhood is approximately 0.5 (\eg\
Woolley 1965), all these authors concluded that buckling modes are
unlikely to be of importance in disc galaxies.

However, the formation of a bar within the disc changes this
conclusion.  Not only does a bar typically make the orbits more
eccentric in the plane, but the direction of their principal axes are
also aligned; \ie\ the majority of stars in the bar move up and down
the bar in highly eccentric orbits (\S4.9).  The creation of organized
streaming motion along the major axis of the bar occurs without
affecting the ``pressure'' normal to the plane, and compromises the
stability to buckling modes (Raha \etal\ 1991).

\nextfig\ shows the development of a bar in a disc which then buckles
out of the plane; the model illustrated is one of the two cases
presented by Raha \etal. The viewing direction for the edge-on frames
is from a point always on the bar intermediate axis.  The bend
develops soon after the bar has formed and becomes quite pronounced by
time 1600, after which the bar regains symmetry about the plane but is
somewhat thicker than before and has a distinctive peanut shape when
viewed from the side.  The instability also weakens the bar amplitude
in the plane (only slightly in this case), and could perhaps destroy
it entirely when very violent (Raha \etal\ 1991).

A similar outcome was originally reported by Combes and Sanders (1981)
and later confirmed and strengthened in simulations of much higher
quality (Combes \etal\ 1990).  However, these authors propose that the
bar thickens because of the narrow instability strip associated with
the vertical Lindblad resonances within the bar.  Pfenniger and
Friedli (1991) attempt to develop this alternative mechanism, which
has to be reconciled with their empirical results that collective
effects and a $z$-asymmetry in the potential are essential.

While there is no doubt that a narrow strip of vertically unstable
orbits exists in their $N$-body bar models, it seems unlikely that the
comparatively small fraction of orbits affected (at any one time)
could lead to the exponentially growing, uniform bend, of the type
shown in \figno0.  On the other hand, Pfenniger and Friedli (1991) and
Raha (1992) show quite convincingly that the final peanut shape of the
bar is caused by a large number of orbits trapped about the stable
2:$2_{\rm a}$:1 periodic orbit family (\S5.2.3); thus the structure of
bar after the instability has saturated, though not the linear
instability itself, is determined by the periodic orbits (\S5.3).

Moreover, the existence of the fire-hose instability in thin
non-rotating bars was predicted by Fridman and Polyachenko (1984).
The $N$-body simulations of Merritt and Hernquist (1991) confirmed
that highly prolate models suffer from buckling modes, which clearly
cannot be caused by unstable orbits in their St\"ackel initial models.
It therefore seems most likely that the buckling modes in both these
and the rapidly rotating bars are of the fire-hose type.

\subsect{Peanut growth}
The final bar in \figno0\ resembles a peanut shape when viewed from
 the side.  A similar shape appeared in every bar which formed in the
 many experiments reported by Combes \etal\ (1990), although it is not
 certain that the formation mechanism was the same.

Their results seem to indicate that any strong bar which forms in a
galaxy must quickly become much thicker than the disc, a conclusion at
variance with those observers who argue that bars seem to be thin
(\eg\ Kormendy 1982).  The simulators have, of course, seized upon
box-peanut shaped bulges sometimes observed (\S2.3.3) and argued that
they are the thickened bars.  Two further pieces of indirect evidence
add credence to this speculation: firstly, it is believed (on
admittedly very slender evidence) that the rotation rate of the stars
in a box-peanut shaped bulge is independent of distance from the
plane, a property which is not found for normal bulges (Kormendy and
Illingworth 1982).  Secondly, the fraction of box-peanut shaped bulges
in nearly edge-on galaxies is some 20\% (\eg\ Shaw 1987) which is not
very different from the expected fraction of strongly barred galaxies,
when allowance is made for those systems in which the randomly
oriented major axis of the bar must lie close to our line of sight.

Unfortunately, there are two drawbacks to this simple association:
Athanassoula (private communication) points out that most, though not
all, box-peanut bulges are much less extensive, relative to the host
galaxy's size, than are strong bars.  Moreover, Shaw (private
communication) raises the further concern that the 30\% of strong bars
expected in edge-on galaxies, have apparently {\it not\/} all
developed a recognisable peanut shape.  A further observational test
of this hypothesis might be to determine whether the luminosity
functions of bars and box/peanut bulges are similar, though dust
obscuration might make this difficult.  Before this last question
acquires the status of a major puzzle, however, it seems necessary for
the simulators to verify that the bars continue to thicken in a much
wider range of mass distributions than have so far been tested,
especially those having a more realistic bulge with a high central
density.

\subsect{Continuing interactions with the disc}
Sellwood (1981) emphasized that when the initial bar in his
simulations was short compared with the total extent of the disc, it
continued to exchange angular momentum with the outer disc through
further spiral activity.  As always, the trailing spirals remove
angular momentum from particles at their inner end; Sellwood found
that this enabled more stars to be trapped into the bar, which
increased its length and also lowered its pattern speed.  In the most
extensive disc he studied, this process continued through a number of
spiral episodes while the bar increased in length by more than 50\%.
Interestingly, the changes in both bar length and pattern speed
conspired to keep co-rotation just beyond the end of the bar.

Combes and Sanders (1981), on the other hand, reported that bars tend
to weaken in the long term.  There are at least two physical reasons
for their different finding: firstly, their models started with much
less extensive discs, precluding the type of secondary bar growth
Sellwood observed; secondly, theirs were three-dimensional models in
which the bar thickened normal to the plane, which caused the strength
in the plane to decline (see \S10.1).  No three-dimensional models
with an extensive outer disc have so far been run to settle the
question of which effect wins.

\subsect{Interations with spheroidal components}
The large majority of simulations to date have employed fixed
additional contributions to the radial force to represent the effects
of spheroidally distributed matter, such as the bulge and halo.  This
computational expedient excludes the possibility of interactions
between the disc population and a responsive bulge/halo component.  In
a direct check on the validity of this approximation, Sellwood (1980)
found it adequate while the disc remained nearly axisymmetric.  Once a
strong bar had formed, however, the (initially non-rotating)
bulge/halo population of particles began to gain angular momentum from
the disc at a prodigious rate.  Thus the coupling between the disc and
spheroidal populations of particles in {\it barred\/} galaxies ought
not to be ignored.

The interaction process can be described as a kind of dynamical
friction (Tremaine and Weinberg 1984a) between the rotating massive
object, the bar, and a background sea of collisionless particles.
Weinberg (1985) calculated that angular momentum should be exchanged
between the bar and a non-rotating spheroid at a rate consistent with
Sellwood's observed value.  Unfortunately, Sellwood's experiment did
not continue for very many bar tumbling periods and did not reveal the
long term consequences for both the bar and the halo population of a
very protracted period of interaction.  We therefore still do not know
whether the bar pattern speed will be reduced (or even raised),
whether its amplitude will be increased or diminished or how the
spheroid will appear once a substantial exchange of angular momentum
has occurred.

The only truly long term experiment to examine this issue has been
conducted by Little and Carlberg (1991), but for computational economy
they again used a two-dimensional model.  Their ``flat halo'' was an
entirely pressure supported population of particles co-planar with the
rotationally supported disc in which the bar formed.  They found that
the bar was indeed slowed by the angular momentum loss to the
``halo'', but a more realistic experiment is clearly warranted.

The predicted efficiency of angular momentum transfer is embarassingly
high, sufficient to arrest a rigid bar in a few rotations, yet a
number of lines of evidence suggest that bars in galaxies rotate
rapidly.  We are not, however, forced to an unattractive conclusion
that bars must either be very young or they have stopped rotating;
there are other ways to avoid this dilemma.  The most plausible seems
to us that Weinberg over-estimates the efficiency of exchange by
assuming a non-rotating spheroid.  The luminous bulges of galaxies
generally rotate sufficiently rapidly for their flattening to be
consistent with a rotating isotropic model (\eg\ Kormendy and
Illingworth 1982).  In the few cases that have been measured, bulge
streaming velocities are typically 40\% of those at the corresponding
distance in the disc and the angular momentum vector lies in the same
direction as that of the disc.  This degree of rotation will alter the
angular momentum exchanges with a bar, and may make them less
efficient.  It is not clear whether the observed angular momentum has
been acquired from the disc component already, or whether it was
primordial.  Weinberg himself also notes that the bar is
simultaneously spun up by interactions with the outer disc, and
speculates that it could act as a catalyst transferring angular
momentum from the disc to the spheroid while not changing its own
angular momentum content by much.  Finally, the bar is not a rigid
body and probably has a much lower moment of inertia (possibly even
negative) than that assumed by Weinberg.  These issues can be
addressed only by high quality, long term experiments.

\subsect{Destruction of bars}
Even three-dimensional bars seem to be long-lived; Pfenniger and
Friedli (1991) report that their bar models cease to evolve after the
thickening phase, and survive essentially unchanged as long as the
calculation is run.  Thus although the buckling instability (\S10.1)
dented the common perception that bars are rugged structures, it does
not seem to have invalidated it.

It is therefore interesting to ask whether a galaxy, having once
acquired a bar, could ever rid itself of it.  A few ideas for
achieving this have been mooted.

The most recent (Raha \etal\ 1991) is that the buckling instability
could be so violent that the bar is destroyed.  Raha \etal\ seemed to
find that the buckling instability was more violent when the bar
formed in a thinner disc, though in both their cases the bar was
merely weakened, not completely destroyed.  If ever the buckling
instability could be violent enough to destroy the bar, therefore, the
bar would have to have formed in an extremely thin, and probably very
young, disc.

Another suggestion is that bars could be destroyed during an
interaction with a companion galaxy.  It has generally been felt that
an interaction violent enough to destroy the bar would also destroy
the disk, but Pfenniger's (1991) report of a simulation in which a
dwarf was accreted by a barred galaxy seems to show that the bar can
be destroyed, and the companion disrupted, without doing very much
damage to the disc.

A further suggestion has been made by Norman and his co-workers
(Pfenniger and Norman 1990, Hasan and Norman 1990) that the growth of
a comparatively light, but dense, object at the centre can destroy the
bar.  They argue that the growth of central mass changes the major
orbit families which support the bar, and thereby threaten its
survival.  Friedli \etal\ (1991) present preliminary results which
seem to indicate that a bar can drive sufficient gas towards the
galactic centre to have this effect -- \ie\ that bars in gas rich
galaxies may self-destruct.

\sect{Conclusions}
It should be clear that we are still far from a complete understanding
of the basic structure of barred galaxies, but progress has been rapid
in recent years -- especially where orbit studies have been connected
to observational or simulated results.  There are a few aspects where
the observed facts seem to fit reasonably well with theoretical ideas.

On the bright side, it seems very likely that strong bars are formed
by the dynamical instability discussed in \S9.  Bars formed in the
$N$-body experiments end near co-rotation and have mass distributions
and kinematic properties which seem to correspond with those observed,
though more detailed comparisons would be desirable.  Moreover, such
bar models appear to have the right strength, and to rotate
sufficiently rapidly to shock gas in places resembling the dust lane
patterns in some barred galaxies.  The flow patterns described in \S6
are understood in terms of orbit theory and fit the observed
kinematics for entirely plausible model parameters.  We also think we
understand the origins of rings (\S7) sufficiently well to be able to
interpret them as signatures of major resonances with the bar pattern.

These successes add up to a compelling, but indirect, case for a bar
pattern speed in galaxies which places the major axis Lagrange point
just beyond the bar's end.  Such a value is consistent with the more
direct observational estimates of pattern speed (\S2.6) which
unfortunately are subject to large uncertainties.

Current observational work is beginning to provide more quantitative
data on the light distributions and kinematics of barred galaxies.  In
particular, the kind of comparison with theoretical models made by
Kent and Glaudell (1989) for NGC~936 should be extended to several
other galaxies in case the structure of that galaxy is special in any
way.  Moreover, the theory of ring formation is in dire need of more
detailed observational comparisons; we need good kinematic maps of
many more such galaxies to confirm that the rings do indeed lie at the
resonances for the bar.

Yet our understanding of the dynamical structure of bars is still far
from complete.  The vast literature on orbits in bar-like potentials
has led to a few major conclusions relevant to the structure of barred
galaxies: the most important is that the majority of stars within the
bar probably follow eccentric orbits which are trapped, or
semi-trapped, about the main family of orbits aligned with the bar
(the $x_1$ family).  Apart from this, knowledge of the main orbit
families has improved our understanding of gas flows and indicates
that it would be hopeless to try to construct a self-consistent bar
having a mean rotational streaming in a sense counter to the pattern
rotation.  It is progress, of a kind, to learn that a simple
generalization of the two-dimensional orbital structure does not
provide a complete description of three-dimensional bars.  Only one
set of approximately self-consistent two-dimensional numerical
solutions has been found (Pfenniger 1984b) and the prospects for
analytic models in the near future are bleak.  Next to this, our
fragmentary understanding of the evolution of barred galaxies, and
almost total blanks on the origins of lenses and the pronounced
asymmetries in some galaxies, seem of secondary concern.

Probably the most pressing need on the theoretical side is for a more
sustained attack on the orbital structure of three-dimensional bars,
preferably through studies of rapidly rotating three-dimensional
objects having density distributions resembling those seen in
galaxies.  It should be possible to address the evolutionary issues as
more powerful computers enable the quality of $N$-body simulations to
rise.

Other fundamental questions also need to be pursued.  The prime
candidate is what determines the $\sim30$\% fraction of galaxies seen
to have strong bars.  Since the bar instability seems able to create
strong bars in nearly all disc galaxies, how can we account for the
current moderate fraction?  The ideas for controlling the bar
instability (\S9.5) and those for destroying bars (\S10.5) do not
explain either why some $\sim70$\% of galaxies manage to avoid a bar
instability, or why the bar was subsequently weakened or destroyed in
that fraction of galaxies.  Are the weak bars in galaxies of
intermediate type (SAB) formed through the partial dissolution of
strong bars, or is some totally different mechanism required?

The bar instability we discuss in \S9 assumed an initially unstable
equilibrium disc without specifying how that could have been created.
A discussion of the formation of disc galaxies is well beyond the
scope of this review, but it does seem likely that the bar instability
would be profoundly affected by the manner in which galactic discs
form.  Sellwood and Carlberg (1984), in a few preliminary experiments
mimicking gradual disc formation, found that the velocity dispersion
of the stars rose sufficiently rapidly to inhibit the formation of a
strong bar as the disc mass built up.  Further experiments of this
kind, especially including gas dynamics seem warranted.

Another issue is whether it is sensible to separate a barred galaxy
into distinct dynamical components.  Most theoretical and
observational work has proceeded on the assumption that the bar and
the bulge are distinct, but we have seen in \S10 that there may be no
dynamical basis for considering them as separable.  Moreover, as it
seems likely that the bar formed from the disc, we may confuse
ourselves by trying to understand these also as unrelated components.

Although we have a lot more to do before we can claim to understand
the structure of these objects, we should be encouraged that progress
over the last few years has been rapid, especially since the knowledge
which enabled us to formulate many of these questions has only
recently been acquired.  The many new telescopes and observational
techniques, particularly operating in the infra-red, are likely to
advance the subject at its currently intense rate.

\blankline
\noindent {\bf Acknowledgments}  We would like to thank J Binney, D Earn and 
D Merritt, as well as E Athanassoula, A Bosma, R Buta, G Contopoulos,
D Friedli, J Gallagher, D Pfenniger, M Shaw, L Sparke and P Teuben for
comments on the manuscript.  We acknowledge support for brief visits
to ST ScI and Manchester as part of the effort to propel this review
forward.

{\refs
Adamson A J, Adams D J and Warwick R S 1987 \MNRAS {\bf 224} 367

Aoki S, Noguchi M and Iye M 1979 \PASJ {\bf 31} 737

Araki S 1985 \PhD MIT

Arsenault R 1989 \AAp {\bf 217} 66

Arsenault R, Boulesteix J, Georgelin Y and Roy J-R 1988 \AAp {\bf 200} 29

Athanassoula E 1983 {\it Internal Kinematics and Dynamics of Galaxies} IAU 
Symposium {\bf 100} ed E Athanassoula \Reidel\ p~243

------ 1984 \PhysRep {\bf 114} 320

------ 1990 {\it Galactic Models} {Annals of the New York Academy of 
Sciences} {\bf 596} ed J R Buchler, S T Gottesman and J H Hunter (New York: NY 
Academy of Sciences) p~181

------ 1992a \AAp in press

------ 1992b \AAp in press

Athanassoula E, Bienaym\'e O, Martinet L and Pfenniger D 1983 \AAp {\bf 127} 
349

Athanassoula E, Bosma A, Cr\'ez\'e M and Schwarz M P 1982 \AAp {\bf 107} 101

Athanassoula E, Bosma A and Papaioannou S 1987 \AAp {\bf 179} 23

Athanassoula E and Martinet L 1980 \AAp {\bf 87} L10

Athanassoula E, Morin S, Wozniak H, Puy D, Pierce M J, Lombard J and Bosma A 
1990 \MNRAS {\bf 245} 130

Athanassoula E and Sellwood J A 1986 \MNRAS {\bf 221} 213

Athanassoula E and Wozniak H 1992 Preprint

Baker N H, Moore D W and Speigel E A 1971 {\it Quart. J Mech. and Applied 
Math.} {\bf 24} 419

Baldwin J E, Lynden-Bell D and Sancisi R 1980 \MNRAS {\bf 193} 313

Ball R 1986 \ApJ {\bf 307} 453

Ball R, Sargent A I, Scoville N Z, Lo K Y and Scott S L 1985 \ApJL {\bf 298} 
L21

Bardeen J M 1975 {\it Dynamics of Stellar Systems} IAU Symposium {\bf 69} ed 
A Hayli \Reidel\ p~297

Barnes J E and Hernquist L E 1991 \ApJL {\bf 370} L65

Baumgart C W and Peterson C J 1986 \PASP {\bf 98} 56

Berry M V 1978 {\it Topics in Non-Linear Dynamics -- A Tribute to Sir Edward 
Bullard\/} AIP Conference series {\bf 46} ed S Jorna (New York: AIP) p~16

Bertin G 1983 {\it Internal Kinematics and Dynamics of Galaxies} IAU 
Symposium {\bf 100} ed E Athanassoula \Reidel\ p~119

Bettoni D 1989 \AJ {\bf 97} 79

Bettoni D, Fasano G and Galletta G 1990 \AJ {\bf 99} 1789

Bettoni D and Galletta G 1988 \Messenger {\bf 54} 51

Bettoni D, Galletta G and Osterloo T 1991 \MNRAS {\bf 248} 544

Bettoni D, Galletta G and Vallenari A 1988 \AAp {\bf 197} 69

Binney J 1981 \MNRAS {\bf 196} 455

Binney J, Gerhard O E and Hut P 1985 \MNRAS {\bf 215} 59

Binney J and Lacey C 1988 \MNRAS {\bf 230} 597

Binney J and Spergel D 1982 \ApJ {\bf 252} 308

------ 1984 \MNRAS {\bf 206} 159

Binney J and Tremaine S 1987 {\it Galactic Dynamics\/} (Princeton: Princeton 
University Press)

Blackman C P 1983 \MNRAS {\bf 202} 379

Blackman C P and Pence W D 1982 \MNRAS {\bf 198} 517

Block D L and Wainscoat R J 1991 \Nature {\bf 353} 48

Bosma A 1983 {\it Internal Kinematics and Dynamics of Galaxies\/} IAU 
Symposium {\bf 100} ed E Athanassoula \Reidel\ p~253

------ 1992 {\it Morphology and Physical Classification of Galaxies\/} ed G 
Longo, M Capaccioli and G Busarello \Kluwer\ p~207

Bottema R 1990 \AAp {\bf 233} 372

Broucke R A 1969 {\it Am. Inst. Aeron. Astronautics J.\/} {\bf 7} 1003

Burstein D 1979 \ApJS {\bf 41} 435

Buta R 1986a \ApJS {\bf 61} 609

------ 1986b \ApJS {\bf 61} 631

------ 1987 \ApJS {\bf 64} 383

------ 1988 \ApJS {\bf 66} 233

------ 1990a {\it Galactic Models\/} Annals of the New York Academy of 
Sciences {\bf 596} ed J R Buchler, S T Gottesman and J H Hunter (New York: NY
Academy of Sciences) p~58

------ 1990b \ApJ {\bf 356} 87

------ 1991 {\it Dynamics of Galaxies and their Molecular Cloud 
Distributions\/} IAU Symposium {\bf 146} ed F Combes and F Casoli \Kluwer\ 
p~251

Buta R and Crocker D A 1991 \AJ {\bf 102} 1715

Canzian B, Mundy L G and Scoville N Z 1988 \ApJ {\bf 333} 157

Casertano S and van Albada T S 1990 {\it Baryonic Dark Matter\/} ed D 
Lynden-Bell and G Gilmore \Kluwer\ p~159

Chandrasekhar S 1941 \ApJ {\bf 94} 511

Chevalier R A and Furenlid I 1978 \ApJ {\bf 225} 67

Chirikov B V 1979 \PhysRep {\bf 52} 265

Cleary P W 1989 \ApJ {\bf 337} 108

Colin J and Athanassoula E 1989 \AAp {\bf 214} 99

Combes F 1992 {\it Physics of Galaxies: Nature or Nurture\/}? ed C Balkowski 
(to appear)

Combes F, Debbasch F, Friedli D and Pfenniger D 1990 \AAp {\bf 233} 82

Combes F and Gerin M 1985 \AAp {\bf 150} 327

Combes F and Sanders R H 1981 \AAp {\bf 96} 164

Contopoulos G 1978 \AAp {\bf 64} 323

------ 1980 \AAp {\bf 81} 198

------ 1983a {\it Physica D\/} {\bf 8} 142

------ 1983b \AAp {\bf 117} 89

------ 1988 \AAp {\bf 201} 44

Contopoulos G and Barbanis B 1989 \AAp {\bf 222} 329

Contopoulos G and Grosb\o l P 1989 \AAp {\it Review\/} {\bf 1} 261

Contopoulos G and Magnenat P 1985 \CelMech {\bf 37} 387

Contopoulos G and Mertzanides C 1977 \AAp {\bf 61} 477

Contopoulos G and Papayannopoulos Th 1980 \AAp {\bf 92} 33

Contopoulos G and Vandervoort P O 1992 \ApJ {\bf 389} 118

Cowie L L 1980 \ApJ {\bf 236} 868

Cox D P and Reynolds R J 1987 \AnnRev {\bf 25} 303

Crane P C 1975 \ApJ {\bf 197} 317

Davidson R C, Chan H-W, Chen C and Lund S 1991 \RMP {\bf 63} 341

de Souza R E and dos Anjos S 1987 \AApS {\bf 70} 465

de Vaucouleurs G 1974 {\it Formation and Dynamics of Galaxies\/} IAU 
Symposium {\bf 58} ed J R Shakeshaft \Reidel\ p~335

------ 1975 \ApJS {\bf 29} 193

de Vaucouleurs G, de Vaucouleurs A and Corwin H G 1976 {\it Second Reference 
Catalogue of Bright Galaxies\/} (Austin: University of Texas Press) (RC2)

de Vaucouleurs G, de Vaucouleurs A and Freeman K C 1968 \MNRAS {\bf 139} 425

de Vaucouleurs G and Freeman K C 1972 \Vistas {\bf 14} 163

de Zeeuw T 1985 \MNRAS {\bf 216} 273 and 599

de Zeeuw T and Franx M 1991 \AnnRev {\bf29} 239

de Zeeuw T, Peletier R and Franx M 1986 \MNRAS {\bf 221} 1001

Disney M, Davies J and Phillipps S 1989 \MNRAS {\bf 239} 939

Dragt A J and Finn J M 1976 {\it J. Geophys. Res.\/} {\bf 81} 2327

Drury L O'C 1980 \MNRAS {\bf 193} 337

Duval M F and Athanassoula E 1983 \AAp {\bf 121} 297

Duval M F and Monnet G 1985 \AApS {\bf 61} 141

Duval M F, Monnet G, Boulesteix J, Georgelin Y, Le Coarer E and Marcelin M 
1991 \AAp {\bf 241} 375

Eddington A S 1915 \MNRAS {\bf 76} 37

Eder J, Giovanelli R and Haynes M P 1991 \AJ {\bf 102} 572

Efstathiou G, Lake G and Negroponte J 1982 \MNRAS {\bf 199} 1069

Elmegreen B G and Elmegreen D M 1985 \ApJ {\bf 288} 438

Elmegreen D M, Elmegreen B G and Bellin A D 1990 \ApJ {\bf 364} 415

England M N 1989 \ApJ {\bf 344} 669

England M N, Gottesman S T and Hunter J H 1990 \ApJ {\bf 348} 456

Erickson S A 1974 \PhD MIT

Ferrers N M 1877 {\it Quart. J Pure Appl. Math.\/} {\bf 14} 1

Freeman K C 1966 \MNRAS {\bf 134} 15

Fridman A M and Polyachenko V L 1984 {\it Physics of Gravitating Systems\/} 
(New York: Springer-Verlag)

Friedli D, Benz W and Martinet L 1991 {\it Dynamics of Disc Galaxies\/} ed B 
Sundelius (Gothenburg: G\"oteborgs University) p~181

Garcia-Barreto J A, Downes D, Combes F, Gerin M, Magri C, Carrasco L and 
Cruz-Gonzalez I 1991a \AAp {\bf 244} 257

Garcia-Barreto J A, Downes D, Combes F, Carrasco L, Gerin M and Cruz-Gonzalez 
I 1991b \AAp {\bf 252} 19

Gerhard O E and Vietri M 1986 \MNRAS {\bf 223} 377

Gerin M, Combes F and Athanassoula E 1990 \AAp {\bf 230} 37

Gerin M, Nakai N and Combes F 1988 \AAp {\bf 203} 44

Gingold R A and Monaghan J J 1977 \MNRAS {\bf 181} 375

Goldreich P and Lynden-Bell D 1965 \MNRAS {\bf 130} 125

Goldreich P and Tremaine S 1982 \AnnRev {\bf 20} 249

Goldstein H 1980 {\it Classical Mechanics} 2nd edition (Addison-Wesley)

Gottesman S T, Ball R, Hunter J H and Huntley J M 1984 \ApJ {\bf 286} 471

Hackwell J A and Schweizer F 1983 \ApJ {\bf 265} 643

Hadjidemetriou J D 1975 \CelMech {\bf 12} 255

Handa T, Nakai N, Sofue Y, Hayashi M and Fujimoto M 1990 \PASJ {\bf 42} 1

Hasan H and Norman C 1990 \ApJ {\bf 361} 69

Hawarden T G, Mountain C M, Leggett S K and Puxley P J 1986 \MNRAS {\it Short 
Communication\/} {\bf 221} 41p

Heisler J, Merritt D and Schwarzschild M 1982 \ApJ {\bf 258} 490

H\'enon M 1976 \CelMech {\bf 13} 267

H\'enon M 1983 {\it Chaotic Behaviour of Deterministic Systems} Les Houches 
Session XXXVI ed G Iooss, R H G Helleman and R Stora (Amsterdam: 
North-Holland) p~55

H\'enon M and Heiles C 1964 \AJ {\bf 69} 73

Hernquist L 1989 \Nature {\bf 340} 687

Hernquist L and Barnes J E 1990 \ApJ {\bf 349} 562

Hernquist L and Katz N 1989 \ApJS {\bf 70} 419

Hietarinta J 1987 \PhysRep {\bf 147} 87

Hill G W 1878 {\it Am. J. Math.\/} {\bf 1} 5

Hockney R W and Brownrigg D R K 1974 \MNRAS {\bf 167} 351

Hockney R W and Hohl F 1969 \AJ {\bf 74} 1102

Hohl F 1975 {\it Dynamics of Stellar Systems\/} IAU Symposium {\bf 69} ed A 
Hayli \Reidel\ p~349

------ 1978 \AJ {\bf 83} 768

Hubble E 1926 \ApJ {\bf 64} 321

Huizinga J E and van Albada T S 1992 \MNRAS {\bf 254} 677

Hummel E, Dettmar R J and Wielebinski R 1986 \AAp {\bf 166} 97

Hummel E, J\"ors\"ater S, Lindblad P O and Sandqvist A 1987a \AAp {\bf 172} 51

Hummel E, van der Hulst J M and Keel W C 1987b \AAp {\bf 172} 32

Hummel E, van der Hulst J M, Kennicutt R C and Keel W C 1990 \AAp {\bf 236} 
333

Hunter C 1970 {\it Studies in Applied Math\/} {\bf 49} 59

------ 1979 \ApJ {\bf 227} 73

------ 1992 {\it Astrophysical Disks} ed S F Dermott,
J H Hunter and R E Wilson (New York Academy of Sciences) (to appear)

Hunter C and de Zeeuw P T 1992 \ApJ {\bf 389} 79

Hunter C, de Zeeuw P T, Park Ch and Schwarzschild M 1990 \ApJ {\bf 363} 367

Hunter J H, Ball R, Huntley J M, England M N and Gottesman S T 1988 \ApJ {\bf 
324} 721

Huntley J M, Sanders R H and Roberts W W 1978 \ApJ {\bf 221} 521

Inagaki S Nishida M T and Sellwood J A 1984 \MNRAS {\bf 210} 589

Ishizuki S, Kawabe R, Ishiguro M, Okamura S K, Morita K-I, Chikada Y and 
Kasuga T 1990 \Nature {\bf 344} 224

James R A and Sellwood J A 1978 \MNRAS {\bf 182} 331

Jarvis B J 1986 \AJ {\bf 91} 65

Jarvis B J, Dubath P, Martinet L and Bacon R 1988 \AApS {\bf 74} 513

J\"ors\"ater S 1979 {\it Photometry, Kinematics and Dynamics of Galaxies\/} 
ed D S Evans (Austin: University of Texas Press) p~197

Julian W H and Toomre A 1966 \ApJ {\bf 146} 810

Kalnajs A J 1965 \PhD Harvard

------ 1971 \ApJ {\bf 166} 275

------ 1977 \ApJ {\bf 212} 637

------ 1978 {\it Structure and Properties of Nearby Galaxies\/} IAU 
Symposium {\bf 77} ed E M Berkhuisjen and R Wielebinski \Reidel\ p~113

------ 1987 {\it Dark Matter in the Universe\/} IAU Symposium {\bf 117} 
ed J Kormendy and G R Knapp \Reidel\ p~289

Kenney J D P 1991 {\it Dynamics of Disc Galaxies\/} ed B Sundelius 
(Gothenburg: G\"oteborgs University) p~171

Kent S M 1987 \AJ {\bf 93} 1062

------ 1990 \AJ {\bf 100} 377

Kent S M and Glaudell G 1989 \AJ {\bf 98} 1588

Kormendy J 1979 \ApJ {\bf 227} 714

------ 1981 {\it The Structure and Evolution of Normal Galaxies\/} Eds 
Fall SM and Lynden-Bell D (Cambridge: Cambridge University Press)

------ 1982 {\it Morphology and Dynamics of Galaxies\/} 12th Advanced 
Course of the Swiss Society of Astronomy and Astrophysics at Saas Fee Eds 
Martinet L and Mayor M (Sauverny: Geneva Observatory)

------ 1983 \ApJ {\bf 275} 529

Kormendy J and Illingworth G 1982 \ApJ {\bf 256} 460

Kulsrud R M, Mark J W-K and Caruso A 1971 \ApSS {\bf 14} 52

Kuz'min G G 1956 {\it Astr. Zh.\/} {\bf 33} 27

Landau L D and Lifshitz E M 1987 {\it Fluid Mechanics\/} 2nd edition (Oxford: 
Pergamon)

Liapunov A M 1907 Reprinted 1947 in {\it Ann. Math Studies} {\bf 17} 
(Princeton)

Lichtenberg A J and Lieberman M A 1983 {\it Regular and Stochastic Motion\/} 
Applied Math Sciences {\bf 38} (New York: Springer-Verlag)

Lin C C 1975 {\it Structure and Evolution of Galaxies} ed G Setti \Reidel\ 
p~119

Lin C C and Bertin G 1985 {\it The Milky Way Galaxy\/} IAU Symposium {\bf 
106} ed H van Woerden, R J Allen and W B Burton \Reidel\ p~513

Lindblad B 1927 \MNRAS {\bf 87} 553

Lindblad P O and J\"ors\"ater S 1987 {\it Evolution of Galaxies\/} 
Proceedings of the 10th IAU Regional Astronomy Meeting vol 4 ed J Palou\u s
{\it Publ. Astr. Inst. Czech. Acad. Sci.\/} {\bf 69} 289

Little B and Carlberg 1991 \MNRAS {\bf 250} 161

Long K 1992 Preprint

Lord S D and Kenney J D P 1991 \ApJ {\bf 381} 130

Louis P D and Gerhard O E 1988 \MNRAS {\bf 233} 337

Lucy L B 1977 \AJ {\bf 82} 1013

Lynden-Bell D 1962a \MNRAS {\bf 124} 1

------ 1962b \MNRAS {\bf 124} 95

------ 1979 \MNRAS {\bf 187} 101

McGill C and Binney J 1990 \MNRAS {\bf 244} 634

MacKay R S, Meiss J D and Percival I C 1984 {\it Physica D\/} {\bf 13} 55

Magnenat P 1982a \AAp {\bf 108} 89

------ 1982b \CelMech {\bf 28} 319

Marcelin M and Athanassoula E 1982 \AAp {\bf 105} 76

Martinet L 1984 \AAp {\bf 132} 381

Martinet L and Pfenniger D 1987 \AAp {\bf 173} 81

Martinet L and Udry S 1990 \AAp {\bf 235} 69

Martinet L and de Zeeuw T 1988 \AAp {\bf 206} 269

Mathews J and Walker R L 1970 {\it Mathematical Methods of
Physics\/} 2nd edition (Menlo Park: Benjamin/ \break Cummings)

Matsuda T and Isaka H 1980 {\it Prog. Theor. Phys.\/} {\bf 64} 1265

Merritt D and Hernquist L 1991 \ApJ {\bf 376} 439

Merritt D and Stiavelli M 1990 \ApJ {\bf 358} 399

Miller R H and Prendergast K H 1968 \ApJ {\bf 151} 699

Miller R H, Prendergast K H and Quirk W J 1970 \ApJ {\bf 161} 903

Miller R H and Smith B F 1979 \ApJ {\bf 227} 785

Miralda-Escud\'e J and Schwarzschild M 1989 \ApJ {\bf 339} 752

Miyamoto M and Nagai R 1975 \PASJ {\bf 27} 533

Moser J 1973 {\it Stable and Random Motions in Dynamical Systems} Annals of 
Math Studies {\bf 77} (Princeton: Princeton University Press)

Mulder W A 1986 \AAp {\bf 156} 354

Mulder W A and Hooimeyer J R A 1984 \AAp {\bf 134} 158

Newton A J and Binney J 1984 \MNRAS {\bf 210} 711

Nilson P 1973 {\it Uppsala General Catalogue of Galaxies\/} (= Acta 
Upsaliensis Ser V: A Vol I)

Noguchi M 1987 \MNRAS {\bf 228} 635

Noguchi M 1988 \AAp {\bf 203} 259

Odewahn S C 1991 \AJ {\bf 101} 829

Ohta K, Hamabe M and Wakamatsu K-I 1990 \ApJ {\bf 357} 71

Okamura S 1978 \PASJ {\bf 30} 91

Ondrechen M P 1985 \AJ {\bf 90} 1474

Ondrechen M P and van der Hulst J M 1989 \ApJ {\bf 342} 29

Ondrechen M P, van der Hulst J M and Hummel E 1989 \ApJ {\bf 342} 39

Ostriker J P and Peebles P J E 1973 \ApJ {\bf 186} 467

Pannatoni R F 1983 {\it Geophys. Astrophys. Fluid Dynamics\/} {\bf 24} 165

Papayannopoulos Th 1979 \AAp {\bf 77} 75

Papayannopoulos T and Petrou M 1983 \AAp {\bf 119} 21

Pence W D and Blackman C P 1984a \MNRAS {\bf 207} 9

------ 1984b \MNRAS {\bf 210} 547

Pence W D, Taylor K, Freeman K C, de Vaucourlers G and Atherton P 1988 \ApJ 
{\bf 326} 564

Percival I C 1979 {\it Nonlinear Dynamics and the Beam-Beam Interaction\/} 
AIP Conference series {\bf 57} ed M Month and J C Herrera (New York: AIP) 
p~302

Peterson C J, Rubin V C, Ford W K and Thonnard N 1978 \ApJ {\bf 219} 31

Petrou M and Papayannopoulos T 1986 \MNRAS {\bf 219} 157

Pfenniger D 1984a \AAp {\bf 134} 373

------ 1984b \AAp {\bf 141} 171

------ 1990 \AAp {\bf 230} 55

------ 1991 {\it Dynamics of Disc Galaxies\/} ed B Sundelius 
(Gothenburg: G\"oteborgs University) p~191

Pfenniger D and Friedli D 1991 \AAp {\bf 252} 75

Pfenniger D and Norman C 1990 \ApJ {\bf 363} 391

Pierce M J 1986 \AJ {\bf 92} 285

Planesas P, Scoville N and Myers S T 1991 \ApJ {\bf 369} 364

Poincar\'e 1892 {\it Les Methods Nouvelles de la Mechanique Celeste\/} 
(Paris: Gauthier-Villars)

Polyachenko V L 1989 \SovAstL {\bf 15} 385

Polyachenko V L and Shukhman I G 1981 \SovAst {\bf 25}, 533

Prendergast K H 1962 {\it Interstellar Matter in Galaxies\/} ed L Woltjer 
(Benjamin) p~217

Prendergast K H 1983 {\it Internal Kinematics and Dynamics of Galaxies\/} IAU 
Symposium {\bf 100} ed E Athanassoula \Reidel\ p~215

Raha N 1992 \PhD University of Manchester

Raha N, Sellwood J A, James R A and Kahn F D 1991 \Nature {\bf 352} 411

Ratcliff S J, Chang K M and Schwarzschild M 1984 \ApJ {\bf 279} 610

Richstone D O 1987 {\it Structure and Dynamics of Elliptical Galaxies\/} IAU 
Symposium {\bf 127} ed T de Zeeuw \Reidel\ p~261

Roberts W W, Huntley J M and van Albada G D 1979 \ApJ {\bf 233} 67

Saha P 1991 \MNRAS {\bf 248} 494

Sancisi R, Allen R J and Sullivan W T 1979 \AAp {\bf 78} 217

Sandage A 1961 {\it Hubble Atlas of Galaxies\/} Publication 618 (Washington: 
Carnegie Institute of Washington)

Sandage A and Brucato R 1979 \AJ {\bf 84} 472

Sandage A and Tammann G A 1981 {\it A Revised Shapley-Ames Catalogue of 
Bright Galaxies\/} Publication 635 (Washington: Carnegie Institute of 
Washington) (RSA)

Sanders R H 1977 \ApJ {\bf 217} 916

Sanders R H and Huntley J M 1976 \ApJ {\bf 209} 53

Sanders R H and Prendergast K H 1974 \ApJ {\bf 188} 489

Sanders R H and Tubbs A D 1980 \ApJ {\bf 235} 803

Sandqvist Aa, Elfhag T and J\"ors\"ater S 1988 \AAp {\bf 201} 223

Schempp W V 1982 \ApJ {\bf 258} 96

Schommer R A, Caldwell N, Wilson A S, Baldwin J A, Phillips M M, Williams T B 
and Turtle A J 1988 \ApJ {\bf 324} 154

Schwarz M P 1979 \PhD Australian National University

------ 1981 \ApJ {\bf 247} 77

------ 1984a \MNRAS {\bf 209} 93

------ 1984b \AAp {\bf 133} 222

Schwarzschild M 1979 \ApJ {\bf 232} 236

Scoville N Z, Matthews K, Carico D P and Sanders D B 1988 \ApJL {\bf 327} L61

Sellwood J A 1980 \AAp {\bf 89} 296

------ 1981 \AAp {\bf 99} 362

------ 1983 \JCP {\bf 50} 337

------ 1987 \AnnRev {\bf 25} 151

------ 1989 \MNRAS {\bf 238} 115

------ 1991 {\it Dynamics of Disk Galaxies\/} ed B Sundelius 
(Gothenburg: G\"oteborgs University) p~123

Sellwood J A and Athanassoula E 1986 \MNRAS {\bf 221} 195 (with a Corrigendum 
in {\bf 229} 707)

Sellwood J A and Carlberg R G 1984 \ApJ {\bf 282} 61

Sellwood J A and Sparke L S 1988 \MNRAS {\it Short Communication\/} {\bf 231} 
25p

Shaw M A 1987 \MNRAS {\bf 229} 691

Shaw M A, Combes F, Axon D and Wright G 1992 \AAp submitted

Shu F H 1970 \ApJ {\bf 160} 89

Simkin S M, Su H J and Schwarz M P 1980 \ApJ {\bf 237} 404

S\o rensen S-A, Matsuda T and Fujimoto M 1976 \ApSS {\bf 43} 491

Sparke L S 1990 {\it Dynamics and Interactions of Galaxies\/} ed R Wielen 
(Berlin: Springer-Verlag) p~338

Sparke L S and Sellwood J A 1987 \MNRAS {\bf 225} 653

Spitzer L 1990 \AnnRev {\bf 28} 71

Statler T S 1987 \ApJ {\bf 321} 113

St\"ackel P 1890 {\it Math. Ann.\/} {\bf 35} 91

Teuben P J 1987 \MNRAS {\bf 227} 815

Teuben P J and Sanders R H 1985 \MNRAS {\bf 212} 257

Teuben P J, Sanders R H, Atherton P D and van Albada G D 1986 \MNRAS {\bf 
221} 1

Thronson H A, Hereld M, Majewski S, Greenhouse M, Johnson P, Spillar E, 
Woodward C E, Harper D A and Rauscher B J 1989 \ApJ {\bf 343} 158

Tilanus R 1990 \PhD Groningen University

Toomre A 1966 {\it Geophysical Fluid Dynamics\/} notes on the 1966 Summer 
Study Program at the Woods Hole Oceanographic Institution, ref no 66-46 p~111

------ 1977 \AnnRev {\bf 15} 437

------ 1981 {\it Structure and Evolution of Normal Galaxies\/} ed S M Fall 
and D Lynden-Bell (Cambridge: Cambridge University Press) p~111

Tremaine S 1989 {\it Dynamics of Astrophysical Discs\/} ed J A Sellwood 
(Cambridge: Cambridge University Press) p~231

Tremaine S and Weinberg M D 1984a \MNRAS {\bf 209} 729

------ 1984b \ApJL {\bf 282} L5

Tsikoudi V 1980 \ApJS {\bf 43} 365 (see also \ApJ {\bf 234} 842)

Tubbs A D 1982 \ApJ {\bf 255} 458

Udry S 1991 \AAp {\bf 245} 99

Udry S and Pfenniger D 1988 \AAp {\bf 198} 135

Valentijn E A 1990 \Nature {\bf 346} 153

van Albada G D, van Leer B and Roberts W W 1982 \AAp {\bf 108} 76

van Albada T S 1986 {\it Use of Supercomputers in Stellar Dynamics\/} ed S 
McMillan and P Hut (New York: Springer-Verlag)

van Albada T S and Sancisi R 1986 \PhilTrans {\bf 320} 447

van Albada T S and Sanders R H 1983 \MNRAS {\bf 201} 303

Vandervoort P O 1979 \ApJ {\bf 232} 91

------- 1980 \ApJ {\bf 240} 478

------ 1991 \ApJ {\bf 377} 49

van Driel W, Rots A H and van Woerden H 1988 \AAp {\bf 204} 39

Wakamatsu K-I and Hamabe M 1984 \ApJS {\bf 56} 283

Weinberg M D 1985 \MNRAS {\bf 213} 451

------ 1991 \ApJ {\bf 368} 66

Whitmore B C, Lucas R A, McElroy D B, Steiman-Cameron T Y, Sackett P D and 
Olling R P 1990 \AJ {\bf 100} 1489

Woolley R 1965 {\it Galactic Structure\/} ed A Blaauw and M Schmidt (Chicago: 
University of Chicago Press) p~85

Wozniak H and Athanassoula E 1992 \AAp (in press)

Wozniak H and Pierce M J 1991 \AApS {\bf 88} 325

Zang T A 1976 \PhD MIT

Zang T A and Hohl F 1978 \ApJ {\bf 226} 521

}

\end